\documentclass[eqsecnum,twocolumn,aps,tightenlines,floats,prd,showpacs]{revtex4-1}%%
\def\bea{\begin{eqnarray}}
\def\eea{\end{eqnarray}}
\def\bal{\begin{align}}
\def\eal{\end{align}}

\def\eqqA{3.30}
\def\eqq2{3.31}

\def\sfrac#1#2{{\textstyle \frac{#1}{#2}}}

\usepackage{graphics}
\usepackage{graphicx}
\usepackage{epsf} 
\usepackage{amsmath}
\usepackage{amssymb}
\usepackage{bbold}
\usepackage{slashed}
\usepackage{appendix}
\usepackage{bm}
\usepackage{dcolumn}
\usepackage{footnote} 
\begin{document}  

%\begin{widetext}

%\preprint{
\phantom{0}
%\vspace{-0.4in}
\hspace{5.5in}\parbox{1.5in}{ \leftline{JLAB-THY-14-1864}
%                \leftline{WM-05-???}
%			             \leftline{nucl-th/05?????}
                \leftline{}\leftline{}\leftline{}\leftline{}
%\vspace{-3.6in}  % moves the preprint box down
}
%}
\title
{\bf  Covariant Spectator Theory of $np$ scattering: \\Deuteron magnetic moment and form factors}

\author{Franz Gross$^{1,2}$ }
\email{email address: gross@jlab.org}
\affiliation{
$^1$Thomas Jefferson National Accelerator Facility, Newport News, VA 23606 \vspace{-0.15in}}
\affiliation{
$^2$College of William and Mary, Williamsburg, Virginia 23185}

\date{\today}

\begin{abstract} 

The deuteron magnetic moment is calculated using two model wave functions obtained from 2007 high precision fits to $np$ scattering data. Included in the calculation are a new class of isoscalar $np$ interaction currents which are  automatically generated by the nuclear force model used in these fits.  After normalizing the wave functions, nearly identical predictions are obtained: model WJC-1, with larger relativistic P-state components, gives 0.863(2), while model WJC-2 with very small P-state components gives 0.864(2)  
These are about 1\% larger than the measured value of the moment, 0.857 n.m., giving a new prediction for the size of the $\rho\pi\gamma$ exchange, and other purely transverse  interaction currents that are largely unconstrained by the nuclear dynamics.   The physical significance of these results is discussed, and  general formulae for the deuteron form factors, expressed in terms of deuteron wave functions and a new class of interaction current wave functions, are given.


\end{abstract}
 
\phantom{0}
\vspace{0.76in}
%\vspace{-6in}
\vspace*{-0.1in}  % sets how far the title is below the preprint box

\maketitle

%%%%%%%%%%%%%%%%%%%%%%%%%%%%%%%%%%%%%%

\section{Introduction, summary, and conclusions} 
\label{sec:intro}

\subsection{Background}

This work is the second in a series of  four planned papers (the first, referred to as Ref.\ I  \cite{RefI}, accompanies this paper) that will present the fourth generation calculation of the deuteron form factors  using what is now called the Covariant Spectator Theory (CST) \cite{Gross:1969rv,Gross:1972ye,Gross:1982nz}.  

This new generation of calculations is required new fits to the 2007 $np$ data base \cite{Gross:2008ps}  obtained using the CST with a one boson exchange (OBE) kernel.  It was found that a high precision fit (one with $\chi^2$/datum $\simeq1$) was possible only if  the $NN\sigma_0$ vertices associated with the exchange of a scalar-isoscalar meson $\sigma_0$  included momentum dependent  terms in the form 
\bea
\Lambda^{\sigma_0}(p,p')= 
g_{\sigma_0}{ \bf 1}-\nu_{\sigma_0}\big[\Theta(p)+\Theta(p')\big]\qquad
\label{eq:1.2}
\eea
where $\nu_{\sigma_0}$ is a new parameter  determined by fitting the $NN$ scattering data, $p$ and $p'$ are the four-momenta of the outgoing and incoming nucleons, respectively, and the $\Theta$ are projection operators 
\bea
\Theta(p)=\frac{m-\slashed{p}}{2m} \, ,\label{eq:theta}
\eea
which are non-zero for off-shell particles, and hence are a feature of Bethe-Salpeter or CST equations.

Two high precision models were found with somewhat different properties. Model WJC-1, designed to give the best fit possible, has 27 parameters, $\chi^2/{\rm datum}\simeq 1.06$, and a large $\nu_{\sigma_0}=-15.2$.  Model WJC-2, designed to give a excellent fit with as few parameters as possible,  has only 15 parameters, $\chi^2/{\rm datum}\simeq 1.12$, and a smaller $\nu_{\sigma_0}=-2.6$.  Both models also predict the correct triton binding energy.  The deuteron wave functions predicted by both of these models \cite{arXiv:1007.0778} have small P-state components of relativistic origin, and the normalization of the wave functions includes a term coming from the energy dependence of the kernel, which contributes $-5.5\%$ for WJC-1 and $-2.3\%$ for WJC-2.

This momentum dependence of the kernel 
implies the existence of a  new class of $np$ isoscalar interaction currents that will contribute to the electromagnetic interaction of the deuteron.  These currents were fixed in Ref.\ I, and this paper completes the derivation started there by decomposing the deuteron current in into three independent form factors \cite{Garcon:2001sz,Gilman:2001yh} and expressing each of these form factors in terms of integrals over bilinear products of four invariant functions, or  alternatively, the two familiar nonrelativistic S and D-state wave functions, $u$ and $w$ and the two small P-state components, $v_t$ and $v_s$ \cite{BC,Buck:1979ff}.  This paper also discusses the contributions of the interaction currents to the charge and the magnetic moment.  Calculation of the quadrupole moment and the dependence of the form factors on the momentum transfer of the scattered electron, $Q^2$, will be discussed in the remaining two papers, under preparation.

\subsection{Organization of the paper}
\label{sec:org}

This paper is long and detailed, so the principal results and conclusions have been extracted and  summarized in this section.  The interaction current makes significant contributions to the wave function normalization (the charge) and these are reviewed in some detail in Sec.~\ref{sec:CandN}.  Then, one of the principal new results of this paper, the calculation of the deuteron magnetic moment including the contributions from the interaction current, are presented in Sec.~\ref{sec:magdis}.  Conclusions are given in Sec.~\ref{sec:conclusions}.

The remainder of the paper includes four more sections and seven appendices where are the details are presented. The two-body current from which all of the results are derived is introduced in Sec.~\ref{sec:wave}.  The entity that contains the relativistic structure of the deuteron is  the $dnp$ vertex function with one nucleon on-shell.  In Sec.~\ref{sec:wave}  this vertex function is written as a sum of products of scalar invariant functions multiplied by  covariant Dirac spin operators.  This  expansion in terms of invariants was first introduced by  Blanckenbecler and Cook in  1960 \cite{BC}, but we use  the notation of Ref.~\cite{Buck:1979ff}.   Appendix \ref{app:H} shows how to expand these invariant functions in terms of  the CST deuteron wave functions $u, w, v_t,$ and $v_s$ (previously reported in the literature), and $\chi_{\ell}=\{z_0^{--}, z_1^{--},z_0^{-+},z_1^{-+}\}$, the negative $\rho$-spin helicity amplitudes for particle 1.  The $\chi_{\ell}$ are not zero even when both particles are on shell (as shown in Appendix \ref{app:offshellterms}), and  are needed for a complete calculation of the magnetic moment.

Next, Sec.~\ref{sec:formfactors} describes how the deuteron form factors are extracted from the helicity amplitudes of the deuteron current, and  general formulae for the form factors, valid to all $Q^2$,  are assembled.  The final results, Eqs.~(\ref{eq:A&V2-3}) and (\ref{eq:B&V1-2}), give the form factors as a sum of products of the nucleon form factors $F_i(Q^2)$ (with $i=1,2,3$, with $F_3$ a new nucleon from factor that contributes to the nucleon current only when both the incoming and outgoing  nucleons are off-shell) multiplied by body form factors expressed an integrals over traces of bilinear products of invariant functions from which the $dnp$ vertex is constructed.  The interaction current contributions are conveniently expressed in terms of two new types of wave functions, $\Psi^{(2)}$ and $\widehat \Psi$, and some details of the computation of these wave functions is given in Appendix \ref{app:truncwf}.    Explicit formulae for the 18 independent traces that appear in the final results are given in Appendix \ref{app:B}.  The formulae are manifestly covariant; once the rest frame wave functions are known these formulae reduce the calculation of the deuteron form factors at any $Q^2$ to quadratures.    These formulae will be used to calculate the form factors in the fourth paper of this series, and are one of the principal new results of this paper.

Finally, the last two sections discuss how the charge (Sec.~\ref{sec:GC0}) and magnetic moment (Sec.~\ref{sec:GM0}) are built up from individual contributions from the wave function components, the off-shell nucleon current, and the interaction current.  These sections assemble details given in Appendices \ref{app:staticdetails} --  \ref{app:magmoment}.  This work is summarized in the following Secs.~\ref{sec:CandN} and \ref{sec:magdis}.

\subsection{Charge and Normalization} 
\label{sec:CandN}

%%%%%%%%%%%%%%%%%%%%%%%%%%%%%%%%%%%%%%%%%%%%%%%%%%
\begin{table}[t]
\begin{minipage}{3in}
\caption{Contributions to the normalization sum (\ref{eq:norm3}) for model WJC-1.  All entries are rounded to three decimal places; all totals are subject to round-off error.  Note that  the total of columns four and five  equals the total in column six, confirming (\ref{eq:Psum}). } 
\label{tab:charge}
\begin{ruledtabular}
\begin{tabular}{lrrrr|r}
%& \multicolumn{2}{c}{$^1S_0$} & \multicolumn{2}{c}{$^3S_1$} \cr
$z_\ell$ & $z_\ell^2\quad$ & $\quad a_\ell z_\ell^2$   & $z_\ell\widehat{z}_\ell$  & $z_\ell z^{(2)}_\ell\quad$   & $\displaystyle{\left<\frac{\partial \tilde V}{\partial P_0}\right>}$   \cr
\tableline
$u$ & 0.974   & 0.014  &  $-$0.035 & $-$0.020$\quad$ &  $-$0.054 \cr  
$w$  & 0.077  &  0.022     &  $-$0.017  & $-$0.002$\quad$ &       $-$0.019    \cr
 $v_t$   &    0.001   &   $-$0.003  &    $-$0.007 &  $-$0.001$\quad$ &    $-$0.007     \cr
  $v_s$ &  0.002 & $-$0.008  &   0.001  &  $-$0.001$\quad$ &   0.000 \cr
  \tableline
  total & 1.055  & 0.025 &  $-$0.057 &  $-$0.023$\quad$  &   $-$0.080
%vspace{1in} 
\end{tabular}
\end{ruledtabular}
\end{minipage}
\end{table}
%
%%%%%%%%%%%%%%%%%%%%%%%%%%%%%%%%%%%%%%%%%%%%%%%%%%

%%%%%%%%%%%%%%%%%%%%%%%%%%%%%%%%%%%%%%%%%%%%%%%%%%
\begin{table}[t]
\begin{minipage}{3in}
\caption{Contributions to the normalization sum (\ref{eq:norm3}) for model WJC-2 (see caption to Table \ref{tab:charge}). } 
\label{tab:charge2}
\begin{ruledtabular}
\begin{tabular}{lrrrr|r}
%& \multicolumn{2}{c}{$^1S_0$} & \multicolumn{2}{c}{$^3S_1$} \cr
$z_\ell$ & $z_\ell^2\quad$ & $\quad a_\ell z_\ell^2$   & $z_\ell\widehat{z}_\ell$  & $z_\ell z^{(2)}_\ell\quad$   & $\displaystyle{\left<\frac{\partial \tilde V}{\partial P_0}\right>}$   \cr
\tableline
$u$ & 0.957   & 0.007  &  $-$0.022 & $-$0.012$\quad$ &  $-$0.034 \cr  
$w$  & 0.065  &  0.011     &  $-$0.010  & 0.001$\quad$ &       $-$0.009    \cr
 $v_t$   &    0.000   &   0.000  &    0.002 &  0.000$\quad$ &    0.002     \cr
  $v_s$ &  0.000 & 0.000  &   0.000  &  0.000$\quad$ &   0.000 \cr
  \tableline
  total & 1.023  & 0.018 &  $-$0.030 &  $-$0.011$\quad$  &   $-$0.041
%vspace{1in} 
\end{tabular}
\end{ruledtabular}
\end{minipage}
\end{table}
%
%%%%%%%%%%%%%%%%%%%%%%%%%%%%%%%%%%%%%%%%%%%%%%%%%%

The normalization condition ensures that the charge of the deuteron is one.  There are many ways to write this condition; Sec.~\ref{sec:GC0} express the contributions from the interaction currents in terms of two new wave functions, $\Psi^{(2)}$, a wave function that depends only on the $\Theta$ contributions from off-shell particle 2, and $\widehat{\Psi}$, a wave function with {\it both\/} particles off-shell, which, because of the interaction current contributions, reduces to $\Psi$ when particle 1 is on-shell.  In this language, the normalization condition (charge) can be expressed as a sum of contributions from the components of $\Psi$, $\widehat{\Psi}$, and $\Psi^{(2)}$:
\bea
1=\int_0^\infty k^2dk\sum_{\ell=1}^4 \big\{1+a_\ell(k)\big\} z_\ell^2  + \left<\frac{\partial\widetilde V}{\partial P_0}\right>\qquad \label{eq:norm3}
\eea
where the notation $z_\ell =z_\ell(k)$ is used generically to denote the  wave functions $u, w, v_t$, or $v_s$ [not to be confused with the {\it helicity amplitudes\/} denoted by $z_\ell^{\rho_1\rho_2}$ and given in Eq.~(\ref{eq:helwf})] with $\ell$ denoting the angular momentum of the state (so that $z_0=u$, $z_2=w$, and $z_1=v_t$ or $v_s$).   In Sec.~\ref{sec:GC0} it is shown how the derivative of the reduced kernel can expressed in terms of products involving the new wave functions
\bea
\left<\frac{\partial\widetilde V}{\partial P_0}\right>&=&\int_0^\infty k^2dk\sum_{\ell=1}^4\Big\{\left<z_\ell z_\ell^{(2)}\right> +\left< z_\ell \widehat{z}'_\ell\right>\Big\}\qquad \label{eq:Psum}
\eea
and the contributions from the derivative of the strong form factor contribute terms proportional to $a_\ell(k)$, with
\bea
a_\ell(k)&=&\begin{cases} -4a(p^2)(E_k-m_d)\delta_k & \ell=0,2\cr +4a(p^2)(E_k-m_d)m_d &\ell=1 \end{cases}
\eea
where $a(p^2)$ was defined in Eq.~(\ref{4.7}) with $p^2=m_d^2+m^2-2m_dE_k$ here, and $\delta_k=2E_k-m_d$.
The budget for these contributions is shown in Tables~\ref{tab:charge} and \ref{tab:charge2}, where all contributions have been rounded to three decimal places.   

Note that, except for the P-state contributions from Model WJC-2, all of these contributions are important at the level of 0.001.  If the magnetic moment is to be calculated to this accuracy (a goal of this paper), then all of these terms must be included.

%%%%%%%%%%%%%%%%%%%%%%%%%%%%%%%%%%%%%%%%%%%%%%%%%%
\begin{table}[t]
\begin{minipage}{3.2in}
\caption{Integrated products of wave functions for Model WJC-1 with the largest P-states.  Entries above the diagonal are the products $z_\ell z_{\ell'}$; those along the diagonal and below are products weighted by $(E_k-m)/E_k$.  } 
\label{tab:moments}
\begin{ruledtabular}
\begin{tabular}{lrrrr}
 & $u\quad$   & $w\quad$  & $v_t$  & $v_s$   \cr
\tableline
$u$ & 0.007    &  0.094 & $-$0.004 & --- \cr
$w$  & $-$0.001       &  0.006  & $-$0.009 &     $-$0.010    \cr
 $v_t$  &    ---        &    $-$0.001 &  ---&   0.001     \cr
  $v_s$ & 0.001  &   $-$0.001  &  --- &  --- 
\end{tabular}
\end{ruledtabular}
\end{minipage}
\end{table}
%
%%%%%%%%%%%%%%%%%%%%%%%%%%%%%%%%%%%%%%%%%%%%%%%%%%

\subsection{Magnetic Moment}\label{sec:magdis}

%%%%%%%%%%%%%%%%%%%%%%%%
\begin{figure*}
%\vspace*{-1in}
%\begin{center}
\centerline{
\mbox{
\includegraphics[width=6.5in]{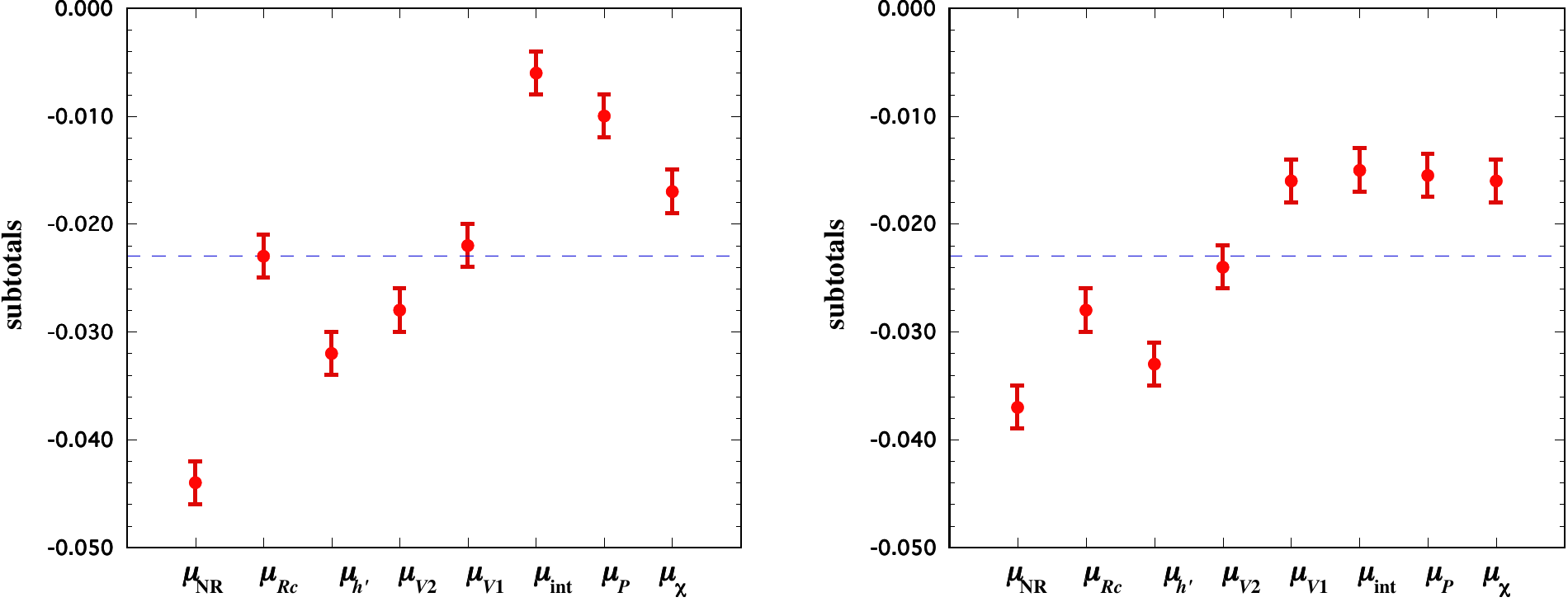}
}
}
%\end{center}
%\vskip 2.0in
%\vspace*{ -0.4in}
\caption{\footnotesize\baselineskip=10pt (Color on line) Running sum of the corrections to the magnetic moment, in the order that they are listed in Tables \ref{tab:magmtnames} and \ref{tab:magmt}.  The dashed line is $\Delta\mu_d=-0.023$, the correction needed to give the experimental value.  The error bars are $\pm0.002$, an estimate of the size of the terms missing from the approximation of Eq.~(\ref{eq:magapprox}).  Model WJC-1 (left panel) and Model WJC-2 (right panel).  % of the magnetic moment.
}
\label{fig:two}
\end{figure*} 
%%%%%%%%%%%%%%%%%%%%%%%%

The algebraic  expression for the magnetic moment  is considerably more complicated than the simple form  (\ref{eq:norm3}) for the charge.  While it is possible to calculate the exact result from the formulae given in the Appendices, this will not give much insight into the underlying physics.  The goal in this paper is to simplify these formulae, retaining all terms that contribute to 1-2 parts per 1000.

Table \ref{tab:moments} will be used to guide the calculation.  It suggests that sufficient accuracy is obtained if the coefficients of all terms but those involving products of the leading wave functions, namely $u$ and $w$, are retained to leading order in the small parameter $\delta_E=(E_k-m)/E_k$ (a few of the other terms  are as large as 0.001, but neglecting all of these corrections is not expected to change the results significantly, and all terms of higher order in $\delta_E$ are negligible). 
Guided by these results  the formulae for the magnetic moment are simplified. 

%%%%%%%%%%%%%%%%%%%%%%%%%%%%%%%%%%%%%%%%%%%%%%%%%%
\begin{table}[t]
\begin{minipage}{2.8in}
\caption{Physical origin of the eight different types of corrections that contribute to the magnetic moment. } 
\label{tab:magmtnames}
\begin{ruledtabular}
\begin{tabular}{c|l}
term  & physical origin   \cr
\tableline  %[0.001in]
$\mu_{\rm NR}$ & nonrelativistic D-state contribution   \cr
 $\mu_{\it Rc}$ & relativistic corrections to S,D terms \cr
 $\mu_{h'}$ & dependence on the strong form factor, $h$ $\quad$ \cr
 $\mu_{V_2}$ & interaction currents: off-shell particle 2 \cr
 $\mu_{V_1}$ &  interaction currents: on-shell particle 1 \cr
$\mu_{\rm int}$  &  S,D and P-state interference  \cr
 $\mu_P$  &P-state squared terms  \cr 
$\mu_{\chi}$  & P-state and  negative $\rho$-spin $z_\ell^{--}$ interference \cr
\end{tabular}
\end{ruledtabular}
\end{minipage}
\end{table}
%
%%%%%%%%%%%%%%%%%%%%%%%%%%%%%%%%%%%%%%%%%%%%%%%%%%

%%%%%%%%%%%%%%%%%%%%%%%%%%%%%%%%%%%%%%%%%%%%%%%%%%
\begin{table}[t]
\begin{minipage}{3in}
\caption{Contributions to the magnetic moment from the eight different types of corrections discussed in the text.  To get the correct experimental value, these corrections must equal $-0.023$.  } 
\label{tab:magmt}
\begin{ruledtabular}
\begin{tabular}{lrrrr}
& \multicolumn{2}{c} {WJC-1} &  
\multicolumn{2}{c} {WJC-2} \cr
\tableline 
 &  $u,w$ only  & ${\rm all}\quad$ &  $u,w$ only  & ${\rm all}\quad$   \cr
\tableline  %[0.001in]
$\mu_{\rm NR}$  & $-0.044$ & $-0.044$  &$-0.037$ & $-0.037$    \cr
 $\mu_{Rc}$ & 0.021 & 0.021 & 0.009 & 0.009 \cr
 $\mu_{h'}$ & $-$0.010 &  $-$0.009  & $-$0.005 &  $-$0.005 \cr
 $\mu_{V_2}$  & 0.001& 0.004  & $-0.001$& $0.009$\cr
 $\mu_{V_1}$ & $0.013$ & 0.006 & 0.008& 0.008 \cr
$\mu_{\rm int}$  & --- & 0.016 & --- & 0.001    \cr
 $\mu_P$  & ---& $-0.004$ & ---& 0.000  \cr 
$\mu_{\chi}$  & --- & $-0.007$ & --- & 0.000 \\[0.02in]
\tableline
total & $-0.019$ & $-0.017$ & $-0.026$ & $-0.016$
\end{tabular}
\end{ruledtabular}
\end{minipage}
\end{table}
%
%%%%%%%%%%%%%%%%%%%%%%%%%%%%%%%%%%%%%%%%%%%%%%%%%%

 If the deuteron is treated as a non-relativistic superposition of S and D-states, normalized to unity so that
\bea
1=\int_0^\infty k^2\,dk \big\{u^2+w^2\big\}=P_S+P_D\, ,
\eea
then the well known result for the magnetic moment is
\bea
\mu_d=\mu_s+\frac34(1-2\mu_s)P_D=\mu_s+\mu_{\rm NR} \label{eq:muNR}
\eea
where $\mu_s=0.880$ is the isoscalar nucleon magnetic moment.  Inserting the measured deuteron magnetic moment, 0.857 (in nuclear magnetons) gives the famous prediction of 4\% for the deuteron D-state, a result too low for most modern models.

The CST results for the leading contributions to the magnetic moment (with an estimated accuracy of $\pm0.002$) were derived in Sec.~\ref{sec:GM0} and Appendix \ref{app:magmoment}.  After some simplification, the results can be written [see Eq.~(\ref{eq:mag1})]
\bea
\mu_d=\mu_s +\Delta\mu_d \label{eq:magapprox}
\eea
where $\Delta\mu_d$ is the sum of eight different types of corrections given in Eqs.~(\ref{eq:munr}) and (\ref{eq:Magcorr2}) and listed in Tables \ref{tab:magmtnames} and \ref{tab:magmt}.   The physical origin of each of these eight corrections is summarized in Table \ref{tab:magmtnames},  and their numerical size for each of the models WJC-1 and WJC-2  are summarized in Table \ref{tab:magmt}.  A running sum of the correction terms is plotted in Fig.~\ref{fig:two}.

From these results we conclude that the  CIA contributions and the interaction currents are not able to explain the magnetic moment precisely.  Within the theoretical errors, the missing contribution is about $\delta\mu_d\simeq-0.006\pm0.002$, less than 1\% of the magnetic moment and closer to the the experimental value than the nonrelativistic D-state contribution (assuming the $P_d\simeq 5$-6\% found in most fits).  This small difference is a new prediction for the total size of the famous $\rho\pi\gamma$ exchange current that has been extensively studied \cite{Casper:1967zz,Chemtob:1974nf,Ito:1993au,Hummel:1989qn,Hummel:1990zz} and other purely transverse contributions not constrained by the $np$ dynamics.   Predictions for these contributions, and comparisons of our results with the many other calculations in the literature, will be the subject of a future paper.

\subsection{Conclusions}
\label{sec:conclusions}

The calculation of the magnetic moment given in this paper is the first precise consequence of interaction current derived in Ref.\ I.  Using this interaction current, and the deuteron wave functions obtained from the  precision CST fits to the $np$ scattering data, Model WJC-1 predicts the magnetic moment to be 0.863(2), while Model WJC-2 predicts it to be 0.864(2), where the theoretical error is an estimate of the size of the many small terms omitted from the calculation.  Taking the value given by  the most precise model (WJC-1) and increasing the error to $\pm0.003$ to allow for the model dependence, our  overall prediction is 0.863(3).  This result is larger than the experimental value by 0.006(3), implying that the total size of the many missing purely transverse interaction currents unconstrained by the $np$ dynamics (including the $\rho\pi\gamma$ and $\omega\sigma\gamma$ currents) is much smaller than previously estimated.  Either these currents are individually quite small, or they tend to cancel when added together.  The CST prediction for the magnetic moment, obtained {\it without any adjustable parameters\/}, is within 1\% of the experimental value.

The  prediction is almost the same for both models, even though the two models have quite different properties.  This is illustrated in Fig.\ \ref{fig:two}, which shows the running sum of the eight contributions, added  in the order listed in Tables \ref{tab:magmtnames} and \ref{tab:magmt}.  For both models the NR correction (\ref{eq:muNR}) is too small and the relativistic corrections ($\mu_{Rc}$) brings the moment up to equal to, or close to its experimental value.  Both of these effects depend on the S and D states only.  Then the contributions from the derivative of the strong nucleon form factor, proportional to $a(p^2)=d\log(h)/dp^2$ [see  Eq.~(\ref{4.7})], reduce the moment again, giving an almost identical value near $-0.032$ for the two models.   The two interaction current contributions, $V_2$ (arising from  the momentum dependence associated with the $\Theta$ attached to the off-shell particle 2) and $V_1$ (arising from  the momentum dependence associated with the $\Theta$ attached to particle 1, which is only contributes when {\it both\/} particles are off-shell), both give positive contributions, pushing the total back up to a value equal, or close to the experimental value.  These interaction current contributions contain significant contributions from the P-states as well as the S and D-states.  Perhaps the most surprizing result comes from the last three terms ($\mu_{\rm int}, \mu_P$, and $\mu_{\chi}$), all of which are zero if the P-states $v_t$ and $v_s$ are zero.  In Model WJC-2 where the P-states are very small, these terms add very little, but their contributions are significant for Model WJC-1, where they give large canceling effects just sufficient to to produce a the same total prediction  as is obtained for Model WJC-2.  Note that even the term $\mu_{\chi}$ which is an interference between the P-states and the negative $\rho$-spin contributions from particle 1 (which contribute only to the diagrams (B) of Fig.~\ref{Fig1} when both particles are off-shell) is important to obtaining agreement between the two models.  As shown in Appendix \ref{app:charge}, these terms cancel in the charge, but make a small but significant contribution to the Model WJC-1 prediction for the magnetic moment. 

 We now turn to the derivation of these results, as as already outlined in Sec.~\ref{sec:org} above.

\section{wave and vertex functions}\label{sec:wave}

%%%%%%%%%%%%%%%%%%%%%%%%
\begin{figure*}
%\vspace*{-1in}
%\begin{center}
\centerline{
\mbox{
\includegraphics[width=5.6in]{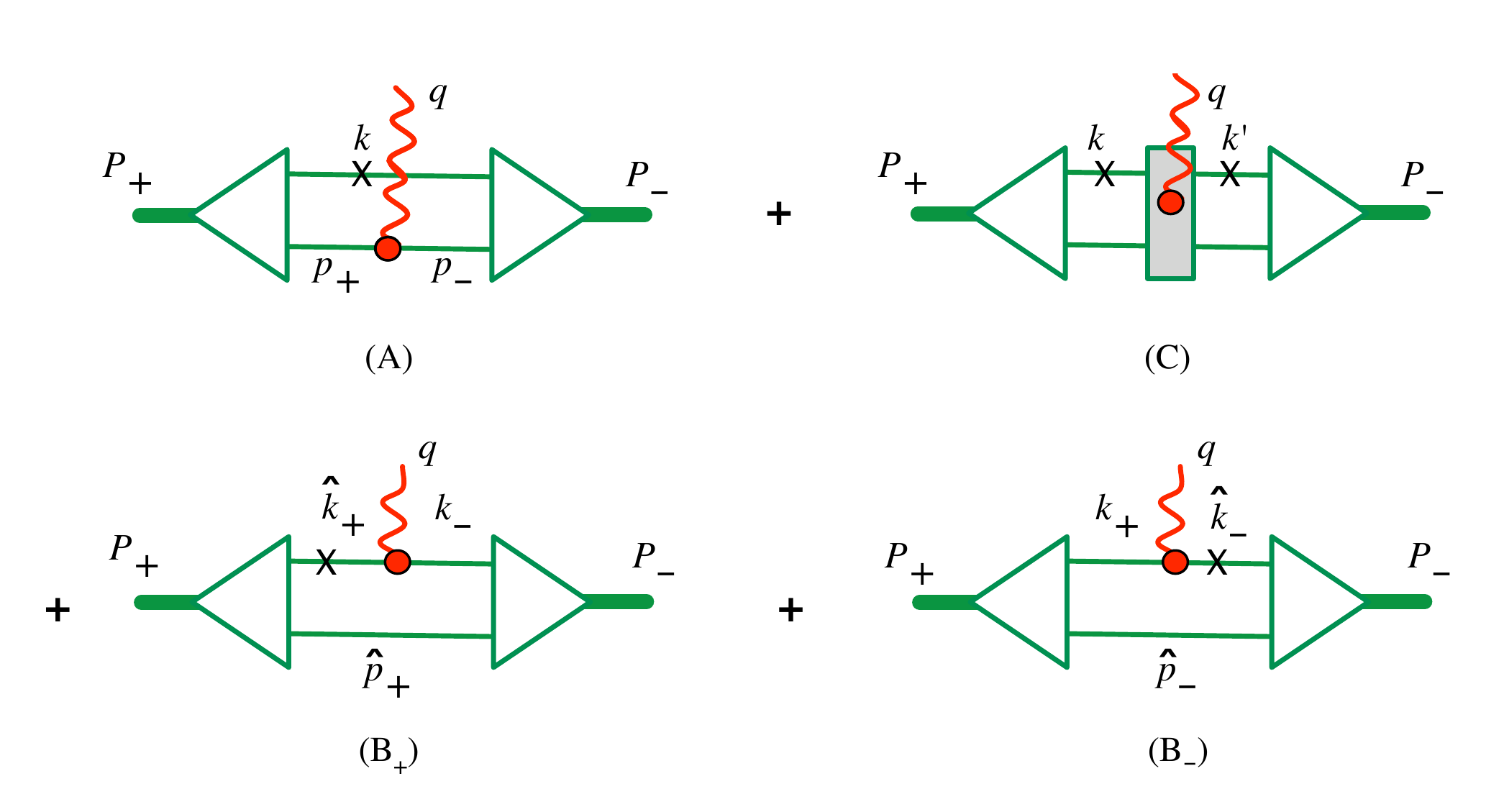}
}
}
%\end{center}
%\vskip 2.0in
%\vspace*{ -0.4in}
\caption{\footnotesize\baselineskip=10pt (Color on line) Diagramatic representation of the current operator of the Covariant Spectator Theory with particle 1 on-shell (the on-shell particle is labeled with a $\times$).  Diagrams (A), (B$_+$), and (B$_-$) are the complete impulse approximation (CIA), while (C) is the interaction current term.  In some cases the sum of the (B) diagrams is approximately equal to the first, and for this reason the relativistic impulse approximation (RIA) is the defined to be two times the first term, (A).  Note that both particles are off-shell in the initial state in diagram B$_+$ and in the final state in diagram B$_-$.
}
\label{Fig1}
\end{figure*} 
%%%%%%%%%%%%%%%%%%%%%%%%

In the CST, the two body current is given by the four diagrams shown in Fig.~\ref{Fig1}.  At the conclusion of Ref.\ I it was shown that these four diagrams can be written as a trace over the product of covariant wave functions of the initial and final deuteron, and a current operator describing the interaction of the virtual photon with the off-shell nucleon.  In this section  the covariant wave and vertex functions will be discussed in detail.

\subsection{General definitions}  

The covariant wave function of the deuteron is defined in terms of the covariant $dnp$ vertex function, ${\cal G}$,
\bea
{\it \Psi}_{\alpha\beta}^{\lambda_d}(k,P) &=&\Psi^{\lambda_d}_{\alpha\beta'}(k,P){\cal C}_{\beta'\beta}
\nonumber\\
&=&S_{\alpha\alpha'}(p){\cal G}_{\alpha'\beta}^{\lambda_d}(k,P)
\, ,\label{eq:27}
\eea
where ${\cal C}$ is the Dirac charge conjugation matrix,  $S$ is the {\it bare\/} nucleon propagator (with the factor of $-i$ removed)
\bea
S(p)=\frac1{m-\slashed{p}}
\eea
and, for an incoming deuteron of four momentum $P$ and polarization four-vector $\xi$,  ${\cal G}$ is  written
\bea
{\cal G}_{\alpha\beta}^{\lambda_d}(k,P)&=&(\Gamma_\nu{\cal C})_{\alpha\beta}(k,P)\xi^\nu_{\lambda_d}(P)
\nonumber\\
&=&\Gamma_{\alpha\beta'}^{\lambda_d}(k,P){\cal C}_{\beta'\beta}\, ,
\label{eq:2.1a}
\eea
with $k$ the four-momentum of particle 1 (with Dirac index $\beta$),  and   $p=P-k$ the four-momentum of particle 2 (with Dirac index $\alpha$).  Care must be taken to distinguish ${\it \Psi}$ (which includes the charge conjugation matrix) from $\Psi$ (which does not).  These wave (or vertex functions) satisfy the bound state CST equation
\bea
&&S^{-1}_{\alpha\alpha'}(p){\it \Psi}^{\lambda_d}_{\alpha'\beta}(k,P)
\nonumber\\
&&\qquad=-\int_{k'} \overline V_{\beta\gamma,\alpha\alpha'}(k,k';P){\it \Psi}_{\alpha'\gamma'}(k',P) \Lambda^T_{\gamma'\gamma}(k)\, , \qquad\quad \label{eq:CST}
\eea
where  $\overline{V}$ is the symmetrised kernel (introduced in Ref.~\cite{Gross:2008ps}), the positive energy Dirac projection operator is  
\bea
\Lambda_{\gamma\gamma'}(k)=\frac{(m+\slashed{k})_{\gamma\gamma'}}{2m}=\sum_{\lambda'} {u}_{\gamma}({\bf k},\lambda')\bar{u}_{\gamma'}({\bf k},\lambda') ,\qquad \label{eq:spindecom}
\eea
with the Dirac spiniors $u=u^+$ discussed in Appendix \ref{app:H}, and the volume integral is
\bea
\int_{k}=\int \frac{d^3 k}{(2\pi)^3}\frac{m}{E_k}\, .
\label{eq:volint1}
\eea
Here particle 1, with four momentum $k=\{E_k,{\bf k}\}$, is on shell (so that $E_k=\sqrt{m^2+{\bf k}^2}$). 

In the OBE models that that are the basis of the work reported here, the strong form factors at the meson-$NN$ vertices are  products of strong from factors for each particle entering or leaving the vertex.   The strong form factor $h(p)$ (where $h(p)$ is a function of $p^2$) associated with each external nucleon line  can be factored our of the $NN$ scattering kernel, leading to
\bea
&&\overline{V}(k,k';P)=h(k)h(p)\widetilde V(k,k';P)h(k')h(p')
 \label{eq:VandVmu}
\eea
where $\widetilde V$ is the {\it reduced\/} kernel, and we recall that, for both primed and unprimed variables, $p=P-k$.  If a particle with momentum $k$ is on-shell, so that  $k^2=m^2$, the strong form factor is defined so that $h(k)=1$.  Note that the expression (\ref{eq:VandVmu}) for the kernel is written allowing for the possibility that any (or all four) of the particles could be off-shell.

The next step in the computation of the form factors is to express the wave and vertex functions in terms of scalar invariant functions, so that when the traces  (\ref{eq:A&V2-3}) and (\ref{eq:B&V1-1}) are computed, the result will be a sum of bilinear products of these scalar functions multiplied by covariant kinematical factors.  The result is manifestly covariant, and the effect of boosting the incoming and outgoing states is easily accounted for by correctly shifting the arguments of the invariant functions.

\subsection{Expansion of the wave or vertex functions}

When particle 1 is on-shell, the covariant $dnp$ deuteron nucleon vertex function defined in Eq.~(\ref{eq:2.1a}) (with the charge conjugation matrix removed) can be expanded into four independent Dirac invariants
\bal
\Gamma^\mu(k,P)&=F\gamma^\mu+\frac{G}{m}k^\mu -2\Theta(p)\left[H\gamma^\mu+\frac{I}{m}k^\mu\right] 
\label{eq:2.3}
\end{align}
where $k$ is the four-momentum of the on-shell particle 1, so that $k^2=m^2$, 
$p=P-k$ is the four-momentum of the off-shell particle 2, and $\Theta(p)$ is  the negative energy projection operator of particle 2 [recall Eq.\ (\ref{eq:theta})].   
The scalar functions $F, G, H,$ and $I$ are all functions of $p^2$, the only free scalar variable.   Note that
\bal
\overline{\Gamma}^\mu(k,P)&=\gamma^0\big[\Gamma^\mu(k,P)\big]^\dagger \gamma^0\cr
&=F\gamma^\mu+\frac{G}{m}k^\mu -\left[H\gamma^\mu+\frac{I}{m}k^\mu\right] 2\Theta(p)\, .\qquad
\label{eq:2.3a}
\end{align}

It is sometimes convenient to work directly with wave function  $\Psi^\mu$ defined in Eq.~(\ref{eq:27}) (with the charge conjugation matrix removed), and the related amplitude $\overline{\Psi}^\mu$,
\bal
&\Psi^\mu(k,P)\equiv S(p) \Gamma^\mu(k,P)
\nonumber\\
&\quad=A\gamma^\mu+\frac{B}{m}k^\mu -2\Theta(p)\left[C\gamma^\mu+\frac{D}{m}k^\mu\right] 
\label{3.5}
\end{align}
where $S(p)$ is the undressed propagator of the off-shell particle, and
\bal
(m^2-p^2)\,C&=m\,F\nonumber\\
(m^2-p^2)\,D&=m\,G\nonumber\\
m\,A&=2m\,C-H\nonumber\\
m\,B&=2m\,D-I \label{eq:AtoF}\, .
\end{align}
The $F, G, H,$ and $I$ are related to the deuteron wave functions, as discussed in Appendix \ref{app:H} and many previous references \cite{Adam:1997rb,Buck:1979ff,Gross:1991pm,arXiv:1007.0778}.  When the spectator is on shell, these invariants depend only on $p^2$, the mass of the off-shell particle.

\subsection{Bethe-Salpeter vertex functions} \label{app:BS}

The (B) diagrams of Fig.\ \ref{Fig1} 
require  Bethe-Salpeter (BS)  vertex functions with {\it both\/} particles off-shell.  These can be expanded in terms of invariant functions that depend on the two invariant variables $p^2$ and $k^2\ne m^2$. To describe these, the expansion (\ref{eq:2.3}) is generalized
\bal
\Gamma^\mu_{\rm BS}(k,&P)= F\gamma^\mu+\frac{G}{m}k^\mu -2\Theta(p)\left[H\gamma^\mu+\frac{I}{m}k^\mu\right] 
\nonumber\\
&-\left[K_1\gamma^\mu+\frac{K_2}{m}k^\mu\right] 2\Theta(-k)
\nonumber\\
&+4\Theta(p)\left[K_3\gamma^\mu+\frac{K_4}{m}k^\mu\right] \Theta(-k)
\nonumber\\
=\; & \Gamma^\mu(k,P)-\Gamma^\mu_{\rm off}(k,P) \,2\Theta(-k) 
\label{eq:BSvertexa}
\end{align}
where the invariants in $\Gamma^\mu$  ($F, G, H, I$) are distinguished from the old only by their arguments (two instead of one).  The appearance of the operator on the right of the last terms, $\Theta(-k)$ (instead of $\Theta(k)$, as might have been expected), comes from moving the charge conjugation matrix  past the projection operator of particle 1: ${\cal C}\,\Theta^T(k)=\Theta(-k)\,{\cal C}$.  Particle interchange symmetry relates  $H$ and $I$ to $K_1$ and $K_2$, but we will ignore this constraint for now; it is a numerical feature of the solutions for the matrix elements. 

As it turns out (see Appendix \ref{app:offshellterms}), all six invariant functions are present in $\Gamma^\mu$, {\it even when particle 1 is on shell\/}.  The $\Gamma^\mu_{\rm off}$ part of the vertex function constructed from the four  invariant functions $K_i$ is not zero when $k^2=m^2$.  However, because of the presence of the the projection operator $\Theta(-k)$ it does not contribute to diagrams where {\it both\/} $k^2=m^2$ {\it and\/} the vertex function is contracted with an on-shell projection operator (or the on-shell $u$ spinor).  Thus it makes no contribution to the (A) diagrams, but a full understanding of the content of the (B) diagrams requires that it be included.

 In the rest frame, when both particles are off-shell,  the covariant variables are  related to ${\bf k}^2$, the square magnitude of the spectator three-momentum,  and $k_0$, the off-shell energy of particle 1, through the relations
\bea
p^2&=&(P-k)^2=k^2 +m_d(m_d-2 k_0)
\nonumber\\
k^2&=&m^{*2}\equiv k_0^2-{\bf k}^2 \, .
\label{eq:kp}
\eea
Solving these relations for $k_0$ and ${\bf k}^2$ gives
\bea
k_0&\to& R_0 %k^P
\equiv\frac{P\cdot k}{m_d} =\frac{m_d^2+m^{*2}-p^2}{2m_d}
\nonumber\\
{\bf k}^2&\to& R^2 %k_P^2
\equiv k_0^2-m^{*2}=\frac{(P\cdot k)^2}{m_d^2}-m^{*2}\, .
\label{eq:kp&eta}
\eea
These relations provide the unique covariant generalization of the rest frame variables $k_0$ and ${\bf k}^2$ (denoted by $R_0$ and $R^2$).   Stated more precisely, if the spectator associated with a deuteron with four-momentum $P$ has energy $k_0$ and a squared three three-momentum ${\bf k}^2$, then the equivalent {\it rest frame\/} values of these quantities are $R_0$ and $R^2$.  Note that $R_0$ and $R=\sqrt{R^2}$ are quite different quantities. 

It is instructive to derive these relations by a direct boost from the moving frame to the rest frame.   To do this, consider (for definiteness) that the moving deuteron has momentum $\{D_0,{\bf 0}_\perp,Q/2\}$, with
\bea
D_0=\sqrt{m_d^2+\frac14Q^2}\, .  \label{eq:D0}
\eea
Then if the spectator has four momentum $k=\{k_0,{\bf k}_\perp,k_z\}$, in the rest frame these values are
\bea
R_0&=&\frac1{m_d}\big(D_0 k_0-\frac12Q k_z\big)=\frac{P\cdot k}{m_d}
\nonumber\\
R_z&=&\frac1{m_d}\big(D_0 k_z-\frac12 Qk_0\big)
\label{eq:k0&kz}
\eea
with the transverse momentum, ${\bf k}_\perp$, unchanged.  The first of the two relations (\ref{eq:kp&eta}) emerges immediately, and to obtain the second simply compute the square of the three momentum in the rest frame
\bea
R^2&=&{\bf k}_\perp^2+R_z^2
\nonumber\\
&=&{\bf k}^2+\eta^2(k_z^2+k_0^2)-\sqrt{\eta}\,k_zk_0D_0
\nonumber\\
&=&\frac{(P\cdot k)^2}{m_d^2}+{\bf k}^2-k_0^2 
\eea
in agreement with (\ref{eq:kp&eta}).  It is also easy to use (\ref{eq:k0&kz}) to confirm that $k^2=m^{*2}$ is covariant by computing
\bea
R_0^2-R_z^2-{\bf k}_\perp^2&=&\frac1{m_d^2}\Big(D_0^2-\frac{Q^2}{4}\Big)(k_0^2-k_z^2)-{\bf k}_\perp^2
\nonumber\\
&=&k_0^2-{\bf k}^2 .
\eea

A word of caution: depending on the context, $k$ is sometimes used to denote either the magnitude of the three-momentum (i.e.~$R$) or the four-momentum (and, when the square of the four momentum is involved, $m^{*2}$ will sometimes be used instead of $k^2$).  Earlier discussions of deuteron wave functions were restricted to cases when particle 1 was on shell, and were evaluated in the rest frame \cite{Buck:1979ff,Gross:1991pm,arXiv:1007.0778} or used wave functions boosted from the rest frame \cite{Adam:2002cn},  where there was no need to make a distinction between $R$ and $k$.

All of the invariants defined in (\ref{eq:BSvertexa}) depend on the {\it two\/} variables $R$ and $R_0$, so that, for example $F= F(R,R_0)$.  However, because of the cancellation between the contributions from the (B) diagram and the $\left<V_1\right>$ interaction currents, discussed in Ref.\ I, the effective BS vertex function of interest reduces to the CST  function when particle 1 is on shell.  The frame independent way to express this on-shell condition is to introduce $E_{R}$, where
\bea
E_{R}\equiv \sqrt{m^2+R^2}
\eea
is the straightforward generalization of $E_k$.  Note that $E_{R}=E_k$ in the rest frame.  Using this notation, the invariant functions satisfy the condition
\bea
 Z(R,E_{R})&=&Z(R)\, .
\label{eq:onshellconditions1}
\eea
where $Z$ is a generic name for any of the eight invariant functions.

\section{Deuteron Form factors}\label{sec:formfactors}

\subsection{Definitions of the Form Factors}

The most general form of the covariant deuteron electromagnetic current can be expressed in terms of three deuteron form factors 
\bal
\left<d\,\lambda\right|J^\mu\left|d'\,\lambda'\right>&=-2D^\mu\bigg\{G_1 \,\xi^*_{\lambda}\cdot\xi'_{\lambda'}-G_3\frac{(\xi^*_{\lambda}\cdot q)(\xi'_{\lambda'}\cdot q)}{2m_d^2}\bigg\} \nonumber\\
&-G_M\Big[\xi'^\mu_{\lambda'}(\xi^*_{\lambda}\cdot q)-\xi^{*\mu}_{\lambda}(\xi'_{\lambda'}\cdot q)\Big]\, ,
\label{eq:G123}
\end{align}
where the form factors $G_1$, $G_3$, and $G_M=G_2$ are all functions of the square of the  momentum transfer $q=P_+-P_-$, with $Q^2=-q^2$, $D^\mu=\frac12(P_++P_-)^\mu$, and $P_-$ ($P_+$) the four-momentum of the incoming (outgoing) deuterons, and $\xi'_{\lambda'}$ ($\xi_{\lambda}$) are the four-vector polarizations of the incoming (outgoing) deuterons with helicities $\lambda'$ ($\lambda$).  The polarization vectors satisfy the well known constraints
\bal
&P_+\cdot \xi_\lambda = P_-\cdot\xi'_{\lambda'}=0 \nonumber\\
&\xi^*_\lambda\cdot \xi_\rho=-\delta_{\lambda\rho} 
\nonumber\\
&\xi'^*_{\lambda'}\cdot \xi'_{\rho'}=-\delta_{\lambda'\rho'}\, .
\end{align}
This notation agrees with that used in  Ref.~\cite{Gilman:2001yh}, except that now $\lambda$ denotes the helicity of the {\it outgoing\/} deuteron and  $\lambda'$ the helicity of the {\it incoming\/} deuteron.

The form factors $G_1$ and $G_3$ are usually replaced by the charge and quadrupole form factors, defined by 
\bea
&&G_C=G_1+\frac{2}{3}\eta G_Q \nonumber\\
&&G_Q=G_1+(1+\eta)G_3-G_M\, , \label{eqLgcgq}
\eea
with $\eta=Q^2/4m^2_d$.  At $Q^2=0$, the three form factors $G_C$, $G_Q$, and $G_M$ give the charge, quadrupole moment, and magnetic moment of the deuteron
%
%\bal
\bea
\begin{array}{ll}
G_C(0)=\;1\;=G_1(0) & ({\rm units\;of}\;e)\cr
G_M(0)=\mu_d=G_2(0) &  ({\rm units\;of}\;e/2m_d)\cr
G_Q(0)=Q_d=G_3(0)+1-\mu_d  &  ({\rm units\;of}\;e/m^2_d)\, .\qquad\quad
\end{array}
%\end{align}
\eea

The form factors can be related to helicity amplitudes.  Working in the Breit frame, and choosing the momenta to be 
\bea
P_\pm^\mu&=&\{D_0,0,0,\pm\sfrac12 Q\}\nonumber\\
q^\mu&=&\{0,0,0,Q\} \label{eq:breit}
\eea
where $D_0$ was defined in Eq.~(\ref{eq:D0}), the helicity four-vector polarizations for the deuteron  and the photon are
\bea
\xi^\mu_{\lambda}
&=&\begin{cases} \{0,\mp1,-i,0\}/\sqrt{2}&\lambda=\pm 1\cr
\{\sfrac12Q,0,0,D_0\} /m_d& \lambda=0\end{cases}\nonumber\\
\xi'^\mu_{\lambda'}
&=&\begin{cases} \{0,\pm1,-i,0\}/\sqrt{2}&\lambda'=\pm 1\cr
\{-\sfrac12Q,0,0,D_0\} /m_d& \lambda'=0\end{cases}
\nonumber\\
\epsilon^\mu_{\lambda_\gamma}
&=&\begin{cases} \{0,\mp1,- i,0\}/\sqrt{2}&\lambda_\gamma=\pm 1\cr
\{1,0,0,0\} & \lambda_\gamma=0\end{cases}
\label{2.4}
\eea
where the polarization vectors for the incoming deuteron (treated as particle 2 in the conventions of Jacob and Wick) have been obtained from those of the outgoing deuteron (particle 1 of Jacob and Wick) by a rotation through $\pi$ about the $\hat y$ axis, multiplied by a phase
\bea
\xi'_{\lambda'}=(-1)^{1+\lambda} R_y(\pi) \xi_{\lambda'}\, .
\eea
These definitions agree with Refs.~\cite{Gilman:2001yh} and \cite{SLAC-PUB-2318} (except that, in Eq.~(2.7) of Ref.~\cite{SLAC-PUB-2318}, the $\xi^\mu(\pm1)$ refer to the spin direction and not the helicity and there is a typo in the expression for $\xi^\mu(0)$).

We will denote the most general helicity amplitude by
\bea
G^{\lambda_\gamma}_{\lambda\lambda'}\equiv
\left<P_+\,\lambda\right|J_\mu\left|P_-\,\lambda'\right>\epsilon_{\lambda_\gamma}^\mu \, . \label{eq:current1}
\eea
Under rotation by $\pi$ about the $\hat z$ axis, all of the helicity four-vectors (\ref{2.4}), represented generically by the vector $\varepsilon$, transform as
\bea
\varepsilon_\lambda=(-1)^{\lambda} \varepsilon_\lambda\, ,
\eea
giving the condition
\bea
\lambda_\gamma+\lambda+\lambda'=0\, , \label{eq:helcons}
\eea
(This relation must be interpreted as arithmetic  modulo 2, and can be written in a variety of ways.)  In addition, the amplitudes are related to each other by $Y$-parity conservation (parity followed by rotation $\pi$ about the $\hat y$ axis), which insures that 
\bea
G^{\lambda_\gamma}_{\lambda\lambda'}=G^{-\!\lambda_\gamma}_{-\!\lambda\,-\!\lambda'}.
\eea
Hence it is sufficient to omit discussion to those nine amplitudes with $\lambda_\gamma=-1$, and of the three amplitudes $G^0_{\lambda-}$ and $G^0_{-0}$.  Of the remaining 14, Eq.~(\ref{eq:helcons}) gives
\bea
G^{+}_{++}&=&G^{+}_{--}=G^+_{00}=G^+_{-+}=G^+_{+-}=0
\nonumber\\ 
G^0_{+0} &=&G^0_{0+} =0 \, ,\qquad
\eea
leaving seven possible amplitudes.

A conserved current must have the form (\ref{eq:G123}), and direct computation using this gives four further relations 
\bea
&&G^+_{+0}=G^+_{0+}
\nonumber\\
&&G^0_{++}=G^+_{-0}=G^+_{0-}=0
\eea
leaving only the three independent amplitudes $G^0_{00}, G^0_{+-}$ and $G^+_{+0}=G^+_{0+}$.  [Note that Eq.~(20) of Ref.~\cite{Gilman:2001yh} states incorrectly that $G^+_{0-}$ and $G^-_{0+}$ are nonzero.]

\vspace{0.1in}

While the sum of all of the individual contributions to the form factors is constructed to give a conserved current, individual terms may not, and for this reason the average of $G^+_{+0}$ and $G^-_{0-}$ (equal to $G^+_{0+}$), which enjoys a desirable symmetry property discussed below, is used to extract the magnetic contributions from individual terms.  The form factors are then extracted from the following combination of helicity amplitudes
\bea
&& {\cal J}_1\equiv G^0_{00}=2D_0\left(G_C+\frac{4}{3}\eta\, G_Q\right)\nonumber\\
&& {\cal J}_2\equiv G^0_{+-}=2D_0\left(G_C-\frac{2}{3}\eta \,G_Q\right)\nonumber\\
&&{\cal J}_3\equiv \frac12(G^+_{+0} +G^-_{0-})=2D_0\sqrt{\eta}\,G_M\, , \label{eq:dffmatrix}
\eea
where ${\cal J}_n$ (with $n=1,2,3$) is a convenient notation for the helicity amplitudes.  To calculate the deuteron form factors, it therefore sufficient to calculate the three independent matrix elements (\ref{eq:dffmatrix}) of the two-body current operator.

The remaining parts of this section assemble the general formulae for the three independent helicity amplitudes,   ${\cal J}_n$ starting from the results of Eqs.\ (\eqqA) and (\eqq2) of Ref.\ I.   From these amplitudes the charge, quadrupole, and magnetic form factors are obtained.  Explicit expressions for the charge will be given in Sec.~\ref{sec:GC0} and for the magnetic moment in Sec.~\ref{sec:GM0}.  Results for the quadrupole moment will be given in a subsequent paper.

\subsection{Off-shell nucleon current}

Following the method of Riska and Gross \cite{Gross:1987bu}, a conserved two-nucleon current can be constructed \cite{Adam:2002cn} using the {\it dressed\/} single nucleon off-shell current 
\bea
j^\mu(p,p')&=&h(p)h(p')j_R^\mu(p,p')
\nonumber\\&=&
e_0\,f_0(p',p){\cal F}_1^\mu+e_0\,g_0(p',p)\Theta(p'){\cal F}_3^\mu\Theta(p)
\nonumber\\
&&\quad+e_0\,f_2(p',p)F_2(Q^2)\frac{i\sigma^{\mu\nu}q_\nu}{2m} \label{3.1}
\eea
where $j_R$ is the reduced current, $f_0, g_0, f_2$ are off-shell functions discussed below,  $e_0=\frac12$ is the isoscalar charge, the off-shell projection operator $\Theta$ was defined in (\ref{eq:theta}),
\bea
{\cal F}_i^\mu&=&[F_i(Q^2)-1]\widetilde\gamma^\mu +\gamma^\mu
\nonumber\\&=&
F_i(Q^2)\widetilde\gamma^\mu+\frac{\slashed{q}q^\mu}{q^2}\, ,\qquad
\label{eq:Fimu}
\eea
and the transverse gamma matrix is
\begin{equation}
\widetilde\gamma^\mu=\gamma^\mu-\frac{\slashed{q}q^\mu}{q^2}\, , \label{eq:gammatilde}
\end{equation}
with $q=p'-p$.  The nucleon form factors are $F_i(Q^2)$, with $Q^2=-q^2$ (and $F_3$, subject to the constraint that $F_3(0)=1$, a new form factor that contributes only when both nucleons are off-shell).   The second from of (\ref{eq:Fimu}) displays the interesting fact that the important physics is contained in the {\it transverse\/} part of the current; the longitudinal part that is constrained by the WT identities will {\it not contribute to any observable\/} since it is proportional to $q^\mu$ which vanishes when contracted into any conserved current or any of the three polarization vectors of an off-shell photon.

The off-shell functions $f_0$ and $g_0$ are determined  from the requirement that the reduced current, $j_R^\mu$, satisfy the Ward-Takahashi (WT) identity
\bea
q_\mu \, j_R^\mu(p,p')&&=e_0\Big[S^{-1}_{d}(p')-S^{-1}_{d}(p)\Big]
\, , \label{eq:210a}
\eea
where  $S_d$ the dressed propagator 
\bea
S_d^{-1}(p)&=&\frac{m-\slashed{p}}{h^2(p)}=\frac{S^{-1}(p)}{h^2(p)} \, ,\label{eq:Sd}
\eea
where  $h$ occurs squared because one comes from the initial and one from the final interactions that connect the propagator.

In all previous references it was assumed that the off-shell function $f_2=f_0$, but since the $\sigma^{\mu\nu}q_\nu$ term is transverse, the WT identity places no constraint on $f_2$.  
Since consistency requires that any variation of $f_2$ also include the overall factors of $hh'$, so that the relationship (\ref{3.1})  between the dressed and reduced currents can be maintained, a simple ans\"atz for possible variations of $f_2$ is
\bea
f_2(p.p')=(1-\omega_2)\,hh' +\omega_2\,f_0(p,p') \label{eq:om2def}
\eea
where $\omega_2=1$ is the choice previously discussed, and $\omega_2=0$ a reasonable alternative.  In this paper it was found that the variation in the results for $\omega_2=0$ and $\omega_2=1$ was less that 0.001, the size of other terms omitted from the calculation.  As a result, $\omega_2$ was set to unity (our original assumption) and is no longer considered a parameter. However, for completeness it was decided to record this dependence in the formulae given in Sec.~\ref{sec:GM0} and Appendix \ref{app:magmoment}.

Using the shorthand notation $h=h(p)$ and $h'=h(p')$, the simplest solution to (\ref {eq:210a}) gives
\bea
f_0(p',p)&=&\frac{h'}{h} \frac{(m^2-p^2)}{p'^2-p^2}+\frac{h}{h'} \frac{(m^2-p'^2)}{p^2-p'^2}
\nonumber\\
g_0(p',p)&=&\frac{4m^2}{p'^2-p^2}\left(\frac{h}{h'}-\frac{h'}{h}\right)
\, .
\eea
An important simplification of the current occurs if it is contracted into the real (or virtual) photon polarization vectors defined in (\ref{2.4}), with the property that $q_\mu \epsilon^\mu_{\lambda_\gamma}=0$.  In this case the $q^\mu$ terms in (\ref{3.1}) can be dropped, and setting $f_2=f_0$ from now on gives
\bea
j^\mu(p',p)&\to&f_0(p',p)j_N^\mu(p',p)
\nonumber\\&&
+e_0g_0(p',p)F_3(Q^2)\Theta(p')\gamma^\mu\Theta(p)
\qquad
\eea
where $j_N^\mu$ is the familiar on-shell nucleon current
\bea
j_N^\mu(p',p)=e_0 F_1(Q^2)\gamma^\mu+e_0 F_2(Q^2)\frac{i\sigma^{\mu\nu} q_\nu}{2m} . \qquad
\eea
In addition, the following limits are useful
\bea
f_{00}\equiv\lim_{{p'^2\to p^2}} f_0(p',p)&=&1+2a(p^2)(m^2-p^2)\nonumber\\
g_{00}\equiv\lim_{{p'^2\to p^2}} g_0(p',p)&=&-8m^2\,a(p^2) \label{eq:FGexpand}
\eea
with 
\bea
a(p^2)=\frac{1}{h}\frac{dh}{dp^2}\, . \label{4.7}
\eea

\subsection{Contributions from diagram A plus the $V_2$ interaction current}

The contributions from diagram (A), and those parts of diagram (C) that arise from the $\nu\Theta(p)$ and $\nu\Theta(p')$  terms in the isosclalar $sNN$ and $vNN$ couplings (denoted by  $\left<V^\mu_2\right>$ in Ref.\ I), were written as a trace in Eq.~(\eqqA) of Ref.\ I.  Using the wave functions and currents introduced above, the  corresponding helicity amplitudes, defined in Eq.~(\ref{eq:current1}), can be written 
\begin{widetext}
\bea
G^{\lambda_\gamma}_{\lambda\lambda'}(q)\Big|_{{\rm A} +V_2}=-\int_k {\rm tr}\Bigg[\Big\{&&\overline{\Psi}^\lambda (k,P_+)\Big[f_0(p_+,p_-)j_N^{\lambda_\gamma}(q)+
g_0(p_+,p_-)\Theta(p_+)F_3(Q^2)e_0 \gamma^{\lambda_\gamma}\Theta(p_-)\Big]\Psi^{\lambda'}(k,P_-)
\nonumber\\&&
-\overline{\Psi}^\lambda (k,P_+)\frac{h_+}{h_-} j_{N}^{\lambda_\gamma}(q) {\Psi}^{(2) \lambda'}(k,P_-)
-\overline{\Psi}^{(2) \lambda}(k,P_+)\,j_{N}^{\lambda_\gamma}(q)\frac{h_-}{h_+}\Psi^{\lambda'}(k,P_-)
\Big\}\Lambda(-k)\Bigg]\, , \qquad\label{eq:A&V2-2}
\eea
\end{widetext}
where $p_\pm=P_\pm-k$, $j_N^{\lambda_\gamma}=j_N^\mu(\epsilon_{\lambda_\gamma})_\mu$ and $\gamma^{\lambda_\gamma}=\gamma_\mu\epsilon^\mu_{\lambda_\gamma}$ are the vector currents $j_N^\mu$ and $\gamma^\mu$ contracted with the photon polarization vector $\epsilon_{\lambda_\gamma}^\mu$.
Part of the interaction current contribution is contained in the new wave function $\Psi^{(2)}$, obtained from a truncated kernel proportional to the off-shell couplings depending on $\Theta(p)$ and $\Theta(p')$ (for details see Ref.\ I).  Calculation of the three independent helicity amplitudes defined in Eq.~(\ref{eq:dffmatrix}), labeled by $n=\{1,2,3\}$, requires the helicity combinations $n\to\{\lambda_\gamma, \lambda, \lambda'\}$ where $1\to\{0,0,0\}, 2\to\{0,+,-\}$ and $3\to\{+,+,0\}+\{-,0,-\}$.  With this correspondence implied in the equations below, six generic traces ${\cal A}_{n,i}$, where and $i=\{1,2\}$ and $n=\{1,3\}$, are defined
\bea
&&{\cal A}_{n,i}(\Psi_1\Psi_2)\equiv-\, {\rm tr} \Big[\overline \Psi_1^\lambda (k,P_+)  \, j_i^{\lambda_\gamma}(q)\,\Psi_2^{\lambda'}(k,P_-)\,\Lambda(-k)\Big]
\nonumber\\
&&\;=-(-1)^{\lambda_\gamma} {\rm tr} \Big[\overline \Psi_2^{\,-\!\lambda'} (k,P_-)   \, j_i^{-\lambda_\gamma}(-q)\,\Psi_1^{-\lambda}(k,P_+)\,\Lambda(-k)\Big]
\nonumber\\
&&\;= (-1)^{\lambda_\gamma}
{\cal A}_{n,i}(\Psi_2\Psi_1)\Big|_{q\to-q}
\, .\qquad \label{eq:genericA}
\eea
where and the transformations in the second line of (\ref{eq:genericA}) follow from the identity tr$[{\cal O}]={\rm tr}[{\cal O}^\dagger]={\rm tr}[\gamma^0{\cal O}^\dagger \gamma^0]$ and the properties $\epsilon^*_{\lambda_\gamma}=(-1)^{\lambda_\gamma}\epsilon_{-\lambda_\gamma}$ and $\xi'^\mu_{-\lambda}(q)=\xi^\mu_\lambda(-q)$.  The third line of (\ref{eq:genericA}) follows immediately from the second line for the $n=1$ or 2 helicity amplitudes (where $\lambda_\gamma=0$, and $\lambda'\leftrightarrow -\lambda$).   However, the second line interchanges the two terms that contribute to the helicity average for the $n=3$ combination,  transforming $\{+,+,0\}\leftrightarrow -\{-,0.-\}$.  Hence choosing the average of the two contributions ensures that  the symmetry relation (\ref{eq:genericA}) holds, even if the individual contribution under study does not, by itself, conserve current.  With this notation the trace (\ref{eq:A&V2-2}), for each independent helicity amplitude, can be written 
\begin{widetext}
\bea
{\cal J}_n(q)\Big|_{{\rm A} +V_2}&&=e_0 F_1(Q^2)\int_k\Bigg\{ f_0(p_+,p_-) {\cal A}_{n,1}(\Psi_+\Psi_-) -\frac{h_+}{h_-}{\cal A}_{n,1}(\Psi_+\Psi^{(2)}_-)-\epsilon_{n3}\frac{h_-}{h_+}{\cal A}_{n,1}(\Psi_+\Psi^{(2)}_-)\Big|_{q\to -q}\Bigg\}
\nonumber\\&&
+e_0 F_2(Q^2)\int_k \Bigg\{ f_0(p_+,p_-) {\cal A}_{n,2}(\Psi_+\Psi_-) -\frac{h_+}{h_-}{\cal A}_{n,2}(\Psi_+\Psi^{(2)}_-)- \epsilon_{n3}\frac{h_-}{h_+}{\cal A}_{n,2}(\Psi_+\Psi^{(2)}_-)\Big|_{q\to -q}\Bigg\}\nonumber\\&&
+e_0 F_3(Q^2)\int_k \frac{g_0(p_+,p_-)}{4m^2} {\cal A}_{n,1}(\Gamma_+\,\Gamma_-)\, , \label{eq:A&V2-3} 
\eea
\end{widetext}
where $\epsilon_{n3}=(1-2\delta_{n3})$ is the extra phase that appears for the $n=3$ helicity amplitudes, as derived in Eq.~(\ref{eq:genericA}), and the last term uses the reduction $\Theta\, \Psi\to \Gamma/(2m)$.  

The formulae for the six ${\cal A}_{n,i}$, expressed in terms of the invariant functions introduced in Sec.\ \ref{sec:wave}, are lengthly and are given in Appendix \ref{app:B}.

\subsection{Contributions from diagrams B plus the $V_1$ interaction current} \label{sec:Bdiagrams}

The contributions from diagram (B), and those parts of diagram (C) that arise from the $\nu\Theta(k)$ and $\nu\Theta(k')$  terms  in the isoscalar $sNN$ and $vNN$ couplings (which can contribute only when $k$ or $k'$ are off-shell, and are denoted by  $\left<V^\mu_1\right>$ in Ref.\ I), were written as a trace in Eq.~(\eqq2) of Ref.\ I.  Contracting these results with the photon polarization vector, and using the notation
\bea
E_\pm&=&\sqrt{m^2 +\left({\bf k}\pm \frac12{\bf q}\right)^2}
\nonumber\\
\widetilde k_\pm&=&\left\{k_0, {\bf k}\pm \frac12 {\bf q}\right\}
\label{eq:Epm}
\eea
gives 
\begin{widetext}
\bea
G^{\lambda_\gamma}_{\lambda\lambda'}(q)\Big|_{{\rm B}+V_1}=\int_k&&\left[\frac{m E_k}{{\bf k}\cdot{\bf q}}\right]{\rm tr} \Bigg\{\frac1{k_0}\,
\overline{\widehat\Gamma}_{BS}^{\lambda}(\widetilde k_+,P_+)\, S_d(\widetilde p)\,\widetilde\Gamma^{\lambda'}(\widetilde k_-,P_-)\Lambda(-\widetilde k_-) \,j_N^{\lambda_\gamma}(q)\,\Lambda(-\widetilde k_+) \Big|_{k_0=E_-}
\nonumber\\&& 
-\frac1{k_0}\, \overline{\widetilde\Gamma}^\lambda(\widetilde k_+,P_+)\, S_d(\widetilde p)\,\widehat\Gamma_{BS}^{\lambda'}(\widetilde k_-,P_-)\Lambda(-\widetilde k_-) \,j^{\lambda_\gamma}_N(q)\,\Lambda(-\widetilde k_+)\Big|_{k_0=E_+} \Bigg\}\, ,\quad
\label{eq:B&V1}
\eea
\end{widetext}
where  $\widetilde p=P_\pm-\widetilde k_\pm$.  When $k_0=E_+$, the {\it outgoing\/} particle is on shell, with $\widetilde k_+=\widehat k_+$ and $\widetilde k_-=k_-=\{E_+, {\bf k}- {\bf q}/2\}$, as labeled in diagram (B$_+$) of Fig.~\ref{Fig1}.  Similarily, when  $k_0=E_-$, the {\it incoming\/} particle is on shell, with $\widetilde k_-=\widehat k_-$ and $\widetilde k_+=k_+=\{E_-, {\bf k}+{\bf q}/2\}$, as labeled in diagram (B$_-$) of Fig.~\ref{Fig1}.   The labeling of momenta in Fig.~\ref{Fig1} and the form of the expression (\ref{eq:B&V1}) show clearly how the singularities in the two diagrams at $E_+=E_-$ cancel, giving a finite result.  Part of the interaction current contribution is contained in the new subtracted wave function $\widehat \Psi(k,P)=S(p)\widehat\Gamma(k,P)$, obtained through a cancellation of the vertex factors $\Theta(\widetilde k_\pm)$ that could be present if particle 1 is off-shell  (for details see Ref.\ I).

Note that the projection operator $\Lambda(-\widetilde k_\pm)$ always accompanies the vertex functions
$\widehat\Gamma_{\rm BS}^\lambda(\widetilde k_\pm,P_\pm)$.     
Following the discussion in Sec.\ \ref{app:BS},  when $\widetilde k_\pm$ is off-shell, the product of the subtracted vertex function and projection operator, $\widehat\Gamma_{\rm BS}^\lambda(\widetilde k_\pm,P_\pm) \Lambda(-\widetilde k_\pm)$,  breaks into two terms
\bea
\widehat\Gamma_{\rm BS}^\lambda(\widetilde k_\pm,P_\pm) \Lambda(-\widetilde k_\pm)&=&\widehat\Gamma^\lambda(\widetilde k_\pm,P_\pm) \Lambda(-\widetilde k_\pm)
\nonumber\\
&&-\frac{(m^2-\widetilde k^2_\pm)}{2m^2} \widehat\Gamma^\lambda_{\rm off}(\widetilde k_\pm,P_\pm),  \qquad\quad\label{eq:hatgamma}
\eea
where  $\widehat\Gamma$ is identical to the on-shell vertex function $\widetilde\Gamma$ when $\widetilde k_\pm$ is on-shell (because the cancellation shown in Ref.\ I  ensures that there is no extra $\widetilde k_\pm$ dependence).   

Introducing the new amplitudes 
\bea
\Upsilon^\lambda(\widetilde k,P)&&={\Gamma}^\lambda(\widetilde k,P)\Lambda(-\widetilde k),
\label{eq:symmamps}
\eea
leads to the following expressions for the independent helicity amplitudes (labeled by the index $n$ as discussed above)
\begin{widetext}
\bea
{\cal J}_n(q)\Big|_{{\rm B}+V_1}=\int_k\left[\frac{m E_k}{{\bf k}\cdot{\bf q}}\right]{\rm tr} \Bigg\{&&\frac1{k_0}\,
\Big[\overline{\Upsilon}^{\lambda}(\widetilde k_+,P_+)-\frac{{\bf k}\cdot{\bf q}}{m^2}\,\overline{\widehat\Gamma}_{\rm off}^{\lambda}(\widetilde k_+,P_+)\Big]\, S_d(\widetilde p)\,\Upsilon^{\lambda'}(\widetilde k_-,P_-) \,j_N^{\lambda_\gamma}(q)\Big|_{k_0=E_-}
\nonumber\\&& 
-\frac1{k_0}\, \overline{\Upsilon}^\lambda(\widetilde k_+,P_+)\, S_d(\widetilde p)\,\Big[\Upsilon^{\lambda'}(\widetilde k_-,P_-) + \frac{{\bf k}\cdot{\bf q}}{m^2}\,\widehat\Gamma_{\rm off}^{\lambda'}(\widetilde k_-,P_-)\Big] \,j_N^{\lambda_\gamma}(q)\Big|_{k_0=E_+} \Bigg\}\, .\quad
\label{eq:B&V1-1}
\eea
%
%\end{widetext}
where the off-shell terms have been reduced using  
\bea
(m^2-\tilde k^2_\pm)\big|_{k_0=E_\mp}=\pm2{\bf k}\cdot{\bf q}\, .
\eea
Equation (\ref{eq:B&V1-1}) is further reduced by  shifting ${\bf k}\pm \frac12{\bf q}\to{\bf k}$ in the terms involving $\Gamma_{\rm off}$, and introducing the generic traces
\begin{subequations}
\bea
{\cal B}_{n,i}(k_0)&\equiv& {\rm tr}\Big[\overline{\Upsilon}^\lambda(\tilde k_+,P_+) S_d(\tilde p) \Upsilon^{\lambda'}(\tilde k_-,P_-) j_i^{\lambda_\gamma}(q)\Big]
\nonumber\\
&=&(-)^{\lambda_\gamma}  {\rm tr}\Big[\overline{\Upsilon}^{\,-\lambda'}(\tilde k_-,P_-) S_d(\tilde p) \Upsilon^{-\lambda}(\tilde k_+,P_+) j_i^{-\lambda_\gamma}(q)\Big]=(-)^{\lambda_\gamma}
{\cal B}_{n,i}(k_0)\Big|_{q\to-q}
 \label{eq:traceB}
\\
{\cal C}_{n,i}(\Gamma\,\Gamma_{\rm off})&=& {\rm tr}\Big[\overline{\Upsilon}^\lambda(k,P_+) S_d(P_+-k) \widehat\Gamma^{\lambda'}_{\rm off}(k-q,P_-)j_i^{\lambda_\gamma}(q)\Big] 
\nonumber\\
&=&(-)^{\lambda_\gamma} {\rm tr}\Big[\overline{\widehat\Gamma}^{\,-\!\lambda'}_{\rm off} (k-q,P_-)S_d(P_+-k) \Upsilon^{-\lambda}(k,P_+)j_i^{-\lambda_\gamma}(-q) \Big] =(-)^{\lambda_\gamma}{\cal C}_{n,i}(\Gamma_{\rm off}\,\Gamma)\Big|_{q\to-q}
 \label{eq:traceB1a}
\eea
\end{subequations}
where, in (\ref{eq:traceB1a}), the four-vector $k$ is always on-shell.     
This allows the B $+\left<V_1\right>$ contributions to the helicity amplitudes  to be written
%\begin{widetext}  
%
\bea
{\cal J}_n(q)\Big|_{{\rm B}+V_1}\!\!=e_0 F_1(Q^2)\int_k&&\Bigg\{\left[\frac{mE_k}{{\bf k}\cdot{\bf q}}\right] \left(\frac{{\cal B}_{n,1}(k_0)}{k_0}\Big|_{-}- \frac{{\cal B}_{n,1}(k_0)}{k_0}\Big|_{+}\right) -
\frac1{m} {\cal C}_{n,1}(\Gamma\,\Gamma_{\rm off})-\frac1{m}\epsilon_{n3}{\cal C}_{n,1}(\Gamma\,\Gamma_{\rm off})\Big|_{q\to-q}\Bigg\}
\nonumber\\
+e_0 F_2(Q^2)\int_k&&\Bigg\{\left[\frac{m E_k}{{\bf k}\cdot{\bf q}}\right] \left(\frac{{\cal B}_{n,2}(k_0)}{k_0}\Big|_{-}- \frac{{\cal B}_{n,2}(k_0)}{k_0}\Big|_{+}\right)   -
\frac1{m} {\cal C}_{n,2}(\Gamma\,\Gamma_{\rm off})-\frac1{m}\epsilon_{n3}{\cal C}_{n,2}(\Gamma\,\Gamma_{\rm off})\Big|_{q\to-q}\Bigg\},\qquad\quad
\label{eq:B&V1-2}
\eea
\end{widetext}
where $|_{\pm} \to |_{k_0=E_\pm}$.  

The formulae for the ${\cal B}$ and ${\cal C}$ traces, when expressed in terms of the invariant functions introduced in Sec.\ \ref{sec:wave}, are lengthly and are given in Appendix \ref{app:B}.

\subsection{Numerical calculation of the Form Factors}

Computation of the form factors involves not only the wave function $\Psi$ and the vertex function $\Gamma$, but also the special wave function $\Psi^{(2)}$ 
and the subtracted vertex functions $\widehat\Gamma$.  
The calculation of the interaction current contributions has been simplified by introducing the special functions $\Psi^{(2)}$ and $\widehat\Gamma$, and their Dirac conjugates.
The kernels that produce the bound state functions $\Psi^{(2)}$ and $\widehat\Gamma$ were already been given in a very general form in Ref.\ I, 
but, for convenience,  are given in a more explicit detail in Appendix \ref{app:truncwf}.

The numerical calculation of the form factors involves the following steps:

(i) Start from the invariant functions $\{F, G, H, I\}$ and $\{A, B, C, D\}$ given in  (\ref{eq:2.3}) and (\ref{eq:AtoF}), or the $K_i$ defined in Eq.~(\ref{eq:BSvertexa}).  In the rest frame these are  functions of $k=|{\bf k}|$ and $k_0$, and are constructed from the eight helicity amplitudes $z_0^{\pm \pm}$, and $z_1^{\pm \pm}$ as described in Appendix \ref{app:H}.

(ii) Replace the rest frame arguments $k$, and $k_0$ by the correctly transformed arguments $R$ and $R_0$ using the general definitions given in Eqs.~(\ref{eq:kp&eta}).  The specific realization of these general definitions depends on the diagram being evaluated; detailed expressions for each diagram are given in Eqs.~(\ref{eq:pp&m}), (\ref{eq:offshellargs}), and (\ref{eq:B4to6args}), (\ref{eq:B4to6argsa}).

(iii) Using the invariants with the proper arguments, evaluate the A+$V_2$ contributions to the helicity amplitudes (\ref{eq:A&V2-3})  
using Eqs.~(\ref{eq:traceA1})--(\ref{eq:traceA2}).  Evaluate the B+$V_1$  contributions (\ref{eq:B&V1-2}) using Eqs.~(\ref{eq:traceB1}) and (\ref{eq:traceB2}) and  (\ref{eq:traceC1})--(\ref{eq:traceC2}).  The total result is the sum of these two contributions.

(iv) Extract the individual form factors using the relations (\ref{eq:dffmatrix}).

These general results do not reduce to simple expressions for the form factors in terms of the the familiar $u, w, v_t,$ and $v_s$ wave functions previously defined in the literature and shown in  Eqs.~(\ref{eq:uz}) and (\ref{eq:Ftou}).  Still, to make connections with the older literature it is useful to express the result for the static moments in terms of leading terms involving integrals over products of $u, w, v_t, v_s$ and ``corrections.''  The charge and magnetic moment will be reduced in this way in the following sections.

\section{Charge} \label{sec:GC0}

The charge and normalization have been previously discussed in many previous references, including Ref.\ I,  so the purpose here is to see how the same result emerges from the general expressions (\ref{eq:A&V2-3}) and (\ref{eq:B&V1-2}).  Using the results of Eqs.\ (\ref{eq:BC}) and (\ref{eq:BC2}), the contributions from (\ref{eq:A&V2-3}) are
\begin{widetext}
\bal
G_c(0)\Big|_{{\rm A}+V_2}=e_0\int_0^\infty k^2dk\Bigg\{&f_{00}\Big[u^2 + w^2+v_t^2+v_s^2\Big] +\frac{g_{00}}{4m^2}\Big[(2E_k-m_d)^2(u^2+w^2)+m_d^2(v_t^2+v_s^2)\Big]
\nonumber\\&-2\Big[uu^{(2)}+ww^{(2)}+v_tv_t^{(2)}+v_sv_s^{(2)}\Big]\Bigg\}
\nonumber\\
=e_0\int_0^\infty k^2dk\Bigg\{&u^2 + w^2+v_t^2+v_s^2-4a(p^2)(E_k-m_d)\Big[(2E_k-m_d)(u^2+w^2)-m_d(v_t^2+v_s^2)\Big]\
\nonumber\\&-2\Big[uu^{(2)}+ww^{(2)}+v_tv_t^{(2)}+v_sv_s^{(2)}\Big]\Bigg\}
 , \label{eq:GcA}
\end{align}
%
%\end{widetext}
%
with $f_{00}$ and $g_{00}$ defined in Eq.~(\ref{eq:FGexpand}) [with $a$ defined in Eq.~(\ref{4.7})], and the second line was obtained by using  $p=P-k$, which reduces $f_{00}$ in the rest frame to
\bea
f_{00}&=&1+a(p^2)\,2m_d(2E_k-m_d) .
\eea
The special wave functions $z^{(2)}$ are obtained from $\Psi^{(2)}$ in precisely the same way that the $z$ are obtained from $\Psi$.

Next,  using the general results (\ref{eq:combinedB}) the contributions to the charge from (\ref{eq:B&V1-2}) are 

\bea
G_c(0)\Big|_{{\rm B}+V_1}=e_0\int_0^\infty k^2dk\Bigg\{&&u^2+w^2+v_t^2+v_s^2 -4a(p^2)(E_k-m_d)\Big[(2E_k-m_d)(u^2+w^2)-m_d(v_t^2+v_s^2)\Big]
\nonumber\\&&\quad-2\Big(u[\delta_+\hat u]_{k_0}+w[\delta_+\hat w]_{k_0}\Big)+2\Big(v_t [\delta_-\hat v_t]_{k_0}+v_s [\delta_-\hat v_s]_{k_0}\Big)\Bigg\} \label{eq:GcB}
\eea
\end{widetext}
where the functions $\delta_+\hat u,\cdots\delta_-\hat v_s$ were defined in (\ref{eq:deltauv}), and if $z=h\,\hat z$,  the derivative is $z_{k_0}=h\,d\hat z(k_0)/dk_0|_{k_0=Ek}$.  

The charge must be  sum of the two contributions (\ref{eq:GcA}) and (\ref{eq:GcB})
\bea
1=G_c(0)\Big|_{{\rm A}+V_2}+\;G_c(0)\Big|_{{\rm B}+V_1}
\label{eq:norm}
\eea
which is also identical to the normalization condition (2.55) of Ref.~I.  

The first lines of (\ref{eq:GcA}) and (\ref{eq:GcB}) are identical; their sum is the RIA contribution.   This contribution arises from diagrams A and B in different ways.  The contribution from the A diagram includes the $f_0$ and $g_0$ factors in the off-shell current; these factors do not appear in the B diagram, but similar contributions arise from the expansion of the dressed propagator $S_d$.  Of course, the fact that these contributions  are identical is not really surprising; it is a consequence of current conservation.  The remaining factors originate from the interaction currents generated by the {\it reduced\/} kernel.

The remaining terms from (\ref{eq:norm}) must equal the contribution from the energy derivative of the reduced kernel, $\partial \widetilde V/\partial P_0$,  which appears ijn the normalization condition discussed in Ref.~I and elsewhere.  This leads to the identity
\begin{widetext}
\bea
-\frac1{2m_d}&&\int_k \int_{k'}
 \overline{\it \Psi}_{\lambda_n\alpha}^{\lambda}(k,P)h(p)\,\frac{\partial}{\partial\,P_0}\widetilde V_{\lambda_n\lambda_n',\alpha\alpha'}(k, k'; P)\,h(p'){\it \Psi}_{\alpha'\lambda_n'}^{\lambda'}(k',P) 
 \nonumber\\
 &&=-\int_0^\infty k^2dk
\Bigg\{uu^{(2)}+ww^{(2)}+v_tv_t^{(2)}+v_sv_s^{(2)}+
u[\delta_+\hat u]_{k_0}+w[\delta_+\hat w]_{k_0}-v_t [\delta_-\hat v_t]_{k_0}-v_s [\delta_-\hat v_s]_{k_0}\Bigg\} \label{eq:identityVp}
\eea
\end{widetext}
where we have set $e_0=\frac12$.  This interesting identity, discussed already in Sec.~\ref{sec:intro}, shows how the energy derivative of $\tilde V$ can be expressed in terms of special wave functions $z^{(2)}$ and $\hat z$.  

\section{Magnetic moment}  \label{sec:GM0}

Predictions for the magnetic moment are presented in this section.  The new interaction current current contributions, which together account for about 5\% of the charge, insure that many new terms not previously encountered will contribute, and the result for the magnetic moment is the first important test of the CST.

The contributions from diagram (A)$+\left<V_2\right>$ were given in Eqs.~(\ref{eq:M1a})-(\ref{eq:M4a}) and from the diagram (B)$+\left<V_1\right>$ in Eqs.~
(\ref{eq:M2B}) and (\ref{eq:M2C}).  Adding these together and keeping the leading $z_\ell^{(2)}$ contributions (those depending on products involving $u$ and $w$), and setting $e_0=\frac12$ gives 
 \bea
 \mu_d\simeq \, \mu_s \Big(P_S-\frac12P_D\Big)+\frac34 P_D +\widetilde\Delta \mu_d\, ,
 \label{eq:nonrelmu}
 \eea
where the correction terms are the sum of several contributions of different origin.  This form resembles the nonrelativistic result, but is misleading because the sum of the S and D-state probabilities is not equal to unity in the relativistic theory.  Instead, it is more instructive to write the result in the form  
\bea
\mu_d=\mu_s +\Delta\mu_d \label{eq:mag1}
\eea
where, for the nonrelativistic theory, the correction is
\bea
\Delta \mu_d\to \mu_{\rm NR}= \frac34P_D(1-2\mu_s) \label{eq:munr}
\eea

To obtain a similar form from the CST, we use the relativistic normalization condition.  In the approximations used to obtain the leading terms for the magnetic moment, the normalization (or charge) is
\begin{widetext}
\bea
1 = \int_0^\infty k^2 dk\Big\{&&
u^2+w^2+v_t^2+v_s^2 +4a(p^2)m\Big[\delta_k(u^2+w^2)-2m(v_t^2+v_s^2)\Big]
\nonumber\\
&&-u[\delta_+\hat u]_{k_0}-w[\delta_+\hat w]_{k_0}+v_t [\delta_-\hat v_t]_{k_0}+v_s [\delta_-\hat v_s]_{k_0}-uu^{(2)}-ww^{(2)}-v_tv_t^{(2)}-v_sv_s^{(2)}\Big\}\, .
\eea
Multiplying this by $\mu_s$, and adding and subtracting it from (\ref{eq:nonrelmu}), gives a a expression of the from (\ref{eq:mag1})  for the magnetic moment, where the correction will be written as a sum of terms
\bea
\Delta\mu_d&=&\mu_{\rm NR}+\mu_{Rc}+\mu_{h'}+\mu_{V_2}+\mu_{V_1}+\mu_{\rm int}+\mu_P+\mu_{\chi}\qquad
\eea
where the individual contributions are
\bea
\mu_{Rc}&=&\int_0^\infty k^2dk \Big[\frac{E_k-m}{E_k}\Big]\Big\{-\mu_s\Big(u^2+\frac12w^2-\sqrt{2}uw\Big)-\frac1{4}\Big(5u^2-\frac{89}{4}w^2+\frac{79}{2\sqrt{2}}uw\Big)\Big\}\nonumber\\
\mu_{h'}&=&\int_0^\infty k^2dk\,a(p^2)m\Big\{4(1-\mu_s)(1-\omega_2)\delta_ku^2-\mu_s\Big[2(2+\omega_2)\,\delta_k w^2-m(6\,v_t^2+4v_s^2+4\sqrt{2}v_tv_s)\Big]
\nonumber\\&&\qquad
+\frac{\delta_k}{2}\Big[(3+4\omega_2)w^2-\sqrt{2}uw\Big]-\frac{m}{2}(9v_t^2+6v_s^2+8\sqrt{2}v_tv_s)\Big\}
\nonumber\\
\mu_{V_2}&=&\int_0^\infty\frac{k^2dk}{2}\Big\{(2\mu_s-1)\frac32ww^{(2)} +\mu_s\Big(
3v_tv_t^{(2)}+2v_s v_s^{(2)}\Big)+(\mu_s-1)\sqrt{2}(v_tv_s^{(2)}+v_sv_t^{(2)})-\frac12v_tv_t^{(2)}+v_sv_s^{(2)}-m^{I(2)}\Big\}
\nonumber\\
\mu_{V_1}&=&\int_0^\infty k^2dk\Big\{(2\mu_s-1)\Big(\frac34w[\delta_+\hat w]_{k_0}-\frac14v_t[\delta_-\hat v_t]_{k_0}-\frac12 v_s[\delta_-\hat v_s]_{k_0}\Big)
%\nonumber\\&&\qquad
-\frac{\mu_s}{\sqrt{2}}\Big(v_t[\delta_-\hat v_s]_{k_0}+v_s[\delta_-\hat v_t]_{k_0}\Big)\Big\}
\nonumber\\
\mu_{\rm int}&=&-\frac{m}{2\sqrt{6}}\int_0^\infty k^2dk\Big\{u'(v_t-\sqrt{2}v_s)
%\nonumber\\&&\qquad
-w(\sqrt{2}v_t+v_s)'+\frac{1}{k}w(\sqrt{2}\,v_t+v_s)\Big\}\, .
\nonumber\\
\mu_P&=&\int_0^\infty k^2dk\Big\{-\mu_s\Big(v^2_t+v_s^2+\sqrt{2} v_tv_s\Big)
%\nonumber\\&&\qquad
-\frac14v^2_t
-\frac12v^2_s+\sqrt{2} v_tv_s
%\nonumber\\&&\qquad
+\frac{3k}{8\sqrt{2}}\Big(v_tv_s'-v_t'v_s\Big)\Big\}
\nonumber\\
\mu_{\chi}&=&-\int_0^\infty k^2dk\Big\{\frac{m\,z_0^{--}}{2k}\Big(\sqrt{2}\,v_s+kv_t'\Big)+\frac{m\,z_1^{--}}{2k}\Big(\sqrt{2}\,v_t+v_s+kv_s'\Big)\Big\}
\, ,
\label{eq:Magcorr2}
\eea
\end{widetext}
where $\omega_2=1$ was defined in Eq.~(\ref{eq:om2def}) and $m^{I(2)}$  in Eq.~(\ref{eq:m2I}).   
Each of these terms has a different physical origin, as discussed in Sec.~\ref{sec:magdis}.

Many remaining details are discussed in the Appendices.

\acknowledgements

This work was partially supported by Jefferson Science
Associates, LLC, under U.S. DOE Contract No. DE-AC05-
06OR23177.

\appendix

\section{Connections between the invariant functions and component wave functions} \label{app:H}

This Appendix shows how to connect the invariant functions defined in Sec.~\ref{sec:wave} to  the helicity amplitudes that are calculated in the
 code described in Refs.~\cite{Gross:2008ps,arXiv:1007.0778}.  The helicity amplitudes are simple linear combinations of the more familiar component wave functions $u, w, v_t$, and $v_s$.  The  traces given in Appendix \ref{app:B} are bilinear products of the invariant functions.

 In the rest frame the relativistic wave function  (\ref{eq:27}) can be expanded in a set of helicity spinors  $u^\rho({\bf k},\lambda)$
\bea
{\it \Psi}_{\alpha\beta}^{\lambda_d}(k,P)&=&\frac1{N_d}\frac{m}{E_k} \sum_{{\lambda_1\rho_1 }\atop
{\lambda_2\rho_2}} u^{\rho_2}_{2\alpha}({\bf k},\lambda_2)\,u^{\rho_1 {\rm T}}_{1\beta'}({\bf k},\lambda_1)\gamma^0_{\beta'\beta}
\nonumber\\&& \qquad\qquad\qquad\times
\phi^{\rho_1\rho_2}_{\lambda_1\lambda_2,\lambda_d}({\bf k})
\nonumber\\
{\it \Psi}_{\alpha\lambda_1}^{\lambda_d}(k,P)&=&\frac1{N_d} \sum_{{\lambda_2\rho_2 }} u^{\rho_2}_{2\alpha}({\bf k},\lambda_2)
%\nonumber\\&& \times
\phi^{\rho_2}_{\lambda_1\lambda_2,\lambda_d}({\bf k}) \label{eq:normhel}
\eea
where $\rho=\pm$ is the rho-spin of the particle (if particle 1 is on-shell,  $\rho_1= +$),  ${\bf k}$ is the three momentum of particle 1 in the deuteron rest frame, and $\phi^{\rho_1\rho_2}_{\lambda_1\lambda_2,\lambda_d}$ are normalized helicity amplitudes  defined by this expansion.  The second form of (\ref{eq:normhel}), obtained from the first using the orthogonality relations (\ref{eq:othrel}) below, will be used only when $\rho_1=+$; reference to $\rho_1$ is suppressed for simplicity.   The transpose symbol is to remind us that, if $\psi_{\alpha\beta}$ is to be viewed as a matrix, then $u_{\beta'}$ must be interpreted as a row vector, but is redundant when the indices are shown explicitly.   
The normalization constant $N_d$ is
\bea
N_d =\frac{1}{\sqrt{(2\pi)^3 2m_d}}
\eea
and the helicity spinors [c.f. Ref.~\cite{Gross:2008ps}, Eqs.~(E1) and (E7)]  are
\bea
u^\rho_ 1({\bf k},\lambda)&=&N_\rho(k,\lambda)
\otimes\chi_\lambda(\theta) 
\nonumber\\
u^\rho_ 2({\bf k},\lambda)&=&N_\rho(k,\lambda)
\otimes\chi_{-\lambda}(\theta)         
\eea
with 
\bea
N_+(k,\lambda)&=&
\left(
\begin{array}{c}
  \cosh\sfrac12\zeta   \\
  \\
  2\lambda\sinh  \sfrac12\zeta
\end{array}
\right)
\nonumber\\
N_-(k,\lambda)&=&
\left(
\begin{array}{c}
 - 2\lambda\sinh  \sfrac12\zeta    \\
  \\
\cosh\sfrac12\zeta 
\end{array}
\right)
\eea
where $\tanh\zeta=k/E_k$, and, for momenta limited to the $\hat x\hat z$ plane, so that ${\bf k}=\{k\sin\theta,0,k\cos\theta\}$, the two-component helicity spinors are
\bea
\chi_{_{1/2}}(\theta)&=&
\left(
\begin{array}{c}
  \cos\sfrac12\theta   \\
  \\
  \sin  \sfrac12\theta
\end{array}
\right) \qquad \chi_{_{1/2}}(\theta)=
\left(
\begin{array}{c}
 - \sin  \sfrac12\theta    \\
  \\
  \cos\sfrac12\theta
\end{array}
\right).\qquad
%\nonumber\\
\eea

These helicity spinors are real, so that $\overline{u}=u^{\rm T}\gamma^0$, and they satisfy the orthogonality relations
\bea
\overline{u}^{\rho' }({\bf k},\lambda') \gamma^0 u^\rho({\bf k},\lambda)&=&\delta_{\lambda'\lambda}\delta_{\rho'\rho}\frac{E_k}{m}
\nonumber\\
\overline{u}^{\rho}({\bf k},\lambda') u^\rho({\bf k},\lambda)&=&\rho\,\delta_{\lambda'\lambda}\, ,
\label{eq:othrel}
\eea
leading to the inverse relation
\bea
\phi^{+\rho_2}_{\lambda_1\lambda_2,\lambda_d}({\bf k})&=&N_d\frac{m}{E_k}\,\overline{u}^{\rho_2}_{2}({\bf k},\lambda_2)
\gamma^0{\it \Psi}^{\lambda_d}(k,P)
%\nonumber\\&&\times
\overline{u}^{+ {\rm T}}_{1}({\bf k},\lambda_1)
\nonumber\\
&=&N_d\frac{m}{E_k}\,\overline{u}^{\rho_2}_{2\alpha}({\bf k},\lambda_2)
\gamma^0_{\alpha\alpha'}{\it \Psi}_{\alpha' \lambda_1}^{\lambda_d}(k,P)\, .\qquad
\label{eq:helamps1}
\eea
This is further reduced by writing the wave function in terms of the vertex function, ${\cal G}$, and the propagator of particle 2, $S(p)$, and decomposing the rest frame propagator  for particle 2 into positive and negative energy parts (or its rho-spin $\pm$ components)
\bea
S_{\alpha\alpha'}(p)=\frac{m}{E_k}\sum_{\rho,\lambda} G^{\rho}(k_0,{\bf k}) u_2^\rho({\bf k},\lambda)\overline{u}_2^\rho({\bf k},\lambda)
\eea
where, if particle 1 is also off-shell so that $k=\{k_0,{\bf k}\}$, the components of the propagator are
\bea
G^+(k_0,{\bf k})&=&\frac1{E_k+k_0-m_d}\equiv \frac1{\;\,\delta_+}
\nonumber\\
G^-(k_0,{\bf k})&=&\frac{-1}{m_d+E_k-k_0}\equiv -\frac1{\;\,\delta_-} \, ,\label{eq:propk0}
\eea
where the arguments of $\delta_\pm$ will be suppressed.  In most cases particle 1 is on shell so that $k_0=E_k$,  and (\ref{eq:propk0}) reduce to (c.f. Eq.~(E14) of Ref.~\cite{Gross:2008ps}) 
\bea
G^+(E_k,{\bf k})&=&\frac1{2E_k-m_d}\equiv \frac1{\delta_k}
\nonumber\\
G^-(E_k,{\bf k})&=&-\frac1{m_d}\, .
\eea
Using the expansion (\ref{eq:propk0}), the helicity amplitudes (\ref{eq:helamps1}) reduce to the previously published form (c.f. Eq.~(3.10) of Ref.~\cite{arXiv:1007.0778}, except here $\phi$ is used in place of $\psi$ and there are other changes in notation) 
\bea
&&\phi^{+\rho_2}_{\lambda_1\lambda_2,\lambda_d}
({\bf k})
=N_d\frac{m}{E_k}G^{\rho_2}\, {\it {\cal G}}^{+\rho_2}_{\lambda_1\lambda_2,\lambda_d}
({\bf k})
\nonumber\\
&&\qquad=N_d\frac{m}{E_k}G^{\rho_2} \overline{u}_2^{\rho_2}({\bf k},\lambda_2) \Gamma^{\lambda_d}(k,P){\cal C}
%\nonumber\\&&\times 
\overline{u}_1^{+{\rm\rm T}}({\bf k},\lambda_1)\qquad \quad
 \label{eq:deuhel}
\eea
where {\it no\/} sum over the repeated index $\rho_2$ is implied.

In the general case (when $k_0\ne E_k$), the projection operators present in the $\Gamma$ of (\ref{eq:deuhel})   can be simplified by recalling that $p=P-k$, $k=\{k_0,{\bf k}\}$, and $\rho=\pm1$, giving
\bea
2m\overline{u}^{\rho_2}_2({\bf k},\lambda)&&\Theta(p)=\overline{u}^\rho_2({\bf k},\lambda)\Big[m-\gamma^0(m_d-k_0)-{\bm \gamma}\cdot {\bf k}\Big]
\nonumber\\
=&&\,-m\, d_2\, \overline{u}^\rho_2({\bf k},\lambda)\gamma^0
\nonumber\\
2m\overline{u}^{\rho_1}_1({\bf k},\lambda)&&\Theta(k)=\overline{u}^\rho_1({\bf k},\lambda)\Big[m-\gamma^0\,k_0+{\bm \gamma}\cdot {\bf k}\Big]
\nonumber\\
=&&\,-m\, d_1\, \overline{u}^\rho_1({\bf k},\lambda)\gamma^0 \label{eq:Thetaiden}
\eea
where
\bea
m\,d_1&=&k_0-\rho E_k
\nonumber\\
m\,d_2&=&m_d-k_0-\rho E_k
\eea
Because the helicity spinors are written as a direct product of $N_\rho$ and $\chi_\lambda$, each operating in its own 2$\times2$ space, it is convenient to similarily decompose the matrix $\Gamma^{\lambda_d}$.  To this end note that 
\bea
\gamma_\mu \xi^\mu_{\lambda_d}&=&
\left(
\begin{array}{cc}
  0  &  \; -{\bm\sigma}\cdot{\bm\xi}_{\lambda_d}   \\
  {\bm\sigma}\cdot{\bm\xi}_{\lambda_d} &     0  
\end{array}
\right) =-i\tau_2\otimes {\bm\sigma}\cdot{\bm\xi}_{\lambda_d} \qquad
\eea
where $\tau_i$ are the $2\times2$ operators operating in the $N_\rho$  Dirac space and $\sigma_i$ operate on the $\chi_\lambda$ spin space, and  the three-component deuteron polarization vectors (for an incoming deuteron), ${\bm \xi}^i_{\lambda_d}$ (with $i=1,2,3$) , are related to the four-vectors by
\bea
\xi^\mu_0&=&\{0,0,0,1\}=\{0,{\bm \xi}^i_0\}
\nonumber\\
\xi^\mu_\pm&=&\frac1{\sqrt{2}}\{0,\pm1,-i,0\}=\{0,{\bm \xi}^i_\pm\}
\eea
Also note that, in $2\times2$ form,
\bea
{\cal C}=-\tau_1\otimes i\sigma_2\, .
\eea

Using this notation, the matrix elements are reduced to the following convenient form
\bea
\phi^{\rho_1\rho_2}_{\lambda_1\lambda_2,\lambda_d}
({\bf k})=&& A^{\rho_1\rho_2}_{\lambda_1\lambda_2}
({k}) \Big(\chi^\dagger_{_{-\lambda_2}} i\sigma_2 \chi_{_{\lambda_1}}\Big)\,(\hat {\bf k}\cdot {\bm\xi}_{\lambda_d})
\nonumber\\
&&+B^{\rho_1\rho_2}_{\lambda_1\lambda_2}
({k}) \Big(\chi^\dagger_{_{-\lambda_2}}{\bm\sigma}\cdot{\bm\xi}_{\lambda_d} i\sigma_2 \chi_{_{\lambda_1}}\Big)\qquad
\nonumber\\
=&&-2\lambda_1\delta_{\lambda_1,\lambda_2}\,d^1_{\lambda_d,0}(\theta)A^{\rho_1\rho_2}_{\lambda_1\lambda_2}
({k}) 
\nonumber\\
&&+\sqrt{2}^{|\lambda|}d^1_{\lambda_d,\lambda}(\theta)\,B^{\rho_1\rho_2}_{\lambda_1\lambda_2}
({k})  \label{eq:psik}
\eea
where the identities (C26) from Ref.~\cite{arXiv:1007.0778} were used to evaluate the angular matrix elements.  The coefficient $A$ contributes only when $\lambda_1=\lambda_2$ and both  of the coefficients are indepent of the deuteron polarization and the angle $\theta$.  Using the definition of $\Gamma$ when both particles are off-shell, Eq.~(\ref{eq:BSvertexa}), and the simplifications (\ref{eq:Thetaiden}), they reduce to
\bea
A^{\rho_1\rho_2}_{\lambda_1\lambda_2}
({k})=&&N_d\frac{m}{E_k}G^{\rho_2}(k)\overline{N}_{\rho_2}(k,\lambda_2)\frac{k}{m}\Bigg\{G-K_4\,d_2\,d_1 
\nonumber\\
&&+({I}d_2-K_2\,d_1) \tau_3
\Bigg\}(\tau_1)\overline{N}^{\rm T}_{\rho_1}(k,\lambda_1)
\nonumber\\
B^{\rho_1\rho_2}_{\lambda_1\lambda_2}
({k})=&&N_d\frac{m}{E_k}G^{\rho_2}(k)\overline{N}_{\rho_2}(k,\lambda_2)\Bigg\{(F+K_3\,d_2\,d_1)i\tau_2 \quad
\nonumber\\
&&
+(H\,d_2+K_1\,d_1)\tau_1
\Bigg\}(\tau_1)\overline{N}^{\rm T}_{\rho_1}(k,\lambda_1)\qquad  \label{eq:A&B}
\eea
where $\gamma^0\to \tau_3$, and ${\cal C}\gamma^0=-\gamma^0{\cal C}$ was used.

Before evaluating the matrix elements (\ref{eq:A&B}), it is convenient to project the partial wave amplitudes from  (\ref{eq:psik}).    Using the definition given in Eq.~(3.21)  of Ref.~\cite{arXiv:1007.0778}, these  are 
\bea
\phi^{\rho_1\rho_2}_{\lambda_1\lambda_2,\lambda_d}
(k)&=&\sqrt{\frac{3}{4\pi}}\int d\Omega_k \, D^{1*}_{\lambda_d,\lambda}(\phi,\theta,0)\phi^{\rho_1\rho_2}_{\lambda_1\lambda_2,
\lambda_d}({\bf k})
\nonumber\\
&=&\sqrt{3\pi}\int_0^1\sin\theta d\theta\, d^1_{\lambda_d, \lambda}(\theta)\phi^{\rho_1\rho_2}_{\lambda_1\lambda_2,\lambda_d}
({\bf k}),\qquad\quad \label{eq:partialwaves}
\eea
where here $\lambda=\lambda_1-\lambda_2$ and  the $d^J_{\lambda', \lambda}(\theta)$ are the rotation matrices, in this case for $J=1$ and $\lambda'=\lambda_d$, where $J$ is the angular momentum and $\lambda_d$ the helicity of the deuteron.  The normalization of the $J=1$ $d$ matrices is independent of helicity
\bea
\int_0^1 \sin\theta\,d\theta \,[d^1_{\lambda_d,\lambda}(\theta)]^2 =\frac23\, , \label{eq:dnorm}
\eea
and hence the result for the partial waves is independent of the deuteron helicity.  Under parity, the amplitudes transform into each other under the substitution $\lambda_1,\lambda_2\to -\lambda_1,-\lambda_2$.  Hence the partial wave amplitudes can conveniently written in terms of a standard helicity with $\lambda_1=\lambda_0=\frac12$.  Writing a separate formula for cases when $\lambda_1=\lambda_2=\lambda_0$ and $\lambda_1=-\lambda_2=\lambda_0$ gives
\bea
&&\phi^{\rho_1\rho_2}_{\lambda_0\lambda_0,\lambda_d}
({k})\equiv z^{\rho_1\rho_2}_{0}
({k})=\sqrt{\frac{4\pi}{3} }\Big[ 
B^{\rho_1\rho_2}_{\lambda_0,\lambda_0}(k)
- A^{\rho_1\rho_2}_{\lambda_0,\lambda_0}(k)\Big]
\nonumber\\
&&\phi^{\rho_1\rho_2}_{\lambda_0,-\lambda_0,\lambda_d}
({k})\equiv z^{\rho_1\rho_2}_{1}
({k})=\sqrt{\frac{8\pi}{3} }\,B^{\rho_1\rho_2}_{\lambda_0,-\!\lambda_0}(k)\, .
\label{eq:helamps}
\eea
There are therefore 8 independent amplitudes from which the 8 invariants that define $\Gamma$ can be determined.  

It is instructive to show explicitly that the parity relation holds.  To this end, introduce the matrix elements
\bea
D^{\rho_1\rho_2}_{i\,\lambda_1\lambda_2}(k)=\overline{N}_{\rho_2}(k,\lambda_2)\bar \tau_i \overline{N}^{\rm T}_{\rho_1}(k,\lambda_1)
\eea
where $i=0,1,2,3$ with $\bar\tau_i=\tau_i$ for $i=1,3$, $\bar\tau_2=i\tau_2$, and $ \bar \tau_0={\bf 1}$.  The entire helicity dependence of the partial waves is contained in the helicity dependence of the $D$'s, and this can be established from  the symmetry property
\bea
\overline{N}_\rho(k,\lambda)&=&\rho\overline{N}_\rho(k,-\lambda)\tau_3
\eea
which leads to the relations
\bea
D^{\rho_1\rho_2}_{j\,\lambda_1\lambda_2}(k)&=&\rho_1\rho_2D^{\rho_1\rho_2}_{j\,-\lambda_1,-\lambda_2}(k) \quad j=0,3
\nonumber\\
D^{\rho_1\rho_2}_{j\,\lambda_1\lambda_2}(k)&=&-\rho_1\rho_2D^{\rho_1\rho_2}_{j\,-\lambda_1,-\lambda_2}(k) \quad j=1,2\, .\qquad\label{eq:Didentities1}
\eea
Examination of the definitions (\ref{eq:A&B}) shows that  only $D_0$ and $D_3$  contribute to $B$, and only $D_1$ and $D_2$  contribute to $A$.  The extra factor of $2\lambda_1$ multiplying $A$ is just sufficient to show that both of the helicity amplitudes (\ref{eq:helamps}) satisfy the expection relation for a $J=1$ amplitude:  $\phi^{\rho_1,\rho_2}=\rho_1\rho_2 \phi^{-\rho_1,-\rho_2}$ (c.f. Eq.~(E22) of Ref.~\cite{Gross:2008ps}), concluding the proof.

Working out the matrix elements (\ref{eq:A&B}) gives the eight independent helicity amplitudes in terms of the eight invariants that define the two-particle off shell vertex function (\ref{eq:BSvertexa}).    Using the notation
\bea
{\cal K}&=&\pi\sqrt{2m_d}
\eea
and recalling the definitions of $\delta_\pm$ and $G^\pm$ from Eq.~(\ref{eq:propk0})
\begin{widetext}
\bea
z^{++}_0&=&\frac{G^+ m}{\sqrt{6}\,{\cal K} E_k}\Big\{F+\frac{k^2}{m^2}G -\frac{E_k}{m^2}\delta_+H-\frac{(E_k-k_0)}{m^2} \Big[E_k K_1-\delta_+\big(K_3-\frac{k^2}{m^2}K_4\big)\Big]\Bigg\}
\nonumber\\
z^{++}_1&=&\frac{G^+ }{\sqrt{3}\,{\cal K}}\Big\{F-\frac{\delta_+}{E_k}H-\frac{(E_k-k_0)}{E_k}\Big[K_1-\frac{E_k\delta_+}{m^2}K_3\Big]\Bigg\}
\nonumber\\
z^{+-}_0&=&-\frac{G^- k}{\sqrt{6}\,{\cal K} E_k}\Big\{F-G+\frac{E_k\delta_-}{m^2} I +\frac{(E_k-k_0)}{m^2}\Big[E_k K_2-\delta_-\big(K_3+K_4\big)\Big]\Bigg\}
\nonumber\\
z^{+-}_1&=&\frac{G^- k}{\sqrt{3}\,{\cal K}\, m E_k}\Big\{\delta_- H-(E_k-k_0)K_1\Big\}
\nonumber\\
z^{-+}_0&=&-\frac{G^+ k}{\sqrt{6}\,{\cal K} E_k}\Big\{F-G
+\frac{E_k\delta_+}{m^2} I +\frac{(E_k+k_0)}{m^2}\Big[E_k K_2-\delta_+\big(K_3+K_4\big)\Big]\Bigg\}
\nonumber\\
z^{-+}_1&=&\frac{G^+ k}{\sqrt{3}\,{\cal K}\, m E_k}\Big\{\delta_+ H-(E_k+k_0)K_1\Big\}
\nonumber\\
z^{--}_0&=&-\frac{G^- m}{\sqrt{6}\,{\cal K} E_k}\Big\{F+\frac{k^2}{m^2}G  -\frac{E_k}{m^2}\delta_-H-\frac{(E_k+k_0)}{m^2} \Big[E_k K_1-\delta_-\big(K_3-\frac{k^2}{m^2}K_4\big)\Big]\Bigg\} 
\nonumber\\
z^{--}_1&=&-\frac{G^- }{\sqrt{3}\,{\cal K}}\Big\{F-\frac{\delta_-}{E_k}H-\frac{(E_k+k_0)}{E_k}\Big[K_1-\frac{E_k\delta_-}{m^2}K_3\Big]\Bigg\}\, . \label{eq:helwf}
\eea
Inverting these equations gives the eight invariants in terms of the helicity amplitudes.  The results are
\bea
F&=&\frac{\sqrt{3}\, {\cal K}}{2 E_km_d}\delta_+\delta_-\Bigg\{(E_k+k_0)\Big[z^{++}_1-\frac{m}{k} z^{+-}_1\Big]-(E_k-k_0)\Big[z^{--}_1+\frac{m}{k}z^{-+}_1\Big]\Bigg\}
\nonumber\\
G&=&\sqrt{\frac32}\frac{m{\cal K}}{E_km_d k^2}\delta_+\delta_-\Bigg\{(E_k+k_0)\Big[E_k\, z^{++}_0-m\frac{z^{++}_1}{\sqrt{2}}-k\frac{z^{+-}_1}{\sqrt{2}}\Big]
\nonumber\\
&&\qquad-(E_k-k_0)\Big[E_k\, z^{--}_0-m\frac{z^{--}_1}{\sqrt{2}}+k\frac{z^{-+}_1}{\sqrt{2}}\Big]\Bigg\}
\nonumber\\
H&=&-\frac{\sqrt{3}m{\cal K}}{2m_d\,k}\Bigg\{(E_k+k_0)\delta_-\,z^{+-}_1+(E_k-k_0)\delta_+\,z^{-+}_1\Bigg\}
\nonumber\\
I&=&\sqrt{\frac32}\frac{m^2{\cal K}}{E_km_dk^2}\Bigg\{(E_k+k_0)\Big[\delta_+\Big( m \,z^{++}_0-\frac{E_k}{\sqrt{2}}z^{++}_1\Big)+\delta_-\,k \,z^{+-}_0\Big]
\nonumber\\&&\qquad
-(E_k-k_0)\Big[\delta_-\Big( m \,z^{--}_0-\frac{E_k}{\sqrt{2}}z^{--}_1\Big)-\delta_+\,k \,z^{-+}_0\Big]\Bigg\}
\nonumber\\
K_1&=&-\frac{\sqrt{3}m{\cal K}}{2m_d\,k}\delta_+\delta_-\Big[z^{+-}_1+  z^{-+}_1\Big]
\nonumber\\
K_2&=&\sqrt{\frac32}\frac{m^2{\cal K}}{E_km_dk^2}\delta_-\delta_+ \Big(m \,z^{++}_0-E_k\frac{z^{++}_1}{\sqrt{2}}-k z^{+-}_0 -m\,z^{--}_0+E_k\frac{z^{--}_1}{\sqrt{2}}-k z^{-+}_0\Big)
\nonumber\\
K_3&=&-\frac{\sqrt{3}m^2{\cal K}}{2E_km_d\,k}\Bigg\{ \delta_+ (k\,z^{++}_1+m\,z^{-+}_1)-\delta_- (k\,z^{--}_1-m\,z^{+-}_1)\Bigg\}
\nonumber\\
K_4&=&\frac{\sqrt{3}m^3{\cal K}}{2m_d\,k^2}\Bigg\{\delta_+ \Big( \sqrt{2} z^{++}_0-\frac{m}{E_k}z^{++}_1 +\frac{k}{E_k} z^{-+}_1\Big)-\delta_- \Big( \sqrt{2} z^{--}_0-\frac{m}{E_k}z^{--}_1 -\frac{k}{E_k} z^{+-}_1\Big)
\Bigg\} \, . \label{eq:AtoK}
\eea
\end{widetext}
When particle 1 is on shell, so that $k_0=E_k$, the first four amplitudes reduce to
\bea
F&=&\sqrt{3}\,{\cal K}\,\delta_k\Big[z^{++}_1-\frac{m}{k}z^{+-}_1\Big]
\nonumber\\
G&=&\sqrt{3} \,{\cal K}\frac{m\,\delta_k}{k^2}\Big[\sqrt{2} E_k z^{++}_0-mz^{++}_1-k\,z^{+-}_1\Big]
\nonumber\\
H&=&-\sqrt{3}\,{\cal K}\,m\,E_k\frac{z^{+-}_1}{k}
\nonumber\\
I&=&\sqrt{6}\,{\cal K}\frac{m^2}{k^2}\Big[\frac{m\delta_k}{m_d}\Big(z^{++}_0-\frac{E_k}{\sqrt{2}m} z^{++}_1\Big) +k\,z^{+-}_0\Big]\, ,\qquad\quad \label{eq:Fz}
\eea
%
%\end{widetext}
with $\delta_k=\delta_+(E_k,{\bf k})=2E_k-m_d$.  
These are uniquely determined by the the four on-shell helicity amplitudes  with $\rho_1=+$.  If these amplitudes are expressed in terms of the $u, w, v_t$, and $v_s$ amplitudes previously defined in the literature (see Eq.~(C31) of Ref.~ \cite{arXiv:1007.0778}), 
\bea
z_0^{++}&=&\frac1{\sqrt{6}}(u+\sqrt{2}w)
\nonumber\\
z_1^{++}&=&\frac1{\sqrt{6}}(\sqrt{2}u-w)
\nonumber\\
z_0^{+-}&=&-\frac1{\sqrt{2}}v_s
\nonumber\\
z_1^{+-}&=&-\frac1{\sqrt{2}}v_t \label{eq:uz}
\eea
the well known expansions of $F,G,H$, and $I$ derived in Ref.~\cite{Buck:1979ff} are obtained, reproduced here for completeness:
\bea
F&=&{\cal K}\delta_k\Big[u-\frac{w}{\sqrt{2}}+\sqrt{\frac32}\frac{m}{k}v_t\Big]
\nonumber\\
G&=&{\cal K}\delta_k\, m\Big[\frac{u}{E_k+m}+(2E_k+m)\frac{w}{\sqrt{2}\,k^2}+\sqrt{\frac32}\frac{v_t}{k}\Big]
\nonumber\\
H&=&{\cal K}E_k\,m\sqrt{\frac32}\frac{v_t}{k}
\nonumber\\
I&=&-\frac{{\cal K}\delta_k m^2}{m_d}\Big[\frac{u}{E_k+m}-(E_k+2m)\frac{w}{\sqrt{2}\,k^2}\Big]
\nonumber\\&&
-\sqrt{3}{\cal K}m^2 \frac{v_s}{k} \, .
 \label{eq:Ftou}
\eea

The on-shell values of the $K_i$ invariants depend on all eight of the helicity amplitudes.  Because of their historical importance, we will continue to express the helicity amplitudes (\ref{eq:uz}) in terms of the $u,w,v_t,v_s$ wave functions, but will use the original notation for the others, giving
\begin{widetext}
\bea
K_1&=&\sqrt{\frac32}\frac{{\cal K}\delta_km}{2k} \Big[v_t-\sqrt{2}\,z_1^{-+}\Big]
\nonumber\\
K_2&=&-\frac{{\cal K}\delta_k m^2}{2E_k}\Big[\frac{u}{E_k+m}-(E_k+2m)\frac{w}{\sqrt{2}\,k^2}- \sqrt{3}\, \frac{v_s}{k}
%\nonumber\\&&\qquad
-\frac{\sqrt{3}}{k^2}(E_k z_1^{--}-\sqrt{2}\,mz_0^{--})+\sqrt{6}\,\frac{z_0^{-+}}{k}\Big]\qquad
\nonumber\\
K_3&=&-\frac{{\cal K} m^2}{2E_k}\Big[\frac{\delta_k}{m_d}\left(u-\frac{w}{\sqrt{2}}\right)-\sqrt{\frac32}\frac{m}{k}v_t 
%\nonumber\\&&\qquad
-\sqrt{3}\left(z_1^{--}-\frac{\delta_k\,m}{k\,m_d}z_1^{-+}\right)\Big]
\nonumber\\
K_4&=& \frac{{\cal K}\,m^3}{2E_k}\Big[\frac{\delta_k}{m_d} \left(\frac{u}{E_k+m}+(2E_k+m)\frac{w}{\sqrt{2}\,k^2}+\sqrt{3}\frac{z_1^{-+}}{k}\right)-\sqrt{\frac32}\frac{v_t}{k}+\frac{\sqrt{3}}{k^2}(m\,z_1^{--}-\sqrt{2}\,E_kz_0^{--})\Big] \label{eq:Ktou}
\eea
\end{widetext}

The wave functions can be transformed into coordinate space (for a full discussion see Ref.\ \cite{arXiv:1007.0778}).  Denoting the typical wave function by $z_\ell$ (so that $z_0=u$, $z_2=w$, and $z_1=v_t$ or $v_s$),  the momentum and position space wave functions are related by the spherical Bessel transforms
\bea
z_\ell(k)&=&\sqrt{\sfrac2{\pi}}\int_0^\infty r d{r}\,j_\ell(kr)\,z_\ell(r)
\nonumber\\
\frac{z_\ell(r)}{r}&=&\sqrt{\sfrac2{\pi}}\int_0^\infty k^2dk\,j_\ell(kr)\,z_\ell(k)
\label{eq:besseltrans}
\eea
where $j_\ell$ is the spherical Bessel function of order $\ell$ with the convenient recursion relation
\bea
j_\ell(z)=z^\ell\left(-\frac{1}{z}\frac{d}{dz}\right)^\ell\frac{\sin z}{z}. \label{eq:recursion}
\eea
The normalization condition for the spherical Bessel functions,
\bea
\int_0^\infty k^2dk j_\ell(kr)j_\ell(kr')=\frac{\pi}{2r^2}\delta(r-r') \label{eq:besselnorm}
\eea
can be used to transform integrals from momentum space to coordinate space.  Another convenient identity is
\bea
\int_0^\infty dk\,&&\frac{d}{dk}(k^2\, z_\ell z_{\ell'})
\nonumber\\
=&&\int_0^\infty k^2\,dk \left(\frac{2z_\ell z_{\ell'}}{k}+z_\ell z_{\ell'}'+z_\ell' z_{\ell'}\right)=0\, . \qquad
\label{eq:H40}
\eea

\section{Appearance of off-shell terms in the calculation of the vertex functions} \label{app:offshellterms}

This Appendix shows that when $k$ is off-shell the full structure of the BS wave function is needed, even though the off shell factor $\Theta(-k)$ cancels from the kernel, used to calculate the subtracted vertex functions $\widehat \Gamma$, as discussed in Ref.\ I. %Sec.~\ref{sec:BScancellation}.  

The equation for the subtracted vertex function given in %[Eq.\ (3.34a) of 
Ref.~I is
\begin{widetext}
\bea
\Big[\widehat{\Gamma}^{\lambda}(k,P)\,{\cal C}\Big]_{\alpha\beta}
=-\int_{k'} \widehat{V}_{\beta\gamma',\alpha\alpha'}( k,   k'; P)\,h(p')
{\it \Psi}^{\lambda}_{\alpha'\gamma}(k',P) \Lambda^T_{\gamma\gamma'}(k')\, .\qquad\quad
\eea
(This has the same form as Eq.\ (\ref{eq:CST}) with the reduced vertex function replaced by the subtracted vertex function and  the reduced kernel $\widetilde V$ replaced by the subtracted kernel $\widehat V$.)  Remember that $k'$ is on-shell but $k$ is not so restricted.   

To show how the factor of $\Theta(-k)$ can emerge, it is sufficient to consider scalar exchanges without the $\Theta(p)$ contributions, so that the kernel can be written
\bea
\widetilde V_{\beta\beta',\alpha\alpha'}(k,p;k',p')=&&\frac12\Big[ g_s^2 {\bf 1}_{\beta\beta'}\otimes{\bf I}_{\alpha\alpha'} 
\widetilde\Delta^s(\tilde q)
\pm {\rm exchange\;term} \Big]
\label{eq:ONEkernel}
\eea
where the dressed scalar meson propagator, $\widetilde \Delta^s(\tilde q)$, is a function of $q^2=(k'-k)^2=(p-p')^2$ only (for details that are not needed here, see Ref.~I). Removing the charge conjugation matrix, the equation can be written as a matrix
\bea
\widetilde{\Gamma}^{\lambda}(k,P)
=&&-\frac12 g_s^2\int_{k'} \widetilde\Delta^s(k'-k)h(p')
{\Psi}^{\lambda}(k',P) \Lambda(-k')  \pm {\rm exchange\;term}\, . \label{eq:matrixgeq}
\eea
Now consider  the $A(p'^2)\gamma^\lambda$ term, which is the simplest  contribution to $\Psi^\lambda$ (where $\gamma^\lambda\equiv \gamma^\mu \xi_\mu^\lambda$).  
Recalling that $h(p')$ is a function of $p'^2$,  that $p'^2= m^2+m_d^2-2P\cdot k'$, and using the identity (which holds for any arbitrary function ${\cal F}$)
\bea
\int d^3 k' {\cal F}[(k-k')^2,P\cdot k']\,k'^\mu&&=\int d^3 k' {\cal F}[(k-k')^2,P\cdot k']\,\left[\frac {P\cdot k}{M_d^2}P^\mu +\left(k'^\mu-\frac {P\cdot k'}{M_d^2}P^\mu\right)\right]
\nonumber\\
&&=\int d^3 k' {\cal F}[(k-k')^2,P\cdot k']\,\left[Y P^\mu+X k^\mu \right]
\eea
with $X$ proportional to the projection of $k'$ on the transverse (with respect to $P$) components of $k$
\bea
X&=&\frac{\left(k-\frac{P\cdot k}{M_d^2}P\right)\cdot k'}{k^2-\frac{(P\cdot k)^2}{M_d^2}} \to \frac{{\bf k}\cdot{\bf k}'}{{\bf k}^2}
\nonumber\\
Y&=&\frac{P\cdot k'}{M_d^2}-X\frac{P\cdot k}{M_d^2}\to \frac{E_{k'}}{m_d}-\frac{{\bf k}\cdot{\bf k}'}{{\bf k}^2}\frac{k_0}{m_d}
\eea
(where the limits are the results in the rest frame), the direct term contribution to (\ref{eq:matrixgeq}) becomes

\bea
\widetilde{\Gamma}^{\lambda}(k,P)
&\sim&-\frac12g_s^2\int_{k'} 
\widetilde\Delta^s(k'-k)h(p') A(p'^2)
\gamma^{\lambda} \left(\frac{m-\slashed{k}'}{2m}\right)  \nonumber\\
&=&-\frac12g_s^2\int_{k'} \widetilde\Delta^b(k'-k)h(p') A(k'_P)
\gamma^{\lambda} \frac1{2m}\left[m-Y\slashed{P}-X\slashed{k}\right]
\nonumber\\
&=& -\frac12g_s^2\int_{k'} \widetilde\Delta^b(k'-k)h(p') A(k'_P) \Big\{Y\Theta(p)\gamma^\lambda -\frac{Y}{2m}k^\lambda+(X-Y)\gamma^\lambda\Theta(-k) +\frac1{2}(1+X)\gamma^\lambda\Big\}\ \, . \label{eq:matrixgeq2}
\eea
%
%\end{widetext}

In this simple example, invariants of the type $F , G, H,$ and $K_1$ are generated (unless any of them accidentally integrate to zero). This shows that the integral equation will generate a factor of $\Theta(-k)$, whether or not $k$ is off-shell.   Even with the cancellations found in Ref.\ I, 
the structure of a full BS wave function must be considered when the form factors are calculated.  However, the strength of the extra invariant functions, $K_i$, are much different from what they would have been if there were no cancellation.

\section{Equations for helicity amplitudes and interaction current kernels}
\label{app:truncwf}

In this Appendix the bound state equations for the helicity amplitudes are extracted from the bound state equations given in the main part of this paper, and the interaction kernels used to calculate the interaction current wave functions, $\widehat \Psi$ and $\Psi^{(2)}$, are given explicitly.

Starting from the bound state Eq.~(\ref{eq:CST}), using the convenient relation 
\bea
\delta_{\alpha\alpha'}=\frac{m}{E_{k'}} u_{2\alpha}^{\rho_2'}({\bf k}',\lambda_2') \,\overline{u}_{2\alpha_1}^{\rho_2'}({\bf k}',\lambda_2')\gamma_{\alpha_1\alpha'}^0\, ,
\eea
and the helicity amplitude definition (\ref{eq:helamps1}), gives
%\begin{widetext}
%
\bea
\phi_{\lambda_1\lambda_2,\lambda_d}^{\rho_2}({\bf k})&=&N_d \frac{m}{E_k}\Big[\overline{u}^{\rho_2}_2({\bf k},\lambda_2)\gamma^0\Big]_\alpha{\it \Psi}^{\lambda_d}_{\alpha\lambda_1}(k,P)
%\nonumber\\
=-N_d \, G^{\rho_2}(k) \frac{m}{E_k} \int_{k'} \overline V^{\rho_2}_{\lambda_1\lambda_1',\lambda_2\alpha'}(k,k';P){\it \Psi}^{\lambda_d}_{\alpha'\lambda'}(k',P)
\nonumber\\
&=&- G^{\rho_2}(k) \int \frac{d^3k'}{(2\pi)^3}\frac{m^2}{E_kE_{k'}} \overline V^{\rho_2\rho'_2}_{\lambda_1\lambda_1',\lambda_2\lambda_2'}(k,k';P)\,{\phi}^{\rho_2'}_{\lambda_1'\lambda_2',\lambda_d}({\bf k}')\, ,
\label{eq:helbseq}
\eea
where ${\it \Psi}^{\lambda_d}_{\alpha\lambda_1}(k,P)={\it \Psi}^{\lambda_d}_{\alpha\beta}(k,P)u^+_{\beta}({\bf k},\lambda_1)$.  In numerical calculations, the factors of $(2\pi)^{-3}$ and $m/E$ are absorbed into the kernel $\overline V$.
The equations for helicity projections $\widehat\phi$, and $\phi^{(2)}$ of the truncated wave functions $\widehat{\it \Psi}$  and ${\it \Psi}^{(2)}$ have the same form.

Only the $\Theta$ factors in the scalar and vector meson vertices contribute to the interaction currents.  
To handle these contributions efficiently, it is convenient to use the truncated wave function $\Psi^{(2)}$ and the subtracted vertex function $\widehat \Gamma$.  Since only scalar and vector exchanges enter the  considerations in this section, we will drop the boson $b$ superscript and use the simplified notation $\Lambda^s\equiv \Lambda$ and $\Lambda _v^\mu\equiv \Lambda^\mu$.  

In the deuteron rest frame, removing the nucleon spinors, the kernels that produce $\Psi^{(2)}$ and $\overline\Psi^{(2)}$ are

\bea
V^{2f}_{\beta\beta',\alpha\alpha'}(k,k'; P)&=&-\epsilon_s\delta \,\nu_s\frac12\Bigg\{g_s\,\delta_{\alpha\alpha'}\Delta^s(k-k')\delta_{\beta\beta'}
\pm\delta_{\alpha\beta'}\Delta^s(k-p')\Big[g_s-\nu_s\Theta(p')\Big]_{\beta\alpha'}\Bigg\}
\nonumber\\
&&+\epsilon_v\delta\, g_v\nu_v\frac12\Bigg\{\gamma^\mu_{\alpha\alpha'}\Delta_{\mu\nu}^v(k-k')\widehat\Lambda^{\nu}_{\beta\beta'}(k,k')
\pm
\gamma^\mu_{\alpha\beta'}\Delta_{\mu\nu}^v(k-p')\widehat\Lambda^{\nu}_{\beta\alpha'}(k,p')\Bigg\}
\nonumber\\
\overline V^{2i}_{\beta\beta',\alpha\alpha'}(k,k'; P)&=&-\epsilon_s\delta\,\nu_s\frac12\Bigg\{g_s\,\delta_{\alpha\alpha'}\Delta^s(k-k')\delta_{\beta\beta'}
\pm\delta_{\beta\alpha'}\Delta^s(p-k')\Big[g_s-\nu_s\Theta(p)\Big]_{\alpha\beta'}\Bigg\}
\nonumber\\
&&+\epsilon_v\delta\, g_v\nu_v\frac12\Bigg\{\gamma^\mu_{\alpha\alpha'}\Delta_{\mu\nu}^v(k-k')\widehat\Lambda^{\nu}_{\beta\beta'}(k,k')
\pm
\gamma^\mu_{\beta\alpha'}\Delta_{\mu\nu}^v(p-k')\widehat\Lambda^{\nu}_{\alpha\beta'}(p,k')\Bigg\}\, .
\eea
where we have explicitly assumed that $k$ and $k'$ are on shell, so that
\bea
\widehat \Lambda^\nu (k,k')&=&g_v\Big[\gamma^\mu +\frac{\kappa_v}{2m}i\sigma^{\nu\kappa}(k-k')_\kappa\Big]
\nonumber\\
\widehat \Lambda^\nu (k,p')&=&g_v\Big[\gamma^\mu \Big(1+\nu_v\Theta(p')\Big)+\frac{\kappa_v}{2m}i\sigma^{\nu\kappa}(k-p')_\kappa\Big]
\nonumber\\
\widehat \Lambda^\nu (p,k')&=&g_v\Big[\Big(1+\nu_v\Theta(p)\Big)\gamma^\mu +\frac{\kappa_v}{2m}i\sigma^{\nu\kappa}(p-k')_\kappa\Big]\, .
\eea
Note that these kernels are not identical.

In the deuteron rest frame the kernels that produce $\widehat\Gamma$ and $\overline{\widehat\Gamma}$ have the same mathematical form, but the kernel for $\widehat\Gamma$ has $k$ off-shell and $k'^2=m^2$, while with kernel for $\overline{\widehat\Gamma}$ has $k'$ off-shell and $k^2=m^2$.  The scalar and vector contributions are the same as the full kernel, but with all factors of $\Theta(k)$ and $\Theta(k')$ removed.  Explicitly
\bea
\widehat V_{\beta\beta',\alpha\alpha'}(k,k'; P)&=&\epsilon_s\delta\frac12\Bigg\{g_s\Lambda_{\alpha\alpha'}(p,p')\Delta_s(k-k')\delta_{\beta\beta'}
\pm \Big[g_s-\nu_s\Theta(p')\Big]_{\beta\alpha'} \Delta_s(p-k') \Big[g_s-\nu_s\Theta(p)\Big]_{\alpha\beta'}
\nonumber\\
&&+\Lambda^{\mu}_{\alpha\alpha'}(p,p')\Delta^{\mu\nu}_v(k-k')\widehat\Lambda^{\nu}_{\beta\beta'}(k,k')
\pm \widehat\Lambda^{\mu}_{\beta\alpha'}(k,p')\Delta^{\mu\nu}_v(p-k')\widehat\Lambda^{\nu}_{\alpha\beta'}(p,k') \Bigg\}
\nonumber\\
&& +{\rm \ all\ other \ meson\ contributions}\, .
\eea
To evaluate these kernels use the methods described in Appendix A of Ref.~\cite{arXiv:1007.0778}.  

Using the scalar currents as an example, we can now demonstrate the correctness of the identity (\ref{eq:identityVp}).  Since there are no factors of $\Theta(k)$ or $\Theta(k')$ in any of the terms, the derivative of the {\it scalar\/} exchange part of $\widetilde V$ is
\bea
\frac{\partial}{\partial P_0} \widetilde V_{\beta\beta',\alpha\alpha'}(k,k';P)=\epsilon_s\delta\frac{\nu_s}{4m}\Bigg\{&&2g_s\Delta_s(q_d)\delta_{\beta\beta'}\gamma^0_{\alpha\alpha'}
\pm \Delta_s(q_e) \Big( \Big[g_s-\nu_s\Theta(p')\Big]_{\beta\alpha'} \gamma^0_{\alpha\beta'}  +\gamma^0_{\beta\alpha'} \Big[g_s-\nu_s\Theta(p)\Big]_{\alpha\beta'}\Big)\Bigg\}
\nonumber\\
&&\pm \epsilon_s\delta\Big[g_s-\nu_s\Theta(p')\Big]_{\beta\alpha'} \Delta'_s(q_e) \Big[g_s-\nu_s\Theta(p)\Big]_{\alpha\beta'}
\eea
where $q_d=k'-k$ and $q_e=p-k'=p'-k$ and $\Delta_s'(q_e)=\partial\Delta_s(q_e)/\partial (q_e)_0$.   Next, observe that, for the {\it scalar\/} exchange part,
\bea
\frac12\Big(\frac{\partial}{\partial k_0}&+&\frac{\partial}{\partial k'_0}\Big)\widehat V_{\beta\beta',\alpha\alpha'}(k,k';P)=\mp\epsilon_s\delta  \Big[g_s-\nu_s\Theta(p')\Big]_{\beta\alpha'} \Delta'_s(q_e) \Big[g_s-\nu_s\Theta(p)\Big]_{\alpha\beta'}
\nonumber\\
&&-\epsilon_s\delta\frac{\nu_s}{4m}\Bigg\{g_s\Delta_s(q_d)\delta_{\beta\beta'} \gamma^0_{\alpha\alpha'} 
\pm\frac12 \Delta_s(q_e) \Big( \Big[g_s-\nu_s\Theta(p')\Big]_{\beta\alpha'} \gamma^0_{\alpha\beta'}  +\gamma^0_{\beta\alpha'} \Big[g_s-\nu_s\Theta(p)\Big]_{\alpha\beta'}\Big)\Bigg\}
\eea
Adding these together and using the relations for the scalar parts of  $V^{2f}$ and $V^{2i}$ gives
\bea
-\frac{\partial}{\partial P_0} \widetilde V =\frac12\Big(\frac{\partial}{\partial k_0}&+&\frac{\partial}{\partial k'_0}\Big)\widehat V+\frac12\left(\frac1{2m}\right)\Big[V^{2i}_{\beta\beta'\alpha\alpha_1}\gamma^0_{\alpha_1\alpha'}+\gamma^0_{\alpha\alpha_1}V^{2f}_{\beta\beta'\alpha_1\alpha'}\Big]\, ,  \label{eq:chargeop}
\eea
which is the operator form of the identity for scalar exchange.  Taking matrix elements of both sides of (\ref{eq:chargeop}) transforms it into a form that leads directly to identity (\ref{eq:identityVp}).   

To simplify the notation, suppress the arguments of the wave functions, denote $h(p)=h$ and $h(p')=h'$, and replace $\lambda_n\to\lambda_1$, etc. 
Then the left hand side reduces to 
\bea
&&-\frac1{2m_d}\int_k \int_{k'}
 \overline{\it \Psi}_{\lambda_1\alpha}^{\lambda}\Big[h\,\frac{\partial}{\partial\,P_0}\widetilde V_{\lambda_1\lambda_1',\alpha\alpha'}\,h'\Big]
 {\it \Psi}_{\alpha'\lambda_1'}^{\lambda'} 
=-\frac1{2m_d}\int_k \int_{k'}\frac{1}{N_d^2}\, {\phi}_{\lambda_1\lambda_2,\lambda}^{\rho_2}({\bf k})\Big[h\,\frac{\partial}{\partial\,P_0}\widetilde V^{\rho_2\rho_2'}_{\lambda_1\lambda_1',\lambda_2\lambda_2'}\,h'\Big]{\phi}_{\lambda_2'\lambda_1',\lambda'}^{\rho_2'} ({\bf k}')
 \nonumber\\
 &&\qquad
 =-\int k^2 dk\int k'^2 dk' \,{\phi}_{\lambda_1\lambda_2,\lambda}^{\rho_2} (k)\Big[\frac{h\,h'}{(2\pi)^3}\frac{m^2}{E_k E_{k'}}\frac{\partial}{\partial\,P_0}\widetilde V^{1\rho_2\rho_2'}_{\lambda_1\lambda_1',\lambda_2\lambda_2'}(k,k';P)\Big]{\phi}_{\lambda_2'\lambda_1',\lambda'}^{\rho_2'}(k')\qquad \label{eq:p0dir}
  \eea
where, in the first line ${\it \Psi}$ has been replaced by the helicity amplitudes using the expansion (\ref{eq:normhel}) (with the sum over repeated indices implied and no longer written explicitly) and the reference to $\rho_1=+$ has been suppressed.   The last step uses the partial wave expansion given in Eq.~(E29) of Ref.~\cite{Gross:2008ps} to carry out the angular integrals.  Explicitly, the generic integral is
\bea
\int d\Omega_k \int d\Omega_{k'} \,\phi_\lambda({\bf k}) \widetilde V({\bf k},{\bf k}';P) \phi_{\lambda'}({\bf k}')&=&\frac{2J+1}{4\pi} \int d\Omega_k \int d\Omega_{k'} \,\phi_\lambda({\bf k}) \widetilde V^J({k},{k}';P) D^{J}_{M_J\lambda}(\Omega_k)D^{J*}_{M_J\lambda'}(\Omega_{k'}) \phi_{\lambda'}({\bf k}')\qquad
\nonumber\\
&=&\phi_\lambda({k})\, \widetilde V^1({k},{k}';P) \, \phi_{\lambda'}({k}')
\eea
where the general result has been specialized to $J=1$ and the definition (\ref{eq:partialwaves})  of the partial wave helicity amplitudes used.   The simple form (\ref{eq:p0dir}) was used in the actual numerical calculations, with the factors of $m^2/[(2\pi)^{3}E_kE_{k'}]$ absorbed into the kernel $\widetilde V^1$.

Recall that the integral equations defining $\widehat \Psi$ and $\Psi^{(2)}$ contribute an extra minus sign.  The term involving $\widehat V$ becomes
\bea
&&\frac1{2m_d}\int_k \int_{k'}
 \overline{\it \Psi}_{\lambda_1\alpha}^{\lambda}(k,P)h\Big[\frac12\Big(\frac{\partial}{\partial k_0}+\frac{\partial}{\partial k'_0}\Big)\widehat V_{\lambda_1\lambda_1',\alpha\alpha'}(k,k';P)\Big]h'
 {\it \Psi}_{\alpha'\lambda_1'}^{\lambda'}(k',P) 
 \nonumber\\
 &&=-\frac1{4m_d} \int_{k}\Big\{ \overline{\it \Psi}_{\lambda_1\alpha}^{\lambda}(k,P)\Big[ \frac{\partial}{\partial k_0} \widehat{{\it {\cal G}}}_{\alpha\lambda_1}^{\lambda'}(k,P)\Big]
 +\Big[\frac{\partial}{\partial k_0} \overline{\widehat{{\it {\cal G}}}}_{\lambda_1\alpha}^{\lambda}(k,P)\Big] {{\it \Psi}}_{\alpha\lambda_1}^{\lambda'}(k,P)\Big\}
 \nonumber\\&&
 =-\frac1{4m_d} \int_{k} \frac1{N_d}\,\Bigg\{ \phi^{\rho_2}_{\lambda_1\lambda_2,\lambda}({\bf k}) \overline{u}^{\rho_2}_{2\alpha}({\bf k},\lambda_2)\Big[ \frac{\partial}{\partial k_0} \widehat{{\it {\cal G}}}_{\alpha\lambda_1}^{\lambda'}(k,P)\Big]
 +\Big[\frac{\partial}{\partial k_0} \overline{\widehat{{\it {\cal G}}}}_{\lambda_1\alpha}^{\lambda}(k,P)\Big] u_{2\alpha}^{\rho_2}({\bf k},\lambda_2) \phi^{\rho_2}_{\lambda_1\lambda_2,\lambda'}({\bf k}) \Bigg\}
 \nonumber\\&&
 =-\frac1{2m_d} \int_{k} \frac{1}{N_d^2}\frac{E_k}{m}\,\phi^{\rho_2}_{\lambda_1\lambda_2,\lambda}({\bf k}) \frac{\partial}{\partial k_0}\Big[\rho_2 \delta_{\rho_2} (k_0)\widehat{\phi}^{\rho_2}_{\lambda_1\lambda_2, \lambda'}(k_0,{\bf k}) \Big]
 \nonumber\\&&
 =-\delta_{\lambda\lambda'}
 \int_0^\infty k^2dk  \,\phi^{\rho_2}_{\lambda_1\lambda_2,\lambda}({k}) \frac{\partial}{\partial k_0}\Big[ \rho_2\delta_{\rho_2} (k_0)\widehat{\phi}^{+\rho_2}_{\lambda_1\lambda_2,\lambda'}(k_0,{k}) \Big]\qquad
 \label{eq:hatterms}
\eea
where the $k_0$ derivatives do {\it not\/}  act on the strong nucleon form factors, and in the fourth line the two $k_0$ derivatives are combined into a single term with the dependence of $\widehat{\phi}$ and $\delta_{\rho_2}$ on $k_0$ is explicitly shown.  In the last line the partial wave helicity amplitudes are introduced using the  partial wave expansions  (\ref{eq:psik}), and the angular integrals are evaluated using the normalization of the $d$ functions (\ref{eq:dnorm}).  Explicitly, for $\lambda=0$
\bea
\int d\Omega\, \phi_{\lambda_0 \lambda_0,0}({\bf k})\widehat{\phi}_{\lambda_0 \lambda_0,0}({\bf k})=\frac{4\pi}{3}[B-A][\widehat{B}-\widehat{A}]=\phi_{\lambda_0 \lambda_0,0}({k})\widehat{\phi}_{\lambda_0 \lambda_0,0}({k})\, ,
\eea
and for $\lambda=1$
\bea
\int d\Omega\, \phi_{\lambda_0 -\lambda_0,1}({\bf k})\widehat{\phi}_{\lambda_0 -\lambda_0,1}({\bf k})=\frac{8\pi}{3}B\widehat{B}=\phi_{\lambda_0 -\lambda_0,1}({k})\widehat{\phi}_{\lambda_0 -\lambda_0,1}({k})\, ,
\eea
showing that the result is independent of $\lambda$.  Note that (\ref{eq:hatterms}) exactly reproduces the corresponding terms in the identity (\ref{eq:identityVp}).

Finally, the terms involving $V^{2i}$ and $V^{2f}$ reduce to
\bea
\frac{1}{4m_d}&&\int_k\int_{k'} \overline{\it \Psi}_{\lambda_1\alpha}^{\lambda}(k,P)\left(\frac1{2m}\right)h\Big[V^{2i}_{\beta\beta'\alpha\alpha_1}\gamma^0_{\alpha_1\alpha'}+\gamma^0_{\alpha\alpha_1}V^{2f}_{\beta\beta'\alpha_1\alpha'}\Big] h'{\it \Psi}_{\alpha' \lambda_1'}^{\lambda'}(k',P)
\nonumber\\&&
=-\frac{1}{4m_d}\int_k \Big[\overline{\it \Psi}_{\lambda_1\alpha}^{\lambda}(k,P)\gamma^0_{\alpha\alpha'}{\it \Psi}_{\alpha' \lambda_1'}^{(2) \lambda'}(k,P) 
+\overline{\it \Psi}_{\lambda_1\alpha}^{(2) \lambda}(k,P)\gamma^0_{\alpha\alpha'}{\it \Psi}_{\alpha' \lambda_1'}^{\lambda'}(k,P)\Big]
\nonumber\\&&
=-\frac{1}{2m_d}\int_k \frac{1}{N_d^2}\,\phi^{\rho_2}_{\lambda_1\lambda_2,\lambda}({\bf k})\Big[\overline{u}_2^{\rho_2}({\bf k},\lambda_2)\gamma^0 u^{\rho_2'}_2({\bf k},\lambda_2')\Big]
\phi^{\rho_2' (2)}_{\lambda_1\lambda_2',\lambda}({\bf k})
\nonumber\\&&
=-\frac{1}{2m_d}\int_k \frac1{N_d^2}\frac{E_k}{m}\phi^{\rho_2}_{\lambda_1\lambda_2,\lambda}({\bf k})
\phi^{\rho_2 (2)}_{\lambda_1\lambda_2,\lambda}({\bf k}) =\int_0^\infty k^2dk\,\phi^{\rho_2}_{\lambda_1\lambda_2,\lambda}({k})
\phi^{\rho_2 (2)}_{\lambda_1\lambda_2,\lambda}({k}) \, .
\eea
This term again reproduces the form of the $zz^{(2)}$ terms in  the identity (\ref{eq:identityVp}), concluding the demonstration.

\section{Results for the Traces}  \label{app:B}

\subsection{Contributions from diagram (A) plus $\left<V^\mu_2\right>$}

In this section the traces (\ref{eq:genericA}) needed for each of the helicity amplitudes defined in Eq.~(\ref{eq:dffmatrix}) are evaluated.  Using the compact notation $Z_\pm=Z(R_\pm)$  (where $Z$ is the generic name for $\{A,B,C,D\}$) with $R_\pm$ defined in Eq.\ (\ref{eq:pp&m}) below, the results are  
\bea
{\cal A}_{n,1}(\Psi_+\Psi_-) &=&\frac2{m}\bigg\{A_+A_-(a_-z_++a_+z_--a_0z_0)+B_+B_-\frac{a_0\,a_+a_-}{m^2}+D_+D_-\frac{a_+a_-}{m^4}\big[2b_0(c_++c_-)-\zeta_0a_0\big]
\nonumber\\
&&+C_+C_-\Big[4m^2(2b_0z_0-b_+z_--b_-z_+)
+8a_+a_-(a_0-b_0)+a_0z_0(\zeta_0-8m^2)-2a_0b_+b_-
\nonumber\\&&\qquad
-2b_0z_0(c_++c_-)+a_+\big(2b_0b_-+4c_-z_--\zeta_0z_-\big)+a_-\big(2b_0b_++4c_+z_+-\zeta_0z_+\big)
\nonumber\\&&\qquad
+2b_+c_+z_-+2b_-c_-z_+\Big]-A_+B_-a_-z_+-B_+A_-a_+z_-
\nonumber\\
&&+A_+C_-\big[z_0(2a_0-b_0)-z_-(2a_+-b_+)\big]
+C_+A_-\big[z_0(2a_0-b_0)-z_+(2a_--b_-)\big]\nonumber\\
&&+A_+D_-\frac{a_-}{m^2}(a_+b_0-a_0b_++z_+c_-)
+D_+A_-\frac{a_+}{m^2}(a_-b_0-a_0b_-+z_-c_+) -(B_+D_-+D_+B_-)\frac{b_0}{m^2}a_+a_-
\nonumber\\
&&+B_+C_-\frac{a_+}{m^2}(2m^2z_--2a_0a_-+b_0a_--z_-c_-)
+C_+B_-\frac{a_-}{m^2}(2m^2z_+ -2a_0a_++b_0a_+ -z_+c_+)
\nonumber\\
&&+C_+D_-\frac{a_-}{2m^2}\big[z_+(\zeta_0-4c_-)+2b_+(2a_0-b_0)\big]
+D_+C_-\frac{a_+}{2m^2}\big[z_-(\zeta_0-4c_+)+2b_-(2a_0-b_0)\big]\bigg\}
\label{eq:traceA1}
\eea
\bea
{\cal A}_{n,2}(\Psi_+&&\Psi_-) =\frac1{m}\bigg\{A_+A_-(b_+z_-+b_-z_+)-A_+B_-\frac{a_-}{m^2}\big[a_0b_+ +z_+(c_+-c_-)\big]
-B_+A_-\frac{a_+}{m^2}\big[a_0b_- +z_-(c_- -c_+)\big]
\nonumber\\&&
+D_+D_-a_+a_-\frac{b_0Q^2}{m^4} +C_+C_-\frac1{m^2}\Big[(4m^2+m_d^2)(b_+z_-+b_-c_+)+(b_0-2a_0)(2a_+b_-+2a_-b_+ +Q^2 z_0)
\nonumber\\&&\qquad
-2(b_0-2a_0)b_+b_- +Q^2(a_+z_-+a_-z_+)+4(a_+z_--a_-z_+)(c_+-c_-)-4(b_+c_+z_- + b_-c_-z_+\Big]
\nonumber\\&&
-A_+C_-\frac1{2m^2}\Big[4m^2(b_+z_-+b_-z_+)-2b_0(a_+b_- - a_-b_+)-2a_0b_+(2a_--b_-)-Q^2(a_0z_0-a_+z_- -a_-z_+)
\nonumber\\&&\qquad+2(c_+-c_-)(b_0z_0-2a_-z_+)-2b_+c_+z_-  -2b_-c_-z_+\Big] +A_+D_-\frac{a_-}{2m^2}\big[z_+Q^2+2b_0\,b_+\big]
\nonumber\\&&
-C_+A_-\frac1{2m^2}\Big[4m^2(b_-z_+ +b_+z_-)-2b_0(a_+b_+ - a_+b_-)-2a_0b_-(2a_+-b_+)-Q^2(a_0z_0-a_-z_+ -a_+z_-)
\nonumber\\&&\qquad-2(c_+-c_-)(b_0z_0-2a_+z_-)-2b_-c_-z_+  -2b_+c_+z_-\Big]+D_+A_-\frac{a_+}{2m^2}\big[z_-Q^2+2b_0\,b_-\big]
\nonumber\\&&
+B_+C_-\frac{a_+}{2m^2}\Big[z_-\Big(Q^2-4(c_+-c_-)\Big)-2b_-(b_0-2a_0)\Big]
+C_+B_-\frac{a_-}{2m^2}\Big[z_+\Big(Q^2+4(c_+-c_-)\Big)-2b_+(b_0-2a_0)\Big]
\nonumber\\&&
+C_+D_-\frac{a_-}{m^4}\Big[a_0(a_+Q^2-b_+m_d^2)+b_0\big(2b_+(c_+-m^2)-a_+(Q^2+2c_+-2c_-)\big]-z_+\big(m^2Q^2+m_d^2(c_+-c_-)\big)
\nonumber\\&&
+D_+C_-\frac{a_+}{m^4}\Big[a_0(a_-Q^2-b_-m_d^2)+b_0\big(2b_-(c_--m^2)-a_+(Q^2-2c_++2c_-)\big]-z_-\big(m^2Q^2-m_d^2(c_+-c_-)\big)\Big]
\nonumber\\&&
-B_+D_-\frac{a_+a_-}{2m^4}\big[a_0Q^2-2b_0(c_+-c_-)\big]
-D_+B_-\frac{a_+a_-}{2m^4}\big[a_0Q^2+2b_0(c_+-c_-)\big]
\label{eq:traceA2}
\eea
%
%\end{widetext}
where the  vector products needed for this expansion  are defined in Table \ref{tab:Acoeffs}.
The results for the  the traces ${\cal A}_{n,3}$ are obtained by the substitutions $A\to F, B\to G, C\to H, D\to I$ in ${\cal A}_{n,1}$.  These expressions are sums of products of invariant functions and four-vector scalar products and hence are  manifestly covariant.

%%%%%%%%%%%%%%%%%%%%%%%%%%%%%%%%%%%%%%%%%%%%%%%%%%
\begin{table*}[t]
\begin{minipage}{6.5in}
\caption{Vector products that depend on $n$ used in the expansions of ${\cal A}_{n,i}$.  All are evaluated in the Breit frame using (\ref{eq:breit}) and (\ref{2.4}).  The helicity amplitude ${\cal A}_{3,i}=\frac12({\cal A}_{3_+,i}+{\cal A}_{3_-,i})$, as explained in Eq.~(\ref{eq:dffmatrix}).  Not shown are $\zeta_0=2m_d^2+Q^2$ and $c_\pm=P_\pm\cdot k=D_0E_k\mp\frac12 k_zQ$ which are the same for all helicity combinations.} 
\label{tab:Acoeffs}
\begin{ruledtabular}
\begin{tabular}{lcccc}
coefficient & $n=1\;(J_{00}^0)$ & $n=2\; (J_{+-}^0)$ & $n=3_+\; (J_{+0}^+)$
& $n=3_-\; (J_{0-}^-)$\cr
\tableline
$a_+=k\cdot \xi^*$&  $(E_kQ-2k_zD_0)/(2m_d)$& $\frac1{\sqrt{2}}(k_x-ik_y)$&  $\frac1{\sqrt{2}}(k_x-ik_y)$ &
$(E_kQ-2k_zD_0)/(2m_d)$ \cr
$a_-=k\cdot \xi'$&$-(E_kQ+2k_zD_0)/(2m_d)$& $\frac1{\sqrt{2}}(k_x+ik_y)$ & $-(E_kQ+2k_zD_0)/(2m_d)$& $\frac1{\sqrt{2}}(k_x+ik_y)$\cr
$a_0=k\cdot \epsilon$&$E_k$ & $E_k$  & $\frac1{\sqrt{2}}(k_x+ik_y)$ & $-\frac1{\sqrt{2}}(k_x-ik_y)$\cr
$b_+=P_-\cdot \xi^*$&$D_0 Q/m_d$&0 & 0 & $D_0Q/m_d$\cr
$b_-=P_+\cdot \xi'$&$-D_0 Q/m_d$&0 & $-D_0 Q/m_d$ & 0\cr
$b_0=P_+\cdot \epsilon=P_-\cdot\epsilon$&$D_0$& $D_0$ & 0 & 0\cr
$z_+=\epsilon\cdot \xi^*$&$Q/(2m_d)$&0& $-1$& 0 \cr
$z_-=\epsilon\cdot \xi'$&$-Q/(2m_d)$&0& 0 & 1\cr
$z_0=\xi^*\cdot \xi'$&$-\zeta_0/(2m_d^2)$&$-1$& 0 & 0\cr
\end{tabular}
\end{ruledtabular}
\end{minipage}
\end{table*}
%
%%%%%%%%%%%%%%%%%%%%%%%%%%%%%%%%%%%%%%%%%%%%%%%%%%

In the terms above, the spectator momentum $k$ is always on shell.   In this case 
the arguments  (\ref{eq:kp&eta}) of the wave functions for the incoming and outgoing deuterons  become 
\bea
R_\pm^2&=&\frac{(P_\pm\cdot  k)^2}{m_d^2}-m^2
={\bf k}^2 %\nonumber\\&&
\mp k_z\,Q \frac{D_0 E_k}{m_d^2} +\eta\left(E_k^2+k_z^2\right)
\nonumber\\
E_{R_\pm}&=&\sqrt{m^2+R_\pm^2}\, . \label{eq:pp&m}
\eea

Careful examination of the formulae for ${\cal A}$ show that they are unchanged under the transformation $+\leftrightarrow -$.   For $n=1$ helicity amplitudes, the plus and minus coefficients transform into each other as $Q\to -Q$ (as do the arguments of the wave functions), so that the ${\cal A}_{1,i}$ satisfy the symmetry property (\ref{eq:genericA}) by inspection.  For the $n=2$ helicity amplitudes, the $a_\pm$ to not change with $Q$, but since the $b_\pm$ and $z_\pm$ coefficients are zero in this case,  the terms that remain contain either no factors of $a_\pm$ or the product  $a_+a_-$, preserving the symmetry in $Q$.  Finally, the separate terms $n=3_\pm$ show no special symmetry, but it can be shown that their sum again satisfies the symmetry (\ref{eq:genericA}) appropriate to the $n=3$  amplitude.

%\begin{widetext}
Although the expressions for ${\cal A}$ are given for identical wave functions in initial and final states, this property has not been used in the derivation of the equations and they can easily be extended to the case when ${\cal A}\to {\cal A}(\Psi_+ \Psi^{(2)}_-)$ needed for the calculation of the interaction current terms.  Consider the operation of changing the sign of $Q$ in  a typical term.  Using the fact that the arguments $R_\pm^2\to R_\mp^2$ when $Q\to-Q$, a typical pair of terms in the expansions for ${\cal A}$ transforms to
\bea
Z_+Y^{(2)}_- C_{ZY}(Q) \pm Y_+Z^{(2)}_- C_{YZ}(Q) 
%\nonumber\\&&\qquad
&\to& Z_-Y^{(2)}_+ C_{ZY}(-Q) \pm  Y_-Z^{(2)}_+ C_{YZ}(-Q)
\nonumber\\
&=&\pm \Big[Z^{(2)}_+Y_-C_{YZ}(-Q)\pm Y^{(2)}_+Z_-C_{ZY}(-Q\Big]\qquad
\eea
Using the symmetry properties just discussed, the coefficients have the property
\bea
C_{ZY}(Q)=\pm\epsilon_{n3}C_{YZ}(-Q)\, ,
\eea
conforming the symmetry properties used in (\ref{eq:A&V2-3}).  This simplifies the calculations of the interaction current contributions.

%%%%%%%%%%%%%%%%%%%%%%%%%%%%%%%%%%%%%%%%%%%%%%%%%%
\begin{table*}%[t]
\begin{minipage}{6.5in}
\caption{Vector products that depend on $n$ used in the expansions of ${\cal B}_{n,i}$.  All are evaluated in the Breit frame using (\ref{eq:breit}) and (\ref{2.4}).  The helicity amplitude ${\cal B}_{3,i}=\frac12({\cal B}_{3_+,i}+{\cal B}_{3_-,i})$, as explained in Eq.~(\ref{eq:dffmatrix}).  Not shown are $\zeta=h^2(\tilde p)/(m^2-\tilde p^2),\;\zeta_1=D^2=D_0^2,\; c_0=D\cdot \tilde k=D_0k_0$, and $c_q=q\cdot \tilde k=-Qk_z$ which are the same for all helicity combinations.  Convenient combinations of these vector products are given in Table \ref{tab:Bcoeffs2}.} 
\label{tab:Bcoeffs}
\begin{ruledtabular}
\begin{tabular}{lcccc}
coefficient & $n=1\;(J_{00}^0)$ & $n=2\; (J_{+-}^0)$ & $n=3_+\; (J_{+0}^+)$
& $n=3_-\; (J_{0-}^-)$\cr
\colrule 
$\tilde a_+=\tilde k\cdot \xi^*$&  $(k_0Q-2k_zD_0)/(2m_d)$& $\frac1{\sqrt{2}}(k_x-ik_y)$&  $\frac1{\sqrt{2}}(k_x-ik_y)$ &
$(k_0Q-2k_zD_0)/(2m_d)$ \cr
$\tilde a_-=\tilde k\cdot \xi'$&$-(k_0Q+2k_zD_0)/(2m_d)$& $\frac1{\sqrt{2}}(k_x+ik_y)$ & $-(k_0Q+2k_zD_0)/(2m_d)$& $\frac1{\sqrt{2}}(k_x+ik_y)$\cr
$\tilde a_0=\tilde k\cdot \epsilon$&$k_0$ & $k_0$  & $\frac1{\sqrt{2}}(k_x+ik_y)$ & $-\frac1{\sqrt{2}}(k_x-ik_y)$\cr
$\tilde b_+=-q\cdot \xi^*$&$D_0 Q/m_d$&0 & 0 & $D_0Q/m_d$\cr
$\tilde b_-=q\cdot \xi'$&$-D_0 Q/m_d$&0 & $-D_0 Q/m_d$ & 0\cr
$\tilde b_0=D\cdot\epsilon$&$D_0$& $D_0$ & 0 & 0\cr
$z_+=\epsilon\cdot \xi^*$&$Q/(2m_d)$&0& $-1$& 0 \cr
$z_-=\epsilon\cdot \xi'$&$-Q/(2m_d)$&0& 0 & 1\cr
$z_0=\xi^*\cdot \xi'$&$-\zeta_0/(2m_d^2)$&$-1$& 0 & 0\cr
\end{tabular}
\end{ruledtabular}
\end{minipage}
\end{table*}
%
%%%%%%%%%%%%%%%%%%%%%%%%%%%%%%%%%%%%%%%%%%%%%%%%%%

%\begin{widetext}

\subsection{Contributions of the on-shell terms for diagrams (B) plus $\left<V^\mu_1\right>$} \label{app:Btraces}

Here the traces (\ref{eq:traceB}) needed for the B $+\left<V^\mu_1\right>$ contributions are evaluated.  
%As above, the magnitude of the three-vector momentum is denoted by $k$ (equal to $|{\bf k}|$).   
In these terms $k_0$ is not fixed until the subtraction shown in Eq.~(\ref{eq:B&V1-2}) is carried out.  The results for the traces that depend on $k_0$ are
%
%\begin{widetext}
%
\bea
{\cal B}_{n,1}(k_0) &=&\frac{\zeta}{16m^2}\bigg\{-2\widetilde F_+\widetilde F_-\big[ 2z_0X_2 +X_3\big]
+\widetilde G_+\widetilde G_-\frac1{m^2}(2\tilde a_--\tilde b_-)(2\tilde a_+-\tilde b_+)\big[X_2 -16m^2 \tilde a_0\big]
\nonumber\\&&
+2\widetilde H_+\widetilde H_-\frac1{m^2}X_1\big[2z_0X_2 +X_3
-16m^2(2a_0z_0-b_+z_--b_-z_+)\big]
-\widetilde I_+\widetilde I_-\frac1{m^4}X_1X_2(2\tilde a_--\tilde b_-)(2\tilde a_+-\tilde b_+)
\nonumber\\&&
+2\widetilde F_+\widetilde G_- (2a_--b_-)\big[Y_1^+ +8\tilde a_0(\tilde a_+-\tilde b_+)
+8\tilde b_0\tilde b_+-4z_+c_q\big]
\nonumber\\&&
+2\widetilde G_+\widetilde F_- (2a_+ -b_+)\big[
Y_1^- +8\tilde a_0(\tilde a_--\tilde b_-)
+8\tilde b_0\tilde b_- +4z_-c_q\big]
\nonumber\\&&
+16(\widetilde F_+\widetilde H_- + \widetilde H_+\widetilde F_-)X_1(2a_0z_0-b_+ z_- -b_-z_+)
%\nonumber\\&&
\nonumber\\&&
-2(\widetilde G_+\widetilde H_-+
\widetilde I_+\widetilde F_-)X_1 \frac{(2\tilde a_+ -\tilde b_+)}{m^2} \big[Y^-_1+4\tilde b_0\tilde b_-\big]
%\nonumber\\&&
-2(\widetilde H_+\widetilde G_- + \widetilde F_+\widetilde I_-) X_1\frac{(2\tilde a_- -\tilde b_-)}{m^2} \big[Y^+_1+4\tilde b_0\tilde b_+\big]
\nonumber\\&&
+8(\widetilde G_+\widetilde I_-+\widetilde I_+\widetilde G_-)X_1\frac{\tilde a_0}{m^2}
(2\tilde a_--\tilde b_-)(2\tilde a_+-\tilde b_+)
\nonumber\\&&
+2\widetilde I_+\widetilde H_- X_1\frac{(2\tilde a_+ -\tilde b_+)}{m^2} \big[Y^-_1- 8\tilde a_0(\tilde a_--\tilde b_-) -4c_qz_-\big]
\nonumber\\&&
+2\widetilde H_+\widetilde I_- X_1\frac{(2\tilde a_- -\tilde b_-)}{m^2} \big[Y^+_1- 8\tilde a_0(\tilde a_+-\tilde b_+) +4c_qz_+ \big]\bigg\}
 \label{eq:traceB1}
\eea
\bea
{\cal B}_{n,2}(k_0) &=&\frac{\zeta}{16m^2}\bigg\{2\widetilde F_+\widetilde F_-(2X_4-X_5) +2\widetilde G_+\widetilde G_-\frac{Q^2}{m^2}(2\tilde a_0-\tilde b_0)(2\tilde a_+-\tilde b_+)(2\tilde a_--\tilde b_-)+2\widetilde H_+\widetilde H_-\frac1{m^2}X_1X_5
\nonumber\\&&
+2\widetilde I_+\widetilde I_-X_1\frac{b_0\,Q^2}{m^4}(2\tilde a_+-\tilde b_+)(2\tilde a_--\tilde b_-) 
-2(\widetilde F_+\widetilde H_- + \widetilde H_+\widetilde F_-)\frac1{m^2}X_1X_4 
\nonumber\\&&
-\widetilde F_+\widetilde G_- \frac1{m^2}(2\tilde a_--\tilde b_-)(Y_+^2+4m^2z_+Q^2)
-\widetilde G_+\widetilde F_-\frac1{m^2}(2\tilde a_+ -\tilde b_+)(Y_-^2+4m^2z_-Q^2)
\nonumber\\&&
+4(\widetilde F_+\widetilde I_- +\widetilde H_+\widetilde G_-)\frac{Q^2}{m^2}z_+(2\tilde a_--\tilde b_-)X_1 +4(\widetilde I_+\widetilde F_- + \widetilde G_+\widetilde H_-)\frac{Q^2}{m^2}z_-(2\tilde a_+-\tilde b_+)X_1
\nonumber\\&&
-2(\widetilde G_+\widetilde I_-+\widetilde I_+\widetilde G_-)X_1\frac{\tilde a_0\,Q^2}{m^4}
(2\tilde a_--\tilde b_-)(2\tilde a_+-\tilde b_+) 
\nonumber\\&&
+\widetilde H_+\widetilde I_-\frac1{m^2}(2\tilde a_--\tilde b_-)(Y_+^2-4m^2z_+Q^2)X_1 +\widetilde I_+\widetilde H_-\frac1{m^2}(2\tilde a_+-\tilde b_+)(Y_-^2-4m^2z_-Q^2)X_1  \label{eq:traceB2}
\eea
%
%%%%%%%%%%%%%%%%%%%%%%%%%%%
\begin{table*}
\begin{minipage}{6.5in}
\caption{Combinations of vector products from Table \ref{tab:Bcoeffs} that simplify Eqs.~(\ref{eq:traceB1}) and (\ref{eq:traceB2}).} 
\label{tab:Bcoeffs2}
\begin{ruledtabular}
\begin{tabular}{l}
$Y^\pm_1= 4z_\pm(m^2-\tilde k^2)-z_\pm Q^2+8\tilde a_0\tilde a_\pm-4\tilde b_0\tilde b_\pm$ \cr
$Y^\pm_2=4(m^2-\tilde k^2)(\tilde a_0 \tilde b_\pm \pm c_q z_\pm) -4\tilde b_0 \tilde b_\pm(m^2+\tilde k^2)+8\tilde a_0\tilde b_\pm c_0 \mp c_q\big[8\tilde b_0\tilde a_\pm-\tilde z_\pm(8c_0-Q^2)\big]+Q^2\big[\tilde b_0\tilde b_\pm+\tilde a_0(4\tilde a_\pm-3\tilde b_\pm)\big]$\cr
$X_1=m^2-\tilde k^2+2c_0-\zeta_1$ \cr   
$X_2=4(m^2-\tilde k^2)(\tilde a_0+\tilde b_0) +Q^2(\tilde a_0-\tilde b_0)+8\tilde a_0c_0$  \cr
$X_3=4(m^2-\tilde k^2)\big[z_+(2\tilde a_--3\tilde b_-)+z_-(2\tilde a_+ -3 \tilde b_+)
\big] -8(2\tilde a_0-\tilde b_0)(\tilde a_+\tilde b_- +\tilde a_-\tilde b_+)   -Q^2\big[z_+(2\tilde a_--\tilde b_-)+z_-(2\tilde a_+-\tilde b_+)\big]$  \cr
$\qquad\qquad +32\tilde a_0\,\tilde a_+\tilde a_- 
+8c_q(\tilde a_+z_--\tilde a_-z_+)
-8 c_0\big(\tilde b_+z_- +\tilde b_-z_+\big) $\cr
$X_4=(\tilde b_+ z_-+\tilde b_- z_+)\big[4(m^2+\tilde k^2)-Q^2\big]-4\tilde a_0(2\tilde a_+\tilde b_-+2\tilde a_-\tilde b_+ -z_0\,Q^2)-8c_q(\tilde a_-z_+-\tilde a_+z_-)$\cr
$X_5=-8(\tilde a_0-\tilde b_0)(\tilde a_+\tilde b_-+\tilde a_-\tilde b_+)-8(c_0-\tilde k^2)(\tilde b_+z_-+\tilde b_-z_+)+Q^2\big[4z_0(\tilde a_0-\tilde b_0)-2z_-(2\tilde a_+-\tilde b_+)-2z_+(2\tilde a_- -\tilde b_-)\big]$\cr
$\qquad\qquad -8c_q(\tilde a_-z_+-\tilde a_+z_-)$
\end{tabular}
\end{ruledtabular}
\end{minipage}
\end{table*}
%%%%%%%%%%%%%%%%%%%%%

\end{widetext}
where
use has been made of the compact notation $\widetilde Z_+=Z(\tilde R_+,R_0^+)$ and  $\widetilde Z_-=Z(\tilde R_-,R_0^-)$ (where here $Z$ is the generic name for $\{F, G, H, I\}$) and the vertex function arguments $\tilde R_\pm$ and $R_0^\pm$ were defined in (\ref{eq:kp&eta}).   These arguments depend on both $k_0$ and $Q$.  Recalling that $\tilde k_\pm=\{k_0, {\bf k}_\pm\}$, with ${\bf k}_\pm={\bf k}\pm {\bf q}/2$,  the arguments of the incoming and outgoing vertex functions are 
\bea
\tilde R_\pm^2&=&\frac{(P_\pm\cdot \tilde k_\pm )^2}{m_d^2}-k_0^2+{\bf k}_\pm^2
\nonumber\\%&&\qquad
&=&{\bf k}_\pm^2
\mp \,({k}_\pm)_z k_0\frac{Q D_0}{m^2_d}
+\frac{Q^2}{4m^2_d}\Big[k_0^2+(k_\pm)_z^2\Big]
\nonumber\\
R_0^\pm&=&\frac{1}{2m_d}\big[2D_0k_0\mp (k_\pm)_z Q \big] .
\label{eq:offshellargs}
\eea
Note that $\tilde R_\pm^2$ [which is not the same as the $R_\pm^2$ of Eq.~({\ref{eq:pp&m})] depends on $Q k_0$, so that all $k_0$ dependence vanishes when $Q=0$, and that in this limit, the arguments reduce to ${\bf k}^2$ and $k_0$.   The denominator of $\zeta$ contains an additional $k_0$ dependence through the factor of 
$m^2-\widetilde p^2 =E_k^2-(D_0-k_0)^2$.

  %%%%%%%%%%%%%%%%%%%%%%%%%%%%%%%%%%%%%%%%%%%%%%%%%%
\begin{table}%[t]
\begin{minipage}{3.5in}
\caption{Combinations of vector products used in the expansions of ${\cal C}_{n,i}$.  The only new terms are $\zeta_B=h^2(p_+)/(m^2-p_+^2)$ and $c_0'=D\cdot k= D_0E_k$; the $\tilde b$'s are taken from Table \ref{tab:Bcoeffs}  and the others from Table \ref{tab:Acoeffs}. } 
\label{tab:Ccoeffs}
\begin{ruledtabular}
\begin{tabular}{l}
$T_1=Q^2+8c_0'+4c_q-4\zeta_1$\cr
$T_2=a_0z_0+a_+z_--a_-z_+$\cr
$T_3= Q^2T_2 -2a_0\,\tilde b_-(2a_++\tilde b_+)+2\tilde b_0(a_+\tilde b_-+a_- \tilde b_++c_qz_0)$\cr
$\qquad-2c_0'(\tilde b_+ z_-+\tilde b_-z_+)+c_q(4a_+z_--\tilde b_+z_--3\tilde b_-z_+)$
\end{tabular}
\end{ruledtabular}
\end{minipage}
\end{table}
%
%%%%%%%%%%%%%%%%%%%%%%%%%%%%%%%%%%%%%%%%%%%%%%%%%%

The symmetry (\ref{eq:traceB}) of the ${\cal B}$'s under the transformation $Q\to-Q$ can be confirmed using arguments similar to those used for the ${\cal A}$'s.

\subsection{Contributions of the off-shell terms for diagrams (B) plus $\left<V^\mu_1\right>$} \label{app:Ctraces}

The results for the ${\cal C}_{n,i}$ traces that involve the four invariant functions $K_i$ (contributing to $\Gamma_{\rm off}$ in the initial state) are
\begin{widetext}
\bea
{\cal C}_{n,1}(\Gamma\,\Gamma_{\rm off}) &=&\frac{\zeta_B}{2m^2}\Big\{-4\widetilde F_+K_{1}m^2(\tilde b_0\,z_0+2a_+ z_--\tilde b_-z_+) +2\widetilde F_+K_2(a_--\tilde b_-)\big[2a_+(2a_0-\tilde b_0)+z_+(2c'_0+c_q)\big]
\nonumber\\&&
+(\widetilde F_+K_3+\widetilde H_+K_{1}) T_1T_2-(\widetilde F_+K_4+\widetilde H_+K_2) T_1z_+(a_--\tilde b_-)
\nonumber\\&&
+2\widetilde G_+K_{1}a_+\big[z_-(4m^2-2c_0'-c_q)+2a_-\tilde b_0-2a_0\,\tilde b_-\big] -4\widetilde G_+K_2a_+(2a_0-\tilde b_0)(a_--\tilde b_-)
\nonumber\\&&
-(\widetilde G_+K_3+\widetilde I_+K_1)T_1a_+z_-
+(\widetilde G_+K_4+\widetilde I_+K_2) \frac{a_0}{m^2}T_1 a_+(a_--\tilde b_-)
\nonumber\\&&
-\widetilde H_+K_3 T_1\big[z_0(2a_0-\tilde b_0)-z_+(2a_--\tilde b_-)\big]
+\widetilde H_+K_4\frac{a_--\tilde b_-}{2m^2}T_1\big[z_+(4m^2-2c_0'-c_q)-2a_+(2a_0-\tilde b_0)\big]
\nonumber\\&&
-\widetilde I_+K_3\frac{a_+}{2m^2}T_1\big[2a_-\tilde b_0-2a_0\,\tilde b_--z_-(2c_0'+c_q)\big]
-\widetilde I_+K_4\frac{\tilde b_0}{m^2}T_1a_+(a_--\tilde b_-)\Big\}\label{eq:traceC1}
\eea

\bea
{\cal C}_{n,2}(\Gamma\,\Gamma_{\rm off}) &=&\frac{\zeta_B}{2m}\Big\{-2\widetilde F_+K_1T_3+2\widetilde F_+K_2(a_- -\tilde b_-)(2\tilde b_0\,\tilde b_+-Q^2z_+)-\widetilde F_+K_3 T_1(\tilde b_+z_-+\tilde b_-z_+)
\nonumber\\&&
-(\widetilde F_+K_4+\widetilde H_+K_2)\frac{a_--\tilde b_-}{m^2}T_1(a_0\,\tilde b_+ +c_q z_+)-2\widetilde G_+K_1 a_+ \big[2\tilde b_-(2a_0-\tilde b_0)-z_-(Q^2+4c_q)\big]
\nonumber\\&&
+2\widetilde G_+K_2 a_+(a_--\tilde b_-)(a_0\,Q^2+2\tilde b_0\,c_q)+(\widetilde G_+K_3+\widetilde I_+K_1) \frac{a_+}{m^2}T_1(a_0\,\tilde b_--c_qz_-) - \widetilde H_+K_1T_1(\tilde b_+z_-+\tilde b_-z_+)
\nonumber\\&&
+\widetilde H_+K_3\frac1{2m^2} T_1\big[4m^2(\tilde b_+z_-+\tilde b_-z_+)+T_3\big]+\widetilde H_+K_4\frac{a_--\tilde b_-}{2m^2} T_1\big[2\tilde b_+(2a_0-\tilde b_0)+z_+(Q^2+4c_q)\big]
\nonumber\\&&
-\widetilde I_+K_3\frac{a_+}{2m^2}T_1(z_-Q^2+2\tilde b_0\,\tilde b_-)-\widetilde I_+K_4\frac{a_+}{2m^4}T_1(a_--\tilde b_-)(a_0\,Q^2+2\tilde b_0\,c_q)\Big\}
\label{eq:traceC2}
\eea
\end{widetext}
where the vector products that enter into these formulae  are defined in Tables \ref{tab:Acoeffs}, \ref{tab:Bcoeffs}, and \ref{tab:Ccoeffs},  $m^2-p_+^2=m^2-(P_+-k)^2=2D_0E_k-m_d^2-Qk_z$, $K_i=K_i(\hat R_-, \hat R_0^-)$,  and the final state is on-shell, so that $\tilde Z$ depends on only one argument $\tilde Z_+=Z(R_+)$.

These terms are finite, so calculations of the static moments require them to order $Q^2$ only.  The arguments of the $K_i$ are
\bea
&&\hat R_-^2
=\frac{1}{m_d^2}\big[D_0E_k+\frac12(k_z-Q)\,Q\big]^2
-(m^2+2k_zQ-Q^2) 
\nonumber\\&&\qquad
\to {\bf k}^2-\frac{k_zQ}{m_d}(2m_d-E_k)+\eta\Big[(2m_d-E_k)^2+k_z^2\Big]
\nonumber\\
&&\hat R_0^-=\frac1{m_d}\big[D_0E_k+\frac12(k_z-Q)\,Q\big]
\nonumber\\&&\qquad
\to E_k +\frac{k_zQ}{2m_d}+ \frac12\eta(E_k-4m_d)\, . \label{eq:B4to6args}
\eea
The argument of the $Z_+$ is
\bea
R_+^2&=&\frac1{m_d^2}\Big[D_0E_k-\frac12k_zQ\Big]^2 - m^2
\nonumber\\
&\to& {\bf k}^2-\frac{k_zQ}{m_d}E_k+\eta(E_k^2+k_z^2)\, . \label{eq:B4to6argsa}
\eea
%

%\end{widetext}

\section{Calculations of the Static Moments}
\label{app:staticdetails}

To calculate the magnetic moment requires the expansion of the traces to order $Q$ (for the (A)$+\left<V^\mu_2\right>$ contributions) and $Q^2$ (for the (B)$+\left<V^\mu_1\right>$ contributions).

\subsection{(A)$+\left<V^\mu_2\right>$ contributions}

The exact arguments of the invariants were given in (\ref{eq:pp&m}).  Here particle 1 is always on shell, the invariants depend only on $R_\pm^2$, and expanding the exact formula to order $Q^2$ reduces to
\bea
R_\pm^2\simeq {\bf k}^2\mp  \frac{k_zQ E_k}{m_d} +\eta(E_k^2+k_z^2).
\label{eq:KpexpanforA}
\eea

For calculations of the charge, the $Q\to0$ limits of ${\cal A}_{1,i}$ or ${\cal A}_{2,i}$ are needed, and can be taken directly.  For the calculation of the magnetic moments we need the limits
\bea
\overline M_{iA}(k)=\frac{m}{m_d}\lim_{Q\to0} \frac{{\cal A}_{3,i}(k)}{Q} \label{eq:Qlim}
\eea
(where $\overline M_{iA}(k)$ is used in Appendix~\ref{app:magmoment} below).  
Using Eqs.~(\ref{eq:traceA1}) and (\ref{eq:traceA2}) the  $\overline M_{iA}(k)$ can be expressed in terms of the invariant functions $A \cdots D$ (or $F\cdots I$ for $\overline M_{3A}$) and their derivatives with respect to $k=|{\bf k}|$, the rest frame value of $R_\pm$ defined in (\ref{eq:KpexpanforA}).  Since ${\cal A}_{3,2}$ is already linear in $Q$, the limit (\ref{eq:Qlim}) is easily taken by letting $Q\to0$ in the remaining terms.

The limit (\ref{eq:Qlim}) of ${\cal A}_{3,1}$ is more subtle, and requires using the symmetry condition (\ref{eq:genericA}),  which implies that the terms in (\ref{eq:traceA1}), for the amplitude $n=3$ near $Q=0$,  have the general form
\bea
I_{XY}&=&(c_0+c_1Q) X^f_+Y^i_- -(c_0-c_1Q)Y^f_+X^i_-
\nonumber\\
&=&c_0(X^f_+Y^i_- -Y^f_+X^i_-)+c_1Q(X^f_+Y^i_-+Y^f_+X^i_-),
\qquad  \label{eq:XYAmag}
\eea
where $I_{XY}$ includes both the $XY$ and $YX$ terms. related because of the symmetry.  (Note that when $X=Y$, the coefficient $c_1$ is {\it one-half\/} of the coefficient of the $X_+X_-$ term in ${\cal A}_{3,i}$.)
The term proportional to $Q$ has the structure expected, but the term proportional to $c_0$ can also be present because it has the correct symmetry.  Consideration of the structure of ${\cal A}_{3,1}$ leads to the conclusion that $c_0=x_1 k_z$.   

%%%%%%%%%%%%%%%%%%%
\begin{table}%[htdp]
\begin{minipage}{3in}
\caption{The $Q=0$ limit of  the coefficients of Table \ref{tab:Bcoeffs}.
Here $\tilde k=\{k_0, {\bf k}\}$, which is appropriate for the (B)-type contributions, but the (A)-type contributions are the same with the replacement $k_0\to E_k$, and in both cases  we use the shorthand $k_\perp^\pm=(k_x\pm i k_y)/\sqrt{2}$. }
%\begin{center}
\begin{tabular}{crrrr}
coeff & $\qquad n=1$ & $\qquad n=2$ & $\qquad n=3_+$
& $\qquad n=3_-$\cr
\colrule 
$\tilde a_+$ &  $-k_z$& $k^-_\perp$&  $k^-_\perp$ &
$-k_z$ \cr
$\tilde a_-$&$-k_z$& $k^+_\perp$ & $-k_z$& $k^+_\perp$\cr
$\tilde a_0$&$k_0$ & $k_0$  & $k^+_\perp$ & $-k^-_\perp$\cr
$\tilde b_+$&$0$&0 & 0 & $0$\cr
$\tilde b_-$&$0$&0 & $0$ & 0\cr
$\tilde b_0$&$m_d$& $m_d$ & 0 & 0\cr
$z_+$&$0$&0& $-1$& 0 \cr
$z_-$&$0$&0& 0 & 1\cr
$z_0$&$-1$&$-1$& 0 & 0\cr
\end{tabular}
%\end{center}
\label{tab:3prods}
\end{minipage}
\end{table}
%%%%%%%%%%%%%%

To prove this refer to Table \ref{tab:3prods}.  Each product of invariants is multiplied by a term which must include one, and {\it only\/} one vector product involving {\it each\/} of the polarization vectors $\epsilon, \xi',$ or $\xi^*$.  Inspection of the Table shows that the only non-zero possibilities at $Q=0$ are the product $a_+a_-a_0$ or the product $z_\pm a_\mp$ (where the upper sign goes with $n=3_+$ and the lower with $n=3_-$).  Each of these terms involves either $a_-=-k_z$ (for $n=3_+$) or $a_+=-k_z$ (for  $n=3_-$), proving our assertion.

The term $c_0$ integrates to zero at $Q=0$.  However, near $Q=0$ it makes a contribution, and  $\overline M_{1A}(k)$ emerges from the limit  
\bea
&&\lim_{Q\to0}\frac{I_{XY}}{Q}\to c_1(X^fY^i + Y^fX^i)
\nonumber\\
&&\qquad-y_0x_1k_z^2\,(X'^fY^i-X^fY'^i-Y'^fX^i+Y^fX'^i)\qquad
\label{eq:XYfotB}
\eea
where all structure functions are evaluated at $Q=0$, the generic derivative is $X'\equiv dX/(dk)$,  and $y_0$ is a shorthand for the factor $k_zQE_k/(2m_dk)=k_zQ\,y_0$ appearing in the expansion of $R_\pm$ derived from Eq.~(\ref{eq:KpexpanforA}).  For identical initial and final states, (\ref{eq:XYfotB}) reduces to
\bea
&&\lim_{Q\to0}\frac{I_{XY}}{Q}\to 2c_1(XY)
%\nonumber\\&&\qquad
-2y_0x_1k_z^2\,(X'Y-XY')\qquad
\eea
The results for $\overline M_{1A}$,  $\overline M_{2A}$, and $\overline M_{3A}$, expressed directly in terms of the wave functions $u, w, v_t$, and $v_s$, will be given in Appendix \ref{app:BMag}.  Note that only the $c_1$ term contributes to $\overline M_{2A}$.   %{\bf check Mathematica!! -- I did; its OK}

\subsection{Singular (B)$+\left<V^\mu_1\right>$ contributions}

To evaluate the singular  terms in (\ref{eq:B&V1-2})  requires study of a typical term of the form
\bea
\delta I_{XY}^B\equiv I_{XY}^B(k_0)\Big|_- - I_{XY}^B(k_0)\Big|_+ \label{eq:Bdiffa}
\eea
where $I_{XY}^B$ is the contribution of the product of the generic invariants $X_+Y_-$ to the traces ${\cal B}_{n,i}$.  The ${\cal B}$ traces have the same symmetry as the ${\cal A}_{n,i}$ traces, so that the expansion of a generic term is a generalization of (\ref{eq:XYAmag}), with a  $k_0$ dependence included, and  in this case it is sufficient to only consider cases where the initial and final states are the same, so that, for both charge and magnetic contributions, 
\bea
I_{XY}^B&=&\Big[c_0+c_1Q+(d_0+d_1Q)(k_0-E_k)\Big] X_+Y_- 
\nonumber\\
&&+\epsilon_{n3}\Big[c_0-c_1Q+(d_0-d_1Q)(k_0-E_k)\Big]Y_+X_-
\nonumber\\
&=&[c_0+d_0(k_0-E_k)](X_+Y_- +\epsilon_{n,3}Y_+X_-)
\nonumber\\
&&+Q[c_1+d_1(k_0-E_k)](X_+Y_--\epsilon_{n,3}Y_+X_-),
\qquad  \label{eq:XYBmag}
\eea
where, again, $I^B_{XY}$ includes both the $XY$ and $YX$ terms, related because of the symmetry, and when $X=Y$ all the coefficients are {\it one-half\/} of the coefficient of the $X_+X_-$ term in ${\cal B}_{n,i}$.  Here $I_{XY}^B$ includes the $1/k_0$ factors in (\ref{eq:B&V1-2}), and $\epsilon_{n3}=1$ for charge terms ($n=1,2$) and $-1$ for magnetic terms ($n=3$) as discussed in Sec.~\ref{sec:Bdiagrams}.  Only the $k_0$ dependent part of $I_{XY}^B$ will contribute to the difference (\ref{eq:Bdiffa}); this dependence comes not only from the coefficients $d_0$ and $d_1$, but also from the arguments of $X$ and $Y$.

Note that, after the cancellation in (\ref{eq:Bdiffa}),  terms of order $k_0-E_k\simeq k_0-E_\pm$ will be of order $Q$.   Therefore, when expanding the arguments of $X$ and $Y$,   terms of order $Q(E_--E_k)^3$, $Q^2(E_--E_k)^2$ and $Q^3(E_--E_k)$ %are of order $Q^4$ and 
can be neglected, and the approximation
\bea
E_--E_+\simeq -\frac{k_zQ}{E_k}\label{eq:epmapprox}
\eea
can be used up to order $Q^2$ (satisfactory for this paper).   In addition, terms of order $Q^3(k_0-E_k)^0$ can be dropped since, to contribute at all, they must accompany a contribution at least of order $(k_0-E_k)$ from the other terms, which would make them of order $Q^4$.  Using this classification, expanding the arguments $\tilde R_+^2$ around its rest frame value $k^2$ and $\tilde R_0^+$ around the rest frame energy $E_k$, and neglecting all terms which will eventually contribute only to ${\cal O}(Q^4)$,  gives
\bea
\tilde R_\pm^2&\simeq& {\bf k}^2\pm\frac{k_zQ}{m_d}(m_d-k_0)
\nonumber\\&&+\frac{Q^2}{4m^2_d}\Big[(m_d-E_k)^2+k_z^2-2(k_0-E_k)(m_d-E_k) \Big]
\nonumber\\
R_0^\pm&\simeq&E_k\mp\frac{k_zQ}{2m_d}+ (k_0-E_k)\Big[1+\frac{Q^2}{8m_d^2}\Big] 
\nonumber\\
&&+\frac{Q^2}{8m^2_d}(E_k -2m_d)\, .\qquad
\label{eq:argexpand}
\eea
Note that, for both arguments, only terms to first order in $(k_0-E_k)$ will contribute to order $Q^4$, and that $R_0^\pm=k_0$ if $Q=0$, as expected.

The derivative of the strong form factor also makes a contribution, and is best included as part of the expansion (\ref{eq:XYBmag}).  Recalling that the form factor depends on $\tilde p^2= (D_0-k_0)^2-{\bf k}^2$, the $k_0$ dependence of the form factor can be expanded around $E_k$.  To order $Q$, the coefficient of the term first order in $k_0-E_k$ is
\bea
d_i^{h}=c_i\frac1{h^2}\frac{d h^2}{dk_0}\Big|_{k_0=E_k}\simeq -4c_i\,a(p^2) (m_d-E_k)\qquad
\label{eq:dh}
\eea
where $a(p^2)$ was defined in Eq.\ (\ref{4.7}), $i=0,1$, and, as suggested by the notation, $d_0^h (d_1^h)$ are a contributions to the $d_0 (d_1)$ terms in (\ref{eq:XYBmag}) that are {\it proportional\/} to $c_0 (c_1)$.  These contributions will be isolated from the other $d_i$ contributions, and identified by their explicit dependence on $a(p^2)$. 

Expanding the invariant functions around $\tilde R_\pm=k$ and $R_0^\pm=E_{k}$, keeping terms only up of order $Q$ or $k_0-E_k$ (sufficient for the charge and magnetic moment), so that $Qk_0\to QE_k$,  
the expansion of the typical product of invariant functions is
\bea
X_+Y_-&\simeq& XY+\frac{k_zQ}{2m_dk}(m_d-E_k)(X_kY-XY_k )
\nonumber\\
&&+(k_0-E_k)(X_{k_0}Y+XY_{k_0})
\nonumber\\
&&-\frac{k_zQ}{2m_d}(X_{k_0}Y-XY_{k_0})
\qquad
\label {eq:Xexpan}
\eea
where the generic derivatives are $X_k =  h\,\partial \tilde X(k,k_0)/(\partial k)$ and $X_{k_0} = h\, \partial\tilde X(k,k_0)/(\partial k_0)$,Ê both evaluated at $Q=0$ and $k_0=E_k$.  Note that $X_k\ne X'$; the $X'$ derivative includes a contribution from the $k_0$ dependence when particle 1 is on-shell (when $k_0=E_k$).  As $Q\to0$, the correct connection is
\bea
X_k=X'-\frac{k}{E_k}X_{k_0}\, . \label{eq:dirkrelation}
\eea

From (\ref{eq:Xexpan}) and (\ref{eq:dirkrelation}), the symmetric and antisymmetric sums become
\bea
X_+Y_--Y_+X_- &\simeq&-\frac{k_zQ}{E_k}(X_{k_0}Y-Y_{k_0}X)
\nonumber\\
&&+\frac{k_zQ}{m_dk}(m_d-E_k)(X'Y-Y'X)\qquad
\nonumber\\
X_+Y_- + Y_+X_- &\simeq& 2XY
\nonumber\\
&&+2(k_0-E_k)(X_{k_0}Y+Y_{k_0}X). \qquad
\eea

Only terms that depend on $k_0$ will contribute to the difference (\ref{eq:Bdiffa}), and  since neither $c_1$ nor $d_1$ can contribute to the charge, the charge terms are proportional to
\bea
\delta I_{XY}^B\Big|_{n=1,2}&=&-\frac{2k_zQ}{E_k}\Big[c_0(X_{k_0}Y+Y_{k_0}X)+d_0XY\Big]. \qquad\quad
\label{eq:chargeB}
\eea
where the approximation (\ref {eq:epmapprox}) has been used.  
Restoring the missing factors, the (B)$+\left<V^\mu_1\right>$ contributions to the charge become %{\bf factor of 2?}
\bea
{\cal Q}^B_{XY}&\equiv& \lim_{Q\to0}\frac{mE_k}{k_zQ}\delta I^B_{XY}\Big|_{n=1,2}
\nonumber\\
&=&-2m\Big[c_0(X_{k_0}Y+Y_{k_0}X)+d_0XY\Big].\qquad
\quad
\label{eq:BQ}
\eea

The magnetic terms are more subtle.  Here the difference is, again to order $Q^2$,
%
%\begin{widetext}
%
\bea
\delta I_{XY}^B\Big|_{n=3}
&=&-\frac{2k_zQ^2}{E_k}\Big[c_1(X_{k_0}Y+Y_{k_0}X) 
+d_1XY\Big]\qquad
\nonumber\\
&&+d_0\frac{(k_zQ)^2}{E_k}\Bigg[\frac1{E_k}(X_{k_0}Y-Y_{k_0}X)
\nonumber\\
&&\quad\quad-\frac{(m_d-E_k)}{m_d\,k}(X'Y-Y'X)\Bigg],
\label{eq:Bmag}
\eea
and now $d_0=x_2k_z$ (recall the discussion of Table \ref{tab:3prods}).   Note that, even if $c_0\ne0$, this term would make no contribution to the magnetic moment.  However, as it turns out, $c_0=0$, so $d_0^h=0$ [recall Eq.~(\ref{eq:dh})] and there is also no $a(p^2)$ contribution from this term.

The last two Appendices use these results to extract formulae for the charge and magnetic moment.

\subsection{Regular (B)$+\left<V^\mu_1\right>$ contributions}

The contribution of the off-shell invariants $K_i$ to the charge is easily calculated by evaluating the ${\cal C}_{n,i}$ traces at $Q=0$.   

To extract the magnetic moment, the ${\cal C}_{3,i}$ traces are divided by $Q$, and the $Q\to0$ limit is taken.  Using the antisymmetry of the magnetic terms, the sum of the two ${\cal C}_{3,i}$ traces near $Q=0$ has the form
\bea
I_{XK}^C&=&\Big[c_0+c_1Q\Big]X_+K_- 
%\nonumber\\&&
-\Big[c_0-c_1Q\Big]X_-K_+\qquad\quad
\label{eq:Cterms}
\eea
where now $c_0=x_1k_z$ and $K$ is the generic name for any of the $K_i$.  The arguments $R_+$ of the on-shell $X_+$ was given in (\ref{eq:B4to6argsa}).  To first order in $Q$ it is 
\bea
R_+\simeq k-\frac{k_z Q}{2m_dk}E_k\, .
\eea
From (\ref{eq:B4to6args}) the arguments of the off-shell $K_i$, to first order in $Q$, are
\bea
\widehat R_-&\simeq& k-\frac{k_zQ}{2m_dk}(2m_d-E_k)
\nonumber\\
\widehat R_0^-&\simeq&E_k+\frac{k_zQ}{2m_d}
\eea
Hence, expanding (\ref{eq:Cterms}) near $Q=0$  keeping only the lowest order terms, and recalling that $c_0=x_1k_z$, gives a result linear in $Q$
\begin{widetext}
\bea
I_{XK}^C&\simeq&x_1k_z (X_+K_- -X_-K_+)+2c_1QXK
%\nonumber\\&\simeq&
\simeq -x_1\frac{k_z^2Q}{m_dk}\Big[E_kX'K-kXK_{k_0}+(2m_d-E_k)XK_k\Big]
%\nonumber\\&&
+2c_1QXK
\nonumber\\
&=&-x_1\frac{k_z^2Q}{m_dk}\Big[E_kX'K +(2m_d-E_k)XK'\Big]
%\nonumber\\&&
+2x_1\frac{k_z^2Q}{E_k}XK_{k_0}+2c_1QXK
\label{eq:Cterms2}
\eea
where (\ref{eq:dirkrelation}) was used to replace $K_k$ by $K'$. 

The derivative of the strong form factor $h$ makes a contribution to the coefficient $c_1$ {\it proportional\/} to $c_0$.  Expanding the argument of the form factor to first order in $Q$,
\bea
p_+^2\simeq p^2 +Qk_z,
\eea
and recalling $a(p^2)$ defined in Eq.\ (\ref{4.7}), this contribution is
\bea
I_{XK}^{C,h}=4c_0a(p^2) Qk_z XK\, .
\eea

%\vspace{0.7in}
%\begin{widetext}

\section{Charge}
\label{app:charge}

In this Appendix, the charge is evaluated by taking the $Q^2=0$ limit of the contributions from Eqs.~(\ref{eq:traceA1}), (\ref{eq:traceB1}), and (\ref{eq:traceC1}).  The results from this Appendix were collected in Sec.~\ref{sec:GC0} and discussed in Sec.~\ref{sec:CandN}.  Here, for simplicity, we return to the notation ${\bf k}^2\to k^2$.

 \subsection{(A)$+\left<V^\mu_2\right>$ contributions}
 
 At $Q^2=0$, $Z_\pm=Z(k^2)$ and all ${\cal A}_{n,2}=0$.  Averaging over $\theta$ using $\left<k_z^2\right>=\left<k_x^2\right>=\left<k_y^2\right>=\frac13\left<k^2\right>$ gives
 %
% \begin{widetext}
  \bea
 {\cal A}_{1,1}= {\cal A}_{2,1}=\frac{2E_k}{m}\bigg\{&&A^2+\frac{k^2}{3m^2}\Big[B^2+m_R^2\,D^2-\frac{2m_d}{E_k}BD\Big]+C^2\Big[4+m_R^2+\frac{4 k^2}{3m^2}-\frac{4m_d}{E_k}\left(1+\frac{k^2}{3m^2}\right)\Big]
 \nonumber\\
 &&+\frac{2m_d \,k^2}{3m^2 E_k}AD -2AC \left(2-\frac{m_d}{E_k}\right)-2BC\frac{k^2}{3m^2} \left(2-\frac{m_d}{E_k}\right)\bigg\}
 \nonumber\\
  {\cal A}_{1,3}= {\cal A}_{2,3}=\frac{E_k}{m}\bigg\{&&F^2+\frac{k^2}{3m^2}\Big[G^2+m_R^2\,I^2-\frac{2m_d}{E_k}GI\Big]+H^2\Big[4+m_R^2+\frac{4 k^2}{3m^2}-\frac{4m_d}{E_k}\left(1+\frac{k^2}{3m^2}\right)\Big]
 \nonumber\\
 &&+\frac{2m_d\, k^2}{3m^2 E_k}FI -2FH \left(2-\frac{m_d}{E_k}\right)-2GH\frac{k^2}{3m^2} \left(2-\frac{m_d}{E_k}\right)\bigg\}\, . \label{eq:charge3}
 \eea
 %
% \end{widetext}

The contributions from the $\left<V^\mu_2\right>$ part of the exchange currents can be easily added.  Using the symmetry relation (\ref{eq:genericA}) at $Q=0$, the generic $XY$ term in the expansions (\ref{eq:charge3}) can be transformed as follows:
\bea
c_0XY&\to& \frac{1}{2}c_0(X^fY^i+Y^fX^i)
%\nonumber\\&\to& 
\to c_0(XY^{(2)}+YX^{(2)})\, ,
\label{eq:2subsitution}
\eea
where $c_0$ is independent of $Q$.  The first step uses the symmetry relation to uncover the structure of the generic $XY$ term in the case when the initial and final wave functions are not identical; the symmetry relation guarantees that this replacement is unique. Then, the second step merely applies the result to the special case when the generic final state functions are $X$  and the generic initial state functions are $X^{(2)}$.  The  two terms in (\ref{eq:A&V2-3}) are identical in this case, giving a factor of 2.

\subsection{(B)$+\left<V^\mu_1\right>$ contributions}\label{app:chargeB}

These contributions are obtained from Eq.~(\ref{eq:B&V1-2}) and the traces ${\cal B}_{n,i}$ (\ref{eq:traceB1}) and the traces ${\cal C}_{n,i}$ (\ref{eq:traceC1}).  The magnetic terms (\ref{eq:traceB2}) and (\ref{eq:traceC2})  are zero and do not contribute.

The singular terms have already been partially reduced.  Using (\ref{eq:BQ}) the result of the expansion of the trace (\ref{eq:traceB1})  is 
\bea
\lim_{Q\to0}\frac{mE_k}{k_z Q}&&\left[\frac{{\cal B}_{1,1}(k_0)}{k_0}\Big|_- - \frac{{\cal B}_{1,1}(k_0)}{k_0}\Big|_+\right]
=\sum_{X,Y}{\cal Q}_{XY}^B\Big|_{n=1}
\nonumber\\
&&={\cal I}_Z+{\cal I}_{Z'}\, , \label {eq:Bdiff}
\eea
where  ${\cal I}_{Z'}$ includes derivatives of the vertex functions $Z$, and ${\cal I}_Z$ all of the rest [including contributions from the $k_0$ expansion of the strong form factors  $h(\tilde p)$].  The results for ${\cal I}_Z$ and ${\cal I}_{Z'}$ are
%
%\begin{widetext}
%
\bea
{\cal I}_Z&=& \frac{2E_k}{m\,\delta_k^2}\Big[F^2 +\frac{k^2}{3m^2}G^2 +\frac{4k^2}{3m_d^2}(F-G)^2\big(1-\frac{m_d}{E_k}\big)-\frac{\delta_k^2}{m^2}\big(H^2+\frac{k^2}{3m^2}I^2\big)\Big] 
-\frac{8 E_k}{m\,\delta_k}a(p^2) (E_k-m_d) \Bigg\{F^2 +\frac{k^2}{3m^2}G^2
\nonumber\\&&
-\frac{2 k^2}{3E_km_d}(F-G)^2-\frac{2\delta_k}{E_k}\Big[FH+\frac{k^2}{3m^2}(FI-H^2+GH-GI)\Big]-\frac{\delta_km_d}{m^2}\Big[\frac{k^2}{3m^2}I^2+\Big(1-\frac{2m^2}{m_dE_k}\Big)H^2\Big]\Bigg\}
\nonumber\\
{\cal I}_{Z'}&=&- \frac{4E_k}{m \,\delta_k}\Bigg\{FF_{k_0} + \frac{k^2}{3m^2}GG_{k_0}-\frac{2 k^2}{3E_km_d}(F-G)(F_{k_0}-G_{k_0}) -\frac{\delta_km_d}{m^2}\Big[\frac{k^2}{3m^2}II_{k_0}+\Big(1-\frac{2m^2}{m_dE_k}\Big)HH_{k_0}\Big]
\nonumber\\&&
-\frac{\delta_k}{E_k}\Big[FH_{k_0}+F_{k_0}H+\frac{k^2}{3m^2}(FI_{k_0}+F_{k_0}I-2HH_{k_0}+GH_{k_0}+G_{k_0}H-GI_{k_0}-G_{k_0}I)\Big]\Bigg\} \label{eq:J6}
\eea
where the strong form factor $h(\tilde p)$ [where $h$ is evaluated at $\tilde p^2=m^2-m_d(2E_k-m_d)]$ has been reabsorbed into the $Z$'s (so that they may be expressed in terms of the $u, w, v_t, v_s$),  $Z_{k_0}=h \,d\tilde Z/(dk_0)$, and the contributions to ${\cal I}_Z$  from the derivative of the strong form factor have been isolated in the term proportional to $a(p^2)$.

The contribution from the regular terms is straightforward:
\bea
{\cal I}_C=-\frac{2}{m}{\cal C}_{1,1}&=&-\frac{4}{m\,\delta_k}\Bigg\{FK_1-\frac{E_k\delta_k}{m^2}(HK_1+FK_3)+\frac{\delta_k^2}{m^2}HK_3
%\nonumber\\&&
+\frac{k^2}{3m^2}GK_1
\nonumber\\
&&+\frac{k^2\delta_k}{3m^4m_d}\Big[m^2(F-G)K_2-m_d^2 I(K_3+K_4)+E_km_d(GK_4+IK_2)-m_d\delta_k HK_4\Big]\Bigg\}
\eea

%\end{widetext}

\subsection{Expressions in terms of the wave functions $z_\ell$}

Expanding the $Z$ in terms of the wave functions $z_\ell(k)$ (where $z_\ell$ is the generic name for $\{u,w,v_t,v_s\}$) using (\ref{eq:AtoF}) and (\ref{eq:Ftou}), reduces (\ref{eq:charge3}) to the following simple forms
%
%\vskip0.2in
\bal
{\cal A}_{n,1}=4\pi^2m_d\frac{E_k}{m}
\Big\{&u^2+w^2+v_t^2+v_s^2\Big\}
\nonumber\\
{\cal A}_{n,3}=4\pi^2m_d\frac{E_k}{m}
\Big\{&\delta_k^2(u^2+w^2)+
m_d^2(v_t^2+v_s^2)\Big\}\, .
\label{eq:BC}
\end{align}
Using (\ref{eq:2subsitution}), the  the result for the interaction current contribution is
%\begin{widetext} 
%
\bea
{\cal A}^{(2)}_{n,1}&=&4\pi^2m_d\frac{E_k}{m}
%\nonumber\\&&\times
\Big\{2uu^{(2)}+2ww^{(2)}+2v_tv_t^{(2)}+2v_sv_s^{(2)}\Big\} \qquad \label{eq:BC2}
\eea
where $z_\ell^{(2)}$ is the generic name for the wave functions that contribute to $\Psi^{(2)}$.  The contribution of these terms to the in normalization condition is discussed in Sec.\  \ref{sec:GC0}.

For the (B)$+\left<V_1\right>$ contributions, the vertex functions $Z$ are expanded in terms of the wave functions $z_\ell(k)$ using (\ref{eq:Fz}) and (\ref{eq:uz}).  This gives 
%\begin{widetext}
%
\bea
{\cal I}_Z&=& 4\pi^2 m_d\frac{E_k}{m}\Bigg\{u^2+w^2-v_t^2-v_s^2+\sqrt{\frac23}\,\frac{E_k+2m}{m_dk}\Big[uv_t\,m_d  + wv_s\,\delta_k\Big] +\frac{2}{\sqrt{3}}\frac{E_k-m}{m_dk}\Big[wv_t\,m_d -uv_s\,\delta_k\Big]
\nonumber\\
&&\qquad\qquad-4 a(p^2)(E_k
-m_d)\Big[\delta_k(u^2+w^2)-m_d(v_t^2+v_s^2)\Big]\Bigg\}\, ,
\eea
To obtain the result for ${\cal I}_{Z'}$, the derivatives of the invariants must be evaluated using the general results (\ref{eq:AtoK}) which give the $k_0$ dependence of the invariants.  These give
\bea
F_{k_0}=\frac{\partial F}{\partial k_0}\Big|_{k_0=E_k}&=&{\cal K}\Big[u-\frac{w}{\sqrt{2}}+\sqrt{\frac32}\frac{m}{k}v_t\Big]\Big(1-\frac{\delta_k^2}{2E_km_d}\Big) +{\cal K}\delta_k\Big[u_{k_0} -\frac1{\sqrt{2}}w_{k_0}+\sqrt{\frac32}\frac{m}{k}v_{tk_0}+\frac{\sqrt{3}}{2E_kk}(kz_1^{--}+mz_1^{-+})\Big]
\nonumber\\
G_{k_0}=\frac{\partial G}{\partial k_0}\Big|_{k_0=E_k}&=& {\cal K}m \Big[\frac{u}{E_k+m} +(2E_k+m)\frac{w}{\sqrt{2}\,k^2} +\sqrt{\frac32}\frac{v_t}{k}\Big] \Big(1-\frac{\delta_k^2}{2E_km_d}\Big)
\nonumber\\
&&\qquad+{\cal K}m\,\delta_k\Big[\frac{u_{k_0}}{E_k+m}+(2E_k+m)\frac{w_{k_0}}{\sqrt{2}\,k^2}+\sqrt{\frac32}\frac{v_{tk_0}}{k}
+\frac{\sqrt{3}}{2E_kk^2}(\sqrt{2}E_kz_0^{--}-mz_1^{--}+kz_1^{-+})\Big]
\nonumber\\
H_{k_0}=\frac{\partial H}{\partial k_0}\Big|_{k_0=E_k}&=&\sqrt{\frac32} \frac{{\cal K}E_k\,m}{k}\,v_{tk_0}-\sqrt{\frac32}\frac{{\cal K}\delta_k\,m}{2k\,m_d}(v_t-\sqrt{2}\,z_1^{-+})
\nonumber\\
I_{k_0}=\frac{\partial I}{\partial k_0}\Big|_{k_0=E_k}&=& -\frac{2{\cal K}E_k\,m^2}{m_d^2} \Big[\frac{u}{E_k+m} -(E_k+2m) \frac{w}{\sqrt{2}\,k^2}\Big] \Big(1-\frac{\delta_k^2}{2E_km_d}\Big)+\frac{\sqrt{3}\,{\cal K} \delta_km^2}{2E_km_dk} v_s
\nonumber\\
&& -\frac{{\cal K}\,m^2}{m_d}\Big[\delta_k\Big(\frac{u_{k_0}}{E_k+m}
-(E_k+2m)\frac{w_{k_0}}{\sqrt{2}\,k^2}\Big)+\sqrt{3}\,m_d\frac{v_{sk_0}}{k}
\Big]
\nonumber\\
&&+\sqrt{\frac32}\frac{{\cal K} m^2}{E_km_dk^2}\Big[m_dmz_0^{--}-\delta_kkz_0^{-+}-\frac{E_km_d}{\sqrt{2}}z_1^{--}\Big]
\eea
where $z_{\ell k_0} = h\,\partial \tilde z_\ell(k,k_0)/(\partial k_0)|_{k_0=E_k}$ and $z_\ell=h\, \tilde z_\ell$.  Note the appearance of the negative $\rho$-spin helicity amplitudes for particle 1, referred to generically as $\chi_\ell$.  Substituting these expressions and the expansions (\ref{eq:Ftou}) into ${\cal I}_{Z'}$ gives
\bea
{\cal I}_{Z'}&=&4\pi^2 m_d\frac{E_k}{m}\Bigg\{
\frac{\delta_k^2}{E_km_d}(u^2+w^2)-\sqrt{\frac23}\,\frac{E_k+2m}{m_dk}\Big[uv_t\,m_d  + wv_s\,\delta_k\Big] 
-\frac{2}{\sqrt{3}}\frac{E_k-m}{m_dk}\Big[wv_t\,m_d -uv_s\,\delta_k\Big]
\nonumber\\
&&\qquad\qquad\quad -2\Big(u[\delta_+u]_{k_0}+w[\delta_+w]_{k_0}\Big)+\Big(2-\frac{\delta_k}{E_k}\Big)(v_t^2+v_s^2) +2\Big(v_t[\delta_-v_t]_{k_0}+v_s[\delta_-v_s]_{k_0}\Big)
\nonumber\\
&&\qquad\qquad\quad-\sqrt{\frac23}\frac{\delta_k}{E_k}\Big[u(z_0^{--}+\sqrt{2}z_1^{--})+w(\sqrt{2}z_0^{--}-z_1^{--})-\sqrt{3}(v_tz_1^{-+}+v_sz_0^{-+})\Big]
\nonumber\\
&&\qquad\qquad\quad+\frac{2\delta_k(E_k-m_d)}{\sqrt{3}\,E_km_dk}z_1^{-+}\Big[(E_k+2m)u+\sqrt{2}(E_k-m)w\Big]+\frac{2(E_k-m_d)}{E_kk}v_s\Big[\sqrt{2}mz_0^{--}-E_kz_1^{--}\Big]\Bigg\}.\qquad\quad
\eea
where the new functions
\bea
\delta_+\{u,w\}&=&(E_k+k_0-m_d)\{u,w\}
\nonumber\\
\delta_-\{v_t,v_s\}&=&(E_k-k_0+m_d)\{v_t,v_s\}\label{eq:deltauv}
\eea
have been introduced.  Finally, the contribution from ${\cal I}_C$ is obtained by substituting the expansions  (\ref{eq:Ftou}) and (\ref{eq:Ktou}), giving
\bea
{\cal I}_{C}&=&-4\pi^2 m_d\frac{E_k}{m}\Bigg\{
\frac{\delta_k^2}{E_km_d}(u^2+w^2)-\frac{\delta_k}{E_k}(v_t^2+v_s^2)
\nonumber\\
&&\qquad\qquad\quad-\sqrt{\frac23}\frac{\delta_k}{E_k}\Big[u(z_0^{--}+\sqrt{2}z_1^{--})+w(\sqrt{2}z_0^{--}-z_1^{--})-\sqrt{3}(v_tz_1^{-+}+v_sz_0^{-+})\Big]
\nonumber\\
&&\qquad\qquad\quad+\frac{2\delta_k(E_k-m_d)}{\sqrt{3}\,E_km_dk}z_1^{-+}\Big[(E_k+2m)u+\sqrt{2}(E_k-m)w\Big]+\frac{2(E_k-m_d)}{E_kk}v_s\Big[\sqrt{2}mz_0^{--}-E_kz_1^{--}\Big]\Bigg\}.\qquad
\eea
Note that all terms with particle 1 in a negative $\rho$-spin state cancel in ${\cal I}_{Z'}$ and ${\cal I}_C$.   {\it The charge is independent of the amplitudes\/}
$z_j^{-\pm}$.
Finally, the sum of all the (B) terms is
\bea
{\cal I}_Z+{\cal I}_{Z'}+{\cal I}_C&=&4\pi^2m_d\frac{E_k}{m}\Bigg\{u^2+w^2+v_t^2+v_s^2 -4a(p^2)(E_k
-m_d)\Big[\delta_k(u^2+w^2)-m_d(v_t^2+v_s^2)\Big]
\nonumber\\
&&\qquad\qquad\qquad -2\Big(u[\delta_+u]_{k_0}+w[\delta_+w]_{k_0}\Big)+2\Big(v_t [\delta_-v_t]_{k_0}+v_s [\delta_-v_s]_{k_0}\Big)\Bigg\} \label{eq:combinedB}
\eea
\vspace{0.2in}
\end{widetext}
This result is discussed further in Sec.\  \ref{sec:GC0}.

%%%%%%%%%%%%%%%%%%%%

\section{Magnetic moment} \label{app:magmoment}

\subsection{(A)$+\left<V^\mu_2\right>$ contributions} \label{app:K1}

The contributions  to the magnetic moment, in units of $e/(2m)$, from diagram (A) $+\left<V^\mu_2\right>$ are obtained from the limit
\bea
\mu_d=\lim_{Q\to0} \frac{m}{m_d} \frac{{\cal J}_3}{Q}&&\Bigg|_{{\rm A} +V_2}
%\nonumber\\
=e_0\int_k\Bigg\{f_{00}\Big[\overline M_{1A}+\kappa_s\,\overline M_{2A}\Big]
\nonumber\\&&
+ \frac{g_{00}}{4m^2}\overline M_{3A}-\overline M_1^{(2)}-\kappa_s\,\overline M_2^{(2)}\Bigg\}
\eea
where $\kappa_s=\kappa_p+\kappa_n$ is the isoscalar anomalous moment of the nucleon and the $\overline{M}_{iA}=\overline{M}_{iA}(k)$ are the limits (\ref{eq:Qlim}).  

The $\overline M_1^{(2)}$ and $\overline M_2^{(2)}$ terms contain the $\left<V^\mu_2\right>$ part of the exchange current.   
Since the anomalous moment term (\ref{eq:traceA2}) is linear in $Q$ (i.e. $c_0=0$), application of (\ref{eq:XYfotB}) gives
\bea
&&\lim_{Q\to0}\frac{I^{(2),2}_{XY}}{Q}\to c_1(XY^{(2)} + YX^{(2)})
\eea
and hence $\overline M_2^{(2)}$ can be obtained directly from $\overline M_{2A}$, just as was done for the charge.   To calculate the  $\overline M_1^{(2)}$ term we need to keep all of the terms in (\ref{eq:XYfotB}), giving 
\bea
&&\lim_{Q\to0}\frac{I^{(2),1}_{XY}}{Q}\to c_1(XY^{(2)} + YX^{(2)})
\nonumber\\
&&\quad-y_0x_1k_z^2\,(X'Y^{(2)}+X^{(2)}{}'Y-XY^{(2)}{}'-X^{(2)}Y'),\qquad
\label{eq:XYfotBa}
\\
&&
\nonumber
\eea
where $X^{(2)}{}'=dX^{(2)}/(dk)$, and the order of the terms proportional to $x_2$ were rearranged to emphasize the substitution rule, which for all terms (i.e. with or without the derivative) is 
\bea
XY'\to XY^{(2)}{}'+X^{(2)}Y'
\label{eq:XYsubst}
\eea

Since the exact result for the magnetic moment does not simplify as it did for the charge, the goal here is to understand the physical content of the ``leading'' terms only (those that are expected to be  larger than about 0.001 nuclear magnetons). These terms are the products of the wave functions, including some products of one large ($u,w$) and one small ($v_t,v_s$) component multiplied by the enhancement $m/k$, and products of one wave function and one derivative, multiplied by $m$ or $k$.   In addition the  leading corrections to order $\delta_E=(E_k-m)/E_k$ to products of $u$ and $w$ are retained.  The results, expressed in terms of the wave functions $z_\ell(k)$ (where $z_\ell$ is the generic name for $\{u,w,v_t,v_s\}$), are 
\begin{widetext}
\bea
\overline M_{1A}(k)=2\pi^2&&\frac{E_k}{m}\Bigg\{
u^2+\frac14w^2-\frac14v_t^2-\frac12v_s^2+\frac{m}{\sqrt{6}}\Big[ m_1(k) +\frac12m_2(k) \Big]
+\Delta M_1(k) \Bigg\} \label{eq:M1a}
\eea
\bea
\overline M_{2A}(k)=2\pi^2&&\frac{E_k}{m}\Bigg\{ u^2-\frac12 w^2-\frac12 v_t^2-\sqrt{2}v_tv_s 
+\Delta M_2(k) \Bigg\} \label{eq:M2a}
\\
\overline M_{3A}(k)=2\pi^2&&\frac{E_k}{m}\Bigg\{ 
m^2\Big(\frac32 v_t^2+v_s^2+2\sqrt{2}v_tv_s\Big) 
\Bigg\} ,
\qquad\label{eq:M3a}
\eea
 the interference terms are
\bea
m_1(k)&=&\frac1{k}\Big[u(v_t-\sqrt{2}v_s)-2w(\sqrt{2}v_t+v_s)\Big]
\nonumber\\
m_2(k)&=&uv_t'-u'v_t-\sqrt{2}(uv_s'-u'v_s-wv_t'+w'v_t)+wv_s'-w'v_s,
\eea
and  the standard notation $z_\ell'=d z_\ell/(d k)$ has been used.  The leading corrections  are
\bea
\Delta M_1(k)&\simeq&-\frac{E_k-m}{3E_k}\Big[ u^2-w^2+\frac1{\sqrt{2}}uw\Big]
\nonumber\\
\Delta M_2(k)
&\simeq&
 -\frac{E_k-m}{3E_k} \Big[ u^2+\frac12w^2-\sqrt{2}uw\Big] .
\eea

The contributions from the $z^{(2)}$  wave functions can be obtained from $\overline M_{1A}$ and  $\overline M_{2A}$ by the substitution (\ref{eq:XYsubst}), but first we transform the expression for $\overline M_{1A}$.  The interference terms can be rearranged and, recalling that the volume integral over $k$ is $k^2dk/E_k$,  integrated by parts, giving 
\bea
\overline M_{1A}^{\;\rm int}(k)&=&2\pi^2\frac{E_k}{m}
\frac{m}{2\sqrt{6}}\Bigg\{\Big[ u(v_t-\sqrt{2}v_s)'-u'(v_t-\sqrt{2}v_s)+\frac{2}{k}u(v_t-\sqrt{2}v_s) \Big]
\nonumber\\
&&\qquad\qquad+\Big[w(\sqrt{2}v_t+v_s)'-w'(\sqrt{2}v_t+v_s)-\frac4{k}w(\sqrt{2}v_t+v_s)\Big]\Bigg\}
\nonumber\\
&=&2\pi^2\frac{E_k}{m}
\frac{m}{\sqrt{6}}\Bigg\{-u'(v_t-\sqrt{2}v_s)+w\Big[(\sqrt{2}v_t+v_s)' -\frac1{k}(\sqrt{2}v_t+v_s)\Big]\Bigg\}\equiv2\pi^2\frac{E_k}{m}\, m^I(k)\, .
\eea
The interference contribution from the  $z^{(2)}$  wave functions can therefore be written
\bea
\overline M_{1A}^{{\;\rm int}(2)}(k)&=&2\pi^2\frac{E_k}{m}
\frac{m}{\sqrt{6}}\Bigg\{-u^{(2)}{}'(v_t-\sqrt{2}v_s)-u'(v_t-\sqrt{2}v_s)^{(2)}+w^{(2)}\Big[(\sqrt{2}v_t+v_s)' -\frac1{k}(\sqrt{2}v_t+v_s)\Big]
\nonumber\\&&\qquad\qquad
+w\Big[(\sqrt{2}v_t+v_s)^{(2)}{}' -\frac1{k}(\sqrt{2}v_t+v_s)^{(2)}\Big]\Bigg\}
\nonumber\\
&=&2\pi^2\frac{E_k}{m}
\frac{m}{\sqrt{6}}\Bigg\{u^{(2)}(v_t-\sqrt{2}v_s)'-u'(v_t-\sqrt{2}v_s)^{(2)}+\frac2{k}u^{(2)}(v_t-\sqrt{2}v_s)
+w^{(2)}(\sqrt{2}v_t+v_s)'
\nonumber\\&&\qquad\qquad
-w'(\sqrt{2}v_t+v_s)^{(2)}-\frac1{k}\Big[w^{(2)}(\sqrt{2}v_t+v_s)+3w(\sqrt{2}v_t+v_s)^{(2)}\Big]\Bigg\}\equiv2\pi^2\frac{E_k}{m}\,m^{I(2)}(k) \label{eq:m2I}
\, .\qquad
\eea
With these definitions, the total contributions from the  $z^{(2)}$ wave functions are
\bea
\overline M_1^{(2)}(k)=2\pi^2&&\frac{E_k}{m}\Big\{ 2uu^{(2)}+\frac12 ww^{(2)}-\frac12 v_tv_t^{(2)}- v_sv_s^{(2)}+m^{I(2)}(k) \Big]   \Big\}
\\
\overline M_2^{(2)}(k)=2\pi^2&&\frac{E_k}{m}\Big\{ 2uu^{(2)}- ww^{(2)}- v_tv_t^{(2)}-\sqrt{2}[v_tv_s^{(2)}+v_t^{(2)}v_s]  \Big\} .\label{eq:M4a}
\eea
There are also small corrections $\Delta M_1^{(2)}$ and $\Delta M_2^{(2)}$ but these can be neglected.

The leading contributions proportional to the derivative of the strong from factor, expressed in terms of $a(p^2)$ defined in Eq.~(\ref{4.7}), are assembled from $\overline M_{1A}$ and $\overline M_{3A}$ using (\ref{eq:FGexpand}).  Dropping all terms proportional to $\delta_k$ except for the large $u^2$ and $w^2$ terms gives
\bea
\overline M_{A1,{\rm a\;term}}&\simeq& a(p^2)\Big[4m\delta_k \overline M_{A1}(k)-2\overline M_{A3}(k)\Big]
%\nonumber\\
\simeq2\pi^2\frac{E_k}{m}a(p^2)\Big\{4m\delta_k \Big(u^2+\frac14w^2\Big)-2m^2\Big(\frac32 v_t^2+v_s^2+2\sqrt{2}v_tv_s\Big) \Big\}
\nonumber\\
\overline M_{A2,{\rm a\;term}}&\simeq& (2\omega_2-1)\,a(p^2)4m\delta_k \overline M_{A2}(k)\simeq 2\pi^2\frac{E_k}{m}a(p^2)\,(2\omega_2-1)\,4m\delta_k \Big( u^2-\frac12 w^2 \Big)\, ,
\eea
where $\omega_2=1$ is the parameter defined in Eq.~(\ref{eq:om2def}).

In view of the rich history and importance of the magnetic moment, it is instructive to rewrite the largest terms in expressions (\ref{eq:M1a}) and (\ref{eq:M2a}) as coordinate space integrals.  In momentum space, the leading terms  for the deuteron magnetic moment can be rearranged into the following form
\bea
\mu_d\Big|_0=e_0\int_0^\infty k^2 dk\Bigg\{&&u^2+\frac14w^2-\frac14v_t^2-\frac12v_s^2+\frac{m}{2\sqrt{6}}\Big[\Big(uv_t'-u'v_t+\frac{2uv_t}{k}\Big)+\sqrt{2}\Big(u'v_s-uv_s'-\frac{2uv_s}{k}\Big)
\nonumber\\
&&
+\sqrt{2}\Big(wv_t'-w'v_t-\frac{4 wv_t}{k}\Big)+\Big(wv_s'-w'v_s-\frac{4wv_s}{k}\Big)\Big]\Bigg\}
\nonumber\\
+e_0\,\kappa_s\int_0^\infty k^2 dk&&\Bigg\{ u^2-\frac12w^2-\frac12 v_t^2-\sqrt{2}\,v_tv_s\Big\}\, .
\eea
These can be cast into integrals over the wave functions in coordinate space, defined by the transforms (\ref{eq:besseltrans}).  The squared terms and the $v_tv_s$ term are straightforwardly reduced using the normalization condition (\ref{eq:besselnorm}).  The interference terms can be reduced by using the  identity (\ref{eq:H40}) to shift  derivatives, giving   
\bea
\int_0^\infty k^2dk\big\{uv'-u'v+\frac{2uv}{k}\big\}&=& -2\int_0^\infty k^2dk \,u'(k)v(k) = 2\int_0^\infty rdr\, u(r)v(r)
\nonumber\\
\int_0^\infty k^2dk\big\{wv'-w'v-\frac{4wv}{k}\big\}
&=&2\int_0^\infty k^2dk \,w(k)\big(v'(k)-\frac{v(k)}{k}\big)=-2\int_0^\infty rdr\, w(r)v(r)
\eea
(where $v$ can be either $v_t$ or $v_s$) and the final integrals are are evaluated using the relations  
\bea
u'(k)&=&\sqrt{\frac2\pi}\int_0^\infty r^2dr\Big( \frac1{r}\frac{d}{dk}\Big)\,j_0(kr)u(r)= -\sqrt{\frac2\pi}\int_0^\infty r^2dr\,j_1(kr)u(r)
\nonumber\\
v'(k)  -\frac{v(k)}{k}&=&\sqrt{\frac2\pi}\int_0^\infty r^2dr\Big(\frac1{r}\frac{d}{dk}-\frac1{kr}\Big)\,j_1(kr)v(r)= -\sqrt{\frac2\pi}\int_0^\infty r^2dr\,j_2(kr)v(r) \label{eq:H43}
\eea
and then using the normalization condition (\ref{eq:besselnorm}).
Writing the final result in terms of the isoscalar magnetic moment, $\mu_s=\kappa_s+1$, gives 
\bea
\mu_d\Big|_0=e_0\,\mu_s\int_0^\infty dr&&\Bigg\{u^2-\frac12w^2- \frac12v_t^2\Big\} - e_0\,\kappa_s\,\sqrt{2}\int_0^\infty dr\, v_tv_s
\nonumber\\
e_0\int_0^\infty dr\Bigg\{&&\frac34w^2+\frac14v_t^2-\frac12v_s^2 +\frac1{\sqrt{6}}mr\Big[u(v_t-\sqrt{2}\,v_s)-w(\sqrt{2}v_t+v_s)\Big]\Bigg\}\, . \label{eq:leadingM}
\eea
%
%\end{widetext}
In Ref.~\cite{SLAC-PUB-2318},  interaction currents were ignored and the (B) diagrams were assumed to be equal, to the (A) diagram (the RIA approximation); in this case the normalization condition was 
\bea
1=\int_0^\infty dr\big\{u^2+w^2+v_t^2+v_s^2\big\}.%\qquad {\rm (RIA\;of\;Ref.\,\cite{SLAC-PUB-2318})}\, .
\eea
With this assumption, the results of Eq.~(\ref{eq:leadingM}) agree with Ref.~\cite{SLAC-PUB-2318}.

\subsection{(B)$+\left<V^\mu_1\right>$ contributions}  \label{app:BMag}

The contributions to the magnetic moment (in nuclear magnetons) from the singular terms (involving the ${\cal B}_{3,i}$ traces) can be written
\bea
\mu_d=&&e_0\int_k\Big\{\overline{M}_{1B}(k) 
+\kappa_s \overline{M}_{2B}(k)\Big\}\, ,
\qquad
\eea
where the $\overline M_{iB}$ are
\bea
\overline M_{iB}&=&\frac{m}{m_d}\lim_{Q\to0}\frac{mE_k}{k_z Q}\left[\frac{{\cal B}_{3,i}(k_0)}{Qk_0}\Big|_- - \frac{{\cal B}_{3,i}(k_0)}{Qk_0}\Big|_+\right]
%\nonumber\\&=&
=\frac{m^2E_k}{m_d\, k_zQ^2}\sum_{XY}\delta I_{XY}^B\Big|_{n=3}.\qquad\; \label {eq:BdiffMag}
\eea
where the terms $\delta I^B_{XY}$ were given in Eq.~(\ref{eq:Bmag}).  The typical term in the sum is therefore 
\bea
\overline M_{iB}&\to&-\frac{2m^2}{m_d}\Big[d_1XY +c_1(X_{k_0}Y+Y_{k_0}X) \Big]
%\nonumber\\&&
+d_0\frac{m^2 k_z}{m_d}\Bigg[\frac{1}{E_k}(X_{k_0}Y-Y_{k_0}X)
%\nonumber\\&&\quad\quad
-\frac{(m_d-E_k)}{m_d\,k}(X'Y-Y'X)\Bigg],\qquad
\eea
where $d_0=x_2k_z$, so that these terms will not be zero when integrated over $k_z$.
The trace $B_{3,2}$ is already linear in $Q$ and hence for this term the $d_0$ terms vanish.  

Reviewing the above discussion, the actual calculation proceeds in two steps.  First, keeping the arguments of the structure functions fixed, expand the traces to first order in $Q$ and $\delta_{k_0}\equiv k_0-E_k$. Then make the following substitutions (for the $d_0, c_1,  d_1$ terms respectively):
% \vspace{0.2in}
%
\bea
k_z\delta_{k_0}(X_+Y_- -Y_+X_-)
&\to&\frac{m^2 k_z^2}{m_dE_k}(X_{k_0}Y-Y_{k_0}X)%\\&&
+{\cal D}_2 (X'Y-Y'X)
%\qquad\qquad
\nonumber\\
%\eea
%\bea
Q(X_+Y_- +Y_+X_-)&\to& -\frac{2m^2}{m_d}(X_{k_0}Y+Y_{k_0}X)
\nonumber\\
Q\delta_{k_0}(X_+Y_- +Y_+X_-)&\to& -\frac{2m^2}{m_d}XY,
\eea
where
\bea
{\cal D}_2&=&-\frac{m^2 k_z^2}{m_d^2 \,k}(m_d-E_k)
\eea
and the factor of $k_z$ that is part of $d_0$ has been shown explicitly.  The substitution for the $d_1^h$ term is special
\bea
Q(X_+Y_- &+&Y_+X_-)\to \frac{8m^2a(p^2)}{m_d}  (m_d-E_k)XY.\qquad
\eea

Using these substitutions, and expressing the $\overline M$'s directly in terms of the wave functions $z_\ell$, gives the following leading order results  
%   
%\begin{widetext}
%
\bea
\overline M_{1B}(k)&=&2\pi^2\frac{E_k}{m}\Bigg\{u^2 -\frac18 w^2-\frac{3\,uw}{4\sqrt{2}}-\Big(2-\frac{3m^2}{4k^2}\Big)v_t^2-\frac12v_s^2-\frac1{4\sqrt{2}}\Big(7-\frac{6m^2}{k^2}\Big)v_tv_s
+\frac{m}{k}(\sqrt{2}v_t-v_s)(\sqrt{2}z_0^{--}-z_1^{--})
\nonumber\\
&&\qquad
 +\frac{k}{4\sqrt{2}}(u'w-uw')+\frac{m^2}{2\sqrt{2}\,k}(v_t v_s'-v_t'v_s) 
\nonumber\\
&&\qquad-2u[\delta_+\hat u]_{k_0}-w[\delta_+\hat w]_{k_0} 
+\sqrt{2}\Big(\frac12u[\delta_+ \hat w]_{k_0}- v_t[\delta_- \hat v_s]_{k_0}-v_s[\delta_- \hat v_t]_{k_0}\Big) 
+2v_t[\delta_- \hat v_t]_{k_0} +v_s[\delta_- \hat v_s]_{k_0}
\nonumber\\
&&\qquad 
 +2\, a(p^2) m\bigg[\delta_k\Big(2u^2+ w^2-\frac1{\sqrt{2}}uw\Big)
-m_d\Big(2v_t^2+v_s^2-\frac3{\sqrt{2}}v_tv_s
\Big)\bigg]
+ \Delta M_{1B}(k)\Bigg\} %\label{eq:M1B}
\nonumber\\
\overline M_{2B}(k)&=&2\pi^2\frac{E_k}{m}\Bigg\{u^2-\frac12 w^2-\frac12 v_t^2 -\sqrt{2} v_t v_s
+\frac{\sqrt{2}\,m}{k}(\sqrt{2}z_0^{--}-z_1^{--})v_t
\nonumber\\
&&\qquad
-2u[\delta_+ \hat u]_{k_0} +w[\delta_+ \hat w]_{k_0}-\sqrt{2}\Big[v_t[\delta_- \hat v_s]_{k_0}+v_s[\delta_- \hat v_t]_{k_0}\Big]
+v_t[\delta_- \hat v_t]_{k_0}
\nonumber\\&&\qquad 
-4a(p^2)(E_k-m_d)\Big[\delta_k\Big(u^2-\frac12 w^2\Big)-m_d\Big(\frac12 v_t^2-\sqrt{2}v_tv_s\Big)\Big] +\Delta M_{2B}(k)
\Bigg\}\qquad
\label{eq:M2B}
\eea
\end{widetext}
where $z_{\ell k}=z_\ell'$ and the leading order correction terms are 
\bea
\Delta M_{1B}(k)&=&\frac{E_k-m}{12E_k}\Big[u^2 +\frac{131}{4} w^2 +\frac{29}{2\sqrt{2}} uw \Big]   %
\nonumber\\
%%%%%
\Delta M_{2B}(k)&=& -\frac{E_k-m}{3E_k}\Big[u^2+\frac12 w^2 -\sqrt{2}uw\Big]\, .
\label{eq:DeltaM12B}
\eea
Note the unexpected presence of a leading  $uw$ contribution to $\overline M_{1B}$.  This term does not reduce the the expected nonrelativistic limit, but is cancelled by a similar contribution from $\overline M_{1C}$ which we discuss now.

It is surprising that significant contributions come from the finite terms that depend on the traces ${\cal C}_{3,i}$.  These give the following additional leading contributions 
\begin{widetext}
\bea
\overline M_{1C}(k)&=&2\pi^2\frac{E_k}{m}\Bigg\{-\frac38 w^2+\frac{9\,uw}{4\sqrt{2}}+\Big(1-\frac{m^2}{4k^2}\Big)v_t^2-\frac1{4\sqrt{2}}\Big(5+\frac{2m^2}{k^2}\Big)v_tv_s
+\frac{m}{k}(v_tz_0^{--}+v_sz_1^{--})
\nonumber\\
&&\qquad
 +\frac{k}{4\sqrt{2}}(u'w+3uw') -\frac12kww'
 -\frac{k}{4\sqrt{2}}(v_tv_s'+7v_t'v_s)
 +\frac{m^2}{2\sqrt{2}\,k}(v_t v_s'+3v_t'v_s) 
 -\Big(\frac12-\frac{m^2}{k^2}\Big)v_t v_t'
\nonumber\\
&&\qquad+\frac12m\left(v_t'z_0^{--}+3v_t{z_0^{--}}'+v_s'z_1^{--}+3v_s{z_1^{--}}'\right)+\frac12w[\delta_+\hat w]_{k_0} -\frac1{\sqrt{2}}\Big(u[\delta_+ \hat w]_{k_0}+v_s[\delta_- \hat v_t]_{k_0}\Big) 
\nonumber\\
&&\qquad 
-\frac12v_t[\delta_- \hat v_t]_{k_0}  -a(p^2) m\,m_d\Big(v_t^2+\sqrt{2}v_tv_s
\Big) + \Delta M_{1C}(k)\Bigg\} %\label{eq:M1B}
\nonumber\\
\overline M_{2C}(k)&=&2\pi^2\frac{E_k}{m}\Big\{v_t^2 -\frac{\sqrt{2}\,m}{k}(\sqrt{2}z_0^{--}-z_1^{--})v_t \Big\}\qquad
\label{eq:M2C}
\eea
with only one correction term
\bea
\Delta M_{1C}(k)=-\frac{E_k-m}{4 E_k}\Big[5u^2-\frac54 w^2+\frac{49}{2\sqrt{2}}uw\Big]\, .
\eea
Adding the (B) and (C) contributions together, and rearranging some terms, gives
%changed 1/md to 1/(2m)
%
\bea
\overline M_{1BC}(k)&=&2\pi^2\frac{E_k}{m}\Bigg\{u^2+\frac14w^2- \frac12\Big\{\frac32w^2+kww'\Big\}+\frac1{2\sqrt{2}}\Big\{3uw  +ku'w+kuw'\Big\}-\frac14v_t^2
-\frac12\Big\{\frac32v_t^2+kv_tv_t'\Big\}-\frac12v_s^2
\nonumber\\&&
+\frac{m^2}{2k^2}\Big\{v_t^2+2kv_tv_t'\Big\}-\frac1{\sqrt{2}}\Big\{3v_tv_s+kv_t'v_s+kv_tv_s'\Big\} +\frac{3k}{4\sqrt{2}}(v_tv_s'-v_t'v_s)
+\frac{m^2}{\sqrt{2}\,k^2}\Big\{v_tv_s+kv_tv_s'+kv_t'v_s\Big\}
\nonumber\\&&
-2u[\delta_+\hat u]_{k_0}-\frac12w[\delta_+\hat w]_{k_0} 
-\sqrt{2}\Big( v_t[\delta_- \hat v_s]_{k_0}+v_s[\delta_- \hat v_t]_{k_0}\Big) 
+\frac32v_t[\delta_- \hat v_t]_{k_0} +v_s[\delta_- \hat v_s]_{k_0}
\nonumber\\&&
-m(v_t'z_0^{--}+v_s'z_1^{--})+\frac32m\Big\{v_t'z_0^{--}+v_t{z_0^{--}}'+\frac2{k}v_tz_0^{--}\Big\} +\frac32m\Big\{v_s'z_1^{--}+v_s{z_1^{--}}'+\frac2{k}v_sz_1^{--}\Big\}
\nonumber\\&&
-\frac{m}{k}\Big[\sqrt{2}v_sz_0^{--} +(\sqrt{2}\,v_t + v_s)z_1^{--}\Big]
+2a(p^2)m\Big[\delta_k\Big(2u^2+w^2-\frac1{\sqrt{2}}uw\Big) -m_d\Big(\frac52v_t^2+v_s^2-\sqrt{2}v_tv_s\Big)\Big]
\nonumber\\&&
+\Delta M_{1B}(k)+\Delta M_{1C}(k)\Bigg\}
\nonumber\\
\overline M_{2BC}(k)&=&2\pi^2\frac{E_k}{m}\Big\{u^2-\frac12w^2+\frac12 v_t^2-\sqrt{2} v_t v_s -2u[\delta_+ \hat u]_{k_0} +w[\delta_+ \hat w]_{k_0}-\sqrt{2}\Big[v_t[\delta_- \hat v_s]_{k_0}+v_s[\delta_- \hat v_t]_{k_0}\Big] 
\nonumber\\&&
+v_t[\delta_- \hat v_t]_{k_0}
%\nonumber\\&& 
-4a(p^2)(E_k-m_d)\Big[\delta_k\Big(u^2-\frac12 w^2\Big)-m_d\Big(\frac12 v_t^2-\sqrt{2}v_tv_s\Big)\Big] +\Delta M_{2B}(k)\Big\}
\eea
\end{widetext}
The expression for $\overline M_{1BC}$ has been arranged so that terms in small curly braces integrate to zero (recalling that the volume of integration is $k^2dk/E_k$).  Note that the correct leading $uw$ term (which equals 0) and $w^2$ term are only obtained by the summing the ${\cal B}$ and ${\cal C}$ traces and retaining the $k$ derivative contributions.   This has been discussed already in %the Introduction and in 
Sec.~\ref{sec:GM0}.

\end{document}